\documentclass[fleqn,usenatbib]{mnras}

\usepackage{mathptmx}

\usepackage[T1]{fontenc}
\usepackage{ae,aecompl}

\pdfoutput=1 
\usepackage{savesym}
\usepackage{amsmath}
\savesymbol{iint}
\usepackage{txfonts}
\restoresymbol{TXF}{iint}
\usepackage{natbib, graphicx}
\usepackage{color}
\usepackage{hyperref}
\usepackage{booktabs}
\usepackage{pdflscape}
\usepackage{multirow}
\usepackage{longtable}
\usepackage{pdflscape}
\usepackage{threeparttable}
\usepackage{subcaption}
\usepackage{float}

\title[Stellar parameters using FASMA]{Atmospheric stellar parameters for large surveys using FASMA, a new spectral synthesis package}

\author[M. Tsantaki et al.]{
M. Tsantaki,$^{1}$\thanks{E-mail: \href{mailto:mtsantaki@crya.unam.mx}{mtsantaki@crya.unam.mx}}
D. T. Andreasen,$^{2,3}$
G. D. C. Teixeira,$^{2,3}$
S. G. Sousa,$^{2,3}$ 
N. C. Santos,$^{2,3}$
\newauthor{E. Delgado-Mena,$^{2}$
G. Bruzual$^{1}$}
\\
$^{1}$Instituto de Radioastronom\'ia y Astrof\'isica, IRyA, UNAM, Campus Morelia, A.P. 3-72, C.P. 58089, Michoac\'an, Mexico \\
$^{2}$Instituto de Astrof\'isica e Ci\^encias do Espa\c{c}o, Universidade do Porto, CAUP, Rua das Estrelas, Porto, 4150-762, Portugal \\ 
$^{3}$Departamento de F\'isica e Astronomia, Faculdade de Ci\^encias, Universidade do Porto, Rua Campo Alegre, Porto, 4169-007, Portugal 
}
\date{Accepted 2017 September 29. Received 2017 September 27; in original form 2017 January 31}

\pubyear{2017}

\begin{document}
\label{firstpage}
\pagerange{\pageref{firstpage}--\pageref{lastpage}}
\maketitle

\begin{abstract}
In the era of vast spectroscopic surveys focusing on Galactic stellar populations, astronomers want to exploit the large quantity and good quality of data to derive their atmospheric 
parameters without losing precision from automatic procedures. In this work, we developed a new spectral package, \textit{FASMA}, to estimate the stellar atmospheric parameters 
(namely effective temperature, surface gravity, and metallicity) in a fast and robust way. This method is suitable for spectra of FGK-type stars in medium and high resolution. 
The spectroscopic analysis is based on the spectral synthesis technique using the radiative transfer code, MOOG. The line list is comprised of mainly iron lines in the optical 
spectrum. The atomic data are calibrated after the Sun and Arcturus. We use two comparison samples to test our method, i) a sample of 451 FGK-type dwarfs from the high resolution 
HARPS spectrograph, and ii) the \textit{Gaia}-ESO benchmark stars using both high and medium resolution spectra. We explore biases in our method from the analysis of synthetic spectra 
covering the parameter space of our interest. We show that our spectral package is able to provide reliable results for a wide range of stellar parameters, different rotational velocities, 
different instrumental resolutions, and for different spectral regions of the VLT-GIRAFFE spectrographs, used among others for the \textit{Gaia}-ESO survey. \textit{FASMA} estimates stellar 
parameters in less than 15\,min for high resolution and 3\,min for medium resolution spectra. The complete package is publicly available to the community.
\end{abstract}



\section{Introduction}\label{intro}

In the last decades, due to the growing number of spectroscopic surveys dedicated to the study of the Galactic stellar populations, the number of high quality spectra has increased 
to several hundreds of thousands. This effort has been achieved mainly owing to ground-based surveys in the optical and near infrared, such as APOGEE \citep{Majewski2015}, 
the \textit{Gaia}-ESO Survey \citep[GES;][]{gilmore2012}, GALAH \citep{desilva2015}, SEGUE \citep{Yanny2009}, RAVE \citep{Steinmetz2006}, and LAMOST \citep{Zhao2012}, to name a few. 
The quality of this data is high enough to provide a detailed stellar characterization in terms of their atmospheric parameters and chemical composition. 
Among the large surveys, \textit{Gaia} \citep{Perryman2001} stands out with the goal to provide a census of one billion 
objects of the Milky Way. \textit{Gaia} is mounted with the Radial Velocity Spectrograph \citep{Wilkinson2005} to provide spectra for 
million of stars suitable for kinematic and chemical characterization. 

The success of the above surveys depends on the efficiency of the spectral analysis techniques to provide precise and accurate spectral information in 
the shortest computation time. The stellar atmospheric parameters one can obtain from such studies are the effective temperature ($T_{\mathrm{eff}}$), surface gravity 
($\log g$), metallicity ($[M/H]$), chemical abundances of individual elements, and in turn we infer the evolution of the star itself by determining the stellar mass and 
radius either via calibrations \citep{torres10_mass, santos13} or stellar evolution models \citep[e.g.][]{girardi2002}. 

Furthermore, precise masses and radii of planet host stars are of paramount importance for planetary science as they are essential for the planetary characterization. 
For instance, \cite{gomez2013} showed how the discrepancies in the stellar atmospheric parameters between different spectroscopic techniques impact the planetary properties in the pilot study of the 
planet host WASP-13. Spectral analysis techniques have to reach a precision for the atmospheric parameters to deduce stellar mass and radius within 10\% and 5\% respectively, 
in order to constrain the bulk composition of their planets \citep[e.g.][]{Wagner2011}.

There are several spectral packages in the literature based on different methods/methodologies {to determine} the atmospheric parameters. 
A standard method for FGK-type stars is based on measuring the equivalent widths (EW) of isolated iron lines and by imposing excitation and ionization balance 
\citep[e.g.][]{magrini2013, Mucciarelli2013, Tabernero2013, sousa2014, Andreasen2017}. 
Other methods rely on matching a grid of synthetic spectra, or spectra synthesized on-the-fly with observations 
under a minimization procedure to obtain the best-fit parameters \citep[e.g.][]{valenti96, recio06, allende2006, Lee2008, blanco2014a}.
Each of the above methods has different limitations depending on the resolution of the spectrograph, on the quality of the data (e.g. signal-to-noise, S/N) but also due to the star itself 
(e.g. rotation, spectral type). We have to consider that all spectroscopic methods are affected by the lack of accuracy in the atomic data of the measured absorption lines 
\citep[e.g.][]{Borrero2002}. Moreover, since the above methods are model dependent, we depend on the reliability of the atmospheric models and on the assumptions of which they are built. 
 Yet, even when using the same spectra, atomic data and atmospheric models, there are recent examples which show large discrepancies between different analysis methods for FGK-type stars 
\citep[e.g.][]{jofre2014, Smiljanic2014, Hinkel2016,jofre2017}. It is therefore, very important to understand the biases of the ``spectral analysis pipelines'' before we interpret the physical 
meaning of their results.

Due to the large amount of available spectra in medium and high resolution, there is a high demand for spectral packages to process the data. 
Motivated by that, 
we developed a new package to derive the fundamental atmospheric parameters using the spectral synthesis technique around iron lines in the optical region of the spectrum. 
We named this package \textit{FASMA}\footnote{Acronym for: Fast Analysis of Spectra Made Automatically} which is built around the spectral synthesis code, MOOG \citep{sneden1973}. 
\textit{FASMA} includes other spectral functionalities, among them is the analysis of spectra using the EW which is described in detail in \cite{Andreasen2017}. In this work, we describe a new 
additional driver of \textit{FASMA} for the derivation of atmospheric parameters using the spectral synthesis technique. 
Many from the aforementioned spectroscopic surveys operate on low or medium resolution spectrographs with short wavelength windows in many cases. 
A common practice in such cases is to anchor their results to studies with homogeneous parameters derived from high quality data. 
In the case of the GES, their atmospheric parameters are scaled to the one set by a group of benchmark stars \citep{Heiter2015}. Among the goals of this work is to use the 
GES benchmark sample and define a single methodology which will provide reliable results in both high and medium resolution regimes, even for narrow wavelength windows without the further 
need to calibrate the lower resolution parameters. In addition, we aim to provide parameters for both giants and dwarfs including stars with high rotational velocities. 

\textit{FASMA} is open access and is suitable for automatic analyses. It includes an easy-to-use graphical interface.

In Sect.~\ref{synthesis}, we describe how we obtain a synthetic spectrum with \textit{FASMA}. We present the models included in our spectral package in Sect.~\ref{models}, the line list used 
for this work in Sect.~\ref{linelist}, and the normalization of the observations in Sect.~\ref{normalization}. The minimization process and the methodology we followed to derive the stellar 
parameters are shown in Sect.~\ref{minimization}. We performed several tests on understanding the limitations of our method in Sect.~\ref{error_analysis}. In Sect.~\ref{451_results}, we derive 
parameters for a sample of FGK-type dwarf stars observed with HARPS spectrograph. In Sect.~\ref{results}, we demonstrate the results for the GES benchmark stars using both high and medium 
resolution spectra. In Sect.~\ref{discussion}, we compare our results with the EW method and show how our parameters change if we constrain surface gravity. 

\section{Spectral synthesis}\label{synthesis}

A complete spectral synthesis package capable to determine stellar parameters should contain the following components: 
1) the radiative transfer code, 2) the grids of atmospheric models, 3) the line list, and 4) the minimization procedure.
The principle of our code is to create a synthetic spectrum using a model atmosphere of a set of initial stellar parameters and a given line list. 
Then, it calculates the $\chi^{2}$ between the synthetic and observed spectrum and yields the best-fit parameters through the minimization process.

The synthetic spectra in \textit{FASMA} are created by the radiative transfer code, MOOG (version 2014)\footnote{For the latest version see:
\url{http://www.as.utexas.edu/~chris/moog.html}} within the pre-defined wavelength intervals (see Sect.~\ref{linelist}). To reproduce a realistic stellar spectrum, we 
need to account for the broadening mechanisms due to velocity fields in the atmosphere of the stars. Macroturbulence ($\upsilon_{mac}$) describes the motion in atmospheric 
cells which are larger than the unit of optical depth driven by convection. Macroturbulence should not be confused 
with microturbulence ($\upsilon_{mic}$) which is used to remove possible trends in the atmospheric parameters due to 1-D model deficiencies \citep[for details see][]{Gray2005}. 
We also assume a uniform rotation of the stellar surface measured as the projected rotational velocity ($\upsilon\sin i$).

The macroturbulent and rotational kernels are defined in \citet{Gray2005} and are convolved separately to the flux spectrum. This process is performed outside the MOOG code. 
The macroturbulent profile has two components corresponding to the radial and the tangential motions projected to the line of sight. In this work, we assume that both components are equal 
and behave in the same way across the stellar disk. 
On the other hand, the Doppler shifts of the spectral lines due to stellar rotation depend on the area of the stellar disk where the light crosses. 
Because of these variations from the disk center to the limb, the rotational profile is defined assuming a linear limb darkening law described by a limb darkening coefficient set at 0.6 
which is a good approximation for solar-type stars \citep[e.g.][]{gimenez2006}. 

Finally, to account for the external broadening due to the resolution of the spectrograph, we convolve the synthetic spectrum with a Gaussian profile of full width half maximum equal to 
$\frac{\lambda}{R}$, $\lambda$ is the mean wavelength of each interval and $R$ the resolution of each instrument. All the above broadening mechanisms are applied to each wavelength interval 
separately.

\section{Model atmospheres}\label{models}

\begin{table*}
\begin{center}
\caption{Grids of the provided model atmospheres included in \textit{FASMA}.}
\label{model_table}
\end{center}
\begin{tabular}{lcccccc}
\hline\hline
Models       & $T_{\mathrm{eff}}$ & step & $\log g$ & step  & $[M/H]$ & step \\
             & (K)                & (K)  & (dex)    & (dex) & (dex)   & (dex) \\
\hline
\multirow{3}{*}{ATLAS-APOGEE$^a$} & 3500  -- 12000  & 250  &            &     & -5.0 -- -3.5 & 0.50 \\
                                  & 12500 -- 20000  & 500  & 0.0 -- 5.0 & 0.5 & -3.0 -- 0.75 & 0.25 \\  
                                  & 21000 -- 30000  & 1000 &            &     &  1.0 -- 1.5  & 0.50 \\ 
\hline
\multirow{3}{*}{ATLAS9$^b$}       & 3500 --  12000  & 250  &            &     & -3.0 -- -0.5 & 0.50 \\
                                  & 12500 -- 20000  & 500  & 0.0 -- 5.0 & 0.5 & -0.3 -- 0.3  & 0.10 \\  
                                  & 21000 -- 30000  & 1000 &            &     &  0.5 -- 1.0  & 0.50 \\ 
\hline
\multirow{3}{*}{MARCS$^c$}        & 2500 --  4000   & 100  &            &     & -5.0 -- -3.0 & 1.00 \\
                                  & 4250 --  8000   & 250  & 0.0 -- 5.0 & 0.5 & -2.5 -- -1.0 & 0.50 \\ 
                                  &                 &      &            &     & -0.75 -- 1.0 & 0.25 \\ 
\hline
\multicolumn{7}{l}{$^a$ The ATLAS-APOGEE models use an updated H$_{2}$O line list and abundances from \cite{asplund2005}.} \\
\multicolumn{7}{l}{$^b$ These ATLAS models use solar abundances from \cite{Anders1989}.} \\
\multicolumn{7}{l}{$^c$ MARCS models offer spherical geometry for $\log g$ $<$\,3.0\,dex and $\alpha$ enhancement for $[M/H]$ $<$\,0.0\,dex.} \\ 
\multicolumn{7}{l}{Their solar abundances are from \cite{Grevesse2007}.}
\end{tabular}
\end{table*}

Most spectroscopic methods require stellar atmospheric models that present how physical quantities (mainly temperature, electron density, mean opacity, gas and radiation pressure) 
change at each layer of the atmosphere, i.e. at each optical depth. Therefore, the derived parameters from these methods are indirect measurements and their accuracy depends on the 
reliability of these models. To calculate a model atmosphere can be computationally expensive and for this reason it is commonly preferred to use pre-computed grids of models 
for a set of atmospheric parameters ($T_{\mathrm{eff}}$, $\log g$, $[M/H]$). 

The model atmospheres included in this work are generated by the ATLAS program\footnote{ATLAS models: \url{kurucz.harvard.edu/grids.html}} \citep{kurucz} assuming 
local thermodynamic equilibrium (LTE). We also include the grid of MARCS models\footnote{MARCS models: \url{marcs.astro.uu.se}} \citep{Gustafsson2008} 
obtained with 'standard abundance composition' and the more extended grid for the APOGEE survey based on ATLAS9\footnote{ATLAS-APOGEE models: \url{www.iac.es/proyecto/ATLAS-APOGEE} 
\citep{meszaros2012}.
All grids are based on 1-D atmosphere in LTE which is a reasonable assumption for FGK-type stars with the exception of very metal-poor stars \citep[e.g.][]{ruchti2012}}, 
but they are calculated with different solar abundances. The parameter space of each grid of models is shown in Table~\ref{model_table}. 

To select a specific model for a set of parameters ($T_{\mathrm{eff}}$, $\log g$, $[M/H]$), we search the eight closest neighboring models from the grid and then, 
we interpolate their physical properties linearly to the parameters of the desired model on-the-fly \citep[the same function is used in][]{Andreasen2017}. 

\section{The line list}\label{linelist}

\begin{figure}
  \centering
   \includegraphics[width=1.05\linewidth]{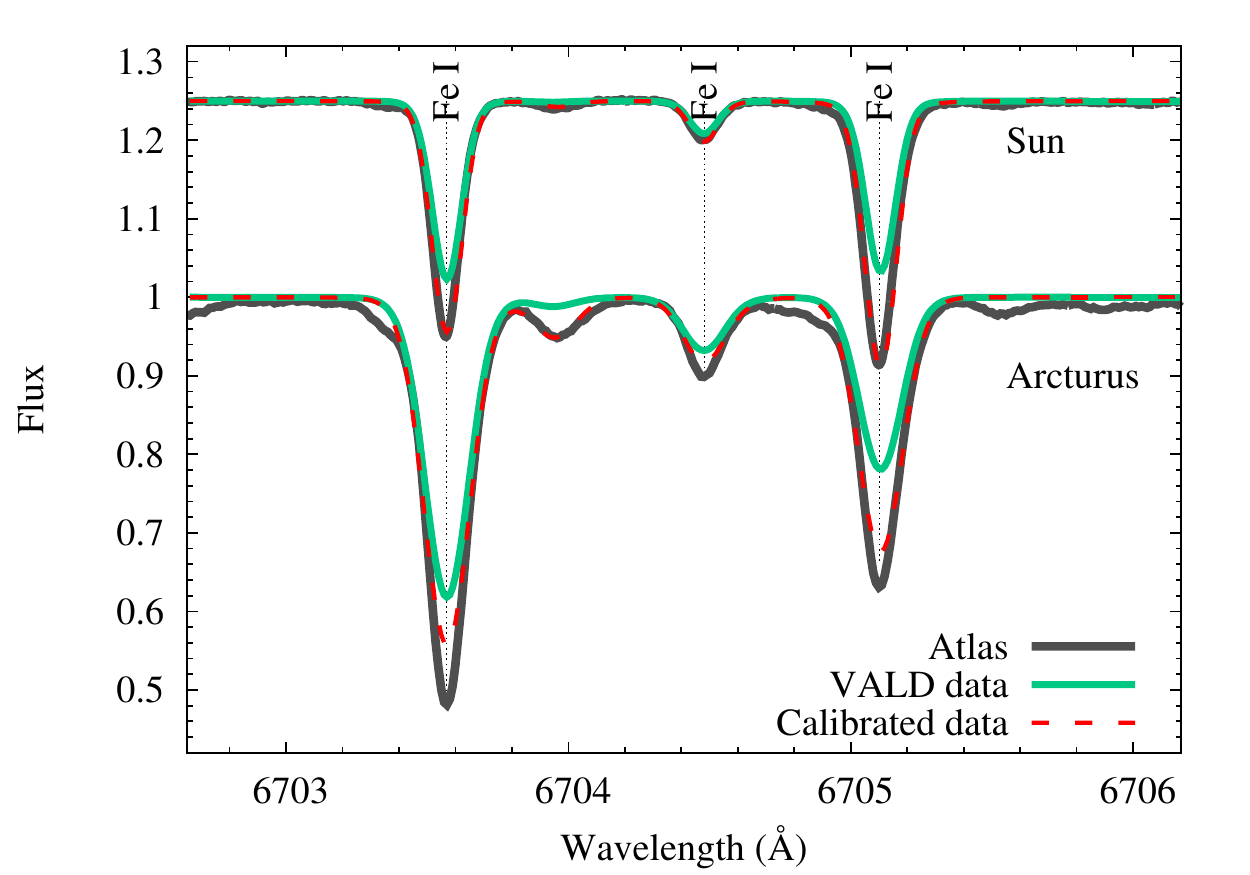}
  \caption{The synthetic spectrum with VALD3 data (green line), the synthetic spectrum after the calibration of the $\log gf$ values (red dashed line) and the corresponding 
  Solar and Arcturus Atlases (black line) for an interval of our line list.}
  \label{calibration}
  \end{figure}

For an accurate spectral synthesis, atomic and molecular data of all lines in the wavelength intervals must be as accurate 
and complete as possible. The lines to be used for the parameter determination must be carefully selected because of their different sensitivity to stellar parameters. 
There are several criteria to consider on how to select the spectral regions for the synthesis. Some authors use large wavelength intervals 
for their analysis \citep[e.g. SME;][]{valenti05}, others mask the areas around individual lines \citep[e.g. iSpec;][]{blanco2014a}. 
In this work, we define regions around iron lines of a few \AA{}ngstr\"{o}ms wide using the same approach as in our previous work \citep{Tsantaki2014}.

One of our goals of this work is to create a line list which covers a wide range in the optical and therefore can be applied to spectra obtained by 
various spectrographs. Moreover, we keep in mind that the majority of the spectra for the GES are obtained for specific set-ups of the VLT-GIRAFFE spectrograph 
(mainly the HR10, HR15n, and HR21 set-up are used for FGK-type stars). Therefore, our wavelength coverage is set to include these spectral areas: 
5399--5619\,\AA{} and 6470--6790\,\AA{}. We exclude the HR21 range (8484--9001\,\AA{}) from this work because it includes the strong \ion{Ca}{ii} triplet and we have to consider 
the absorption from the triplet to affect neighboring lines which significantly suppress the continuum. In this case, dividing the spectrum into small wavelength intervals is not optimal 
because in this region only a few lines will not be affected by the absorption of the triplet and a different approach from this methodology should be used such as in \cite{Kordopatis2011} 
where the whole wavelength region was used.

For our line list, we first selected all iron lines in the regions 5399--5619\,\AA{} and 6470--6790\,\AA{} from the Vienna Atomic Line 
Database\footnote{VALD3: \url{http://vald.inasan.ru/~vald3/php/vald.php}} 
\citep[VALD3;][]{piskunov95, kupka99, Ryabchikova2015} and excluded the very weak ones that are present in the Sun with a line depth smaller than 1\% relative to the continuum, calculated from 
a synthetic solar spectrum.

We then queried for all atomic and molecular lines inside intervals of $\pm$\,2\,\AA{} around the iron lines. 
We included the hyper-fine splitting components for the elements Mn and Co from the Kurucz line lists\footnote{\url{http://kurucz.harvard.edu/linelists}}. 
The extracted atomic data were obtained for all the predicted transitions for a star with solar parameters and for a K-type star ($T_{\mathrm{eff}}$\,=\,4400\,K) to include lines 
from both spectral types. We merged both line lists into one after removing duplicates. The overlapping intervals were also merged into larger ones.

The length of the intervals is wide enough to include lines broadened by the instruments and by stellar rotation. For instance, a Doppler velocity of 50\,km\,s$^{-1}$ will broaden 
a line in the middle of the optical spectrum by $\Delta\lambda$ $\sim$\,1\,\AA{}. On the other hand, an instrumental resolution of 17\,000 (e.g. GIRAFFE) cannot resolve lines separated less than 
$\Delta\lambda$ $\sim$\,0.3\,\AA{} which corresponds to a velocity broadening of $\sim$\,18\,km\,s$^{-1}$. 
The regions were later cut or expanded at the edges by eye in order to discard neighboring lines or to keep enough continuum points using the solar spectrum as a reference. 
An example of our intervals is shown in Fig.~\ref{calibration}.

Molecular data of the most abundant molecules in solar-type stars (C$_{2}$, CN, OH, and MgH) were also obtained from VALD3 using the same requests as for 
the atomic data. The final line list contains 1187 lines mostly neutral and singly ionized atoms, as well as diatomic molecules. 
However, the strongest lines which dominate the intervals are mainly iron. From the 249 unique lines bigger than 10\,m\AA{} (of the Sun) of our line list, 159 are iron (Table~\ref{line_data}).

\subsection{Calibration of the atomic and molecular data}

There are several broadening mechanisms (namely natural, collisional, and thermal) which contribute to the final line profile. Each of these processes has its own coefficient which we generally 
refer as atomic data. Among them, the most essential data to simulate an atomic line are the transition probabilities (oscillator strengths, $\log gf$). 
This data is usually calculated from laboratory or semi-empirical estimates. Even though many improvements have occurred recently, large discrepancies still appear when comparing to the atomic data 
derived from calibrations, i.e. astrophysical data. We determine astrophysical $\log gf$ values for our line list to avoid uncertainties that may arise from the aforementioned estimations 
but also such calibrations will mitigate systematic errors due to imperfections of the model atmospheres.

Moreover, we consider the broadening due to the collisional interaction between the atoms and hydrogen known as the van der Waals damping ($\Gamma_{6}$). 
The damping coefficients for most lines are taken from \cite{barklem2000} and had to be adjusted by hand only in a couple of cases. In lack of these values, $\Gamma_{6}$ are taken 
from VALD and if VALD does not provide these values, they are then calculated using the Unsold formula \citep{unsold1955} (MOOG option: \textit{damping=1}).    

A common practice to calibrate atomic data is to use high resolution spectra from stars with very well constrained parameters. The Sun is a standard choice. However, if the sample 
contains stars far from solar parameters, the solar calibrated atomic data may introduce uncertainties. Thus, in various works the authors use more than one star, usually of different 
spectral type and luminosity class \citep{Shetrone2015, boeche2016}. 

We select Arcturus (K-type giant) apart from the Sun to improve the transition probabilities in an inverted analysis meaning that we vary the $\log gf$ values to fit the observations. 
We use the National Solar Observatory Atlas \citep{wallace2011} and a Kurucz model atmosphere with the typical solar parameters ($T_{\mathrm{eff}}$ = 5777\,K, 
$\log g$ = 4.44\,dex, $[Fe/H]$ = 0.0\,dex, $\upsilon_{mic}$ = 1.0\,km\,s$^{-1}$, $\upsilon_{mac}$ = 3.21\,km\,s$^{-1}$, and $\upsilon sini$ = 1.9\,km\,s$^{-1}$). 
The solar chemical abundances used in this work are taken from \cite{Anders1989} ($\log_{\epsilon}(Fe)$\,=\,7.47\,dex). For Arcturus, we use the atlas spectrum from \cite{Hinkel2000} 
and the atmospheric parameters as provided by the GES benchmark stars analysis ($T_{\mathrm{eff}}$ = 4286\,K, $\log g$ = 1.64\,dex, $[Fe/H]$ = --0.53\,dex, 
$\upsilon_{mic}$ = 1.25\,km\,s$^{-1}$, $\upsilon_{mac}$ = 5.07\,km\,s$^{-1}$, and $\upsilon sini$ = 3.8\,km\,s$^{-1}$). The chemical abundances of other elements for Arcturus are 
taken from \cite{jofre2015}. 

We used \textit{FASMA} to adjust the atomic data as free parameters to match both atlases at the same time under a $\chi^{2}$ minimization (the same algorithm 
as in Sect.~\ref{minimization}). For each line, the minimization is performed not in the whole interval but around a smaller one of $\pm$0.5\,\AA{} where all lines inside are minimized 
at the same time to account for the blending. Because our knowledge of the solar parameters is far more accurate than for Arcturus, we give higher weights to the solar 
flux points (50\% higher weights) during the minimization process. Finally, we performed extra corrections to the lines which were not fit properly after visual inspection. 
For 50 lines we obtained $\log gf$ values smaller than -15 which indicates that these lines are too weak and they are not detectable for our analysis. Therefore, we excluded these lines reaching a 
total of 1137 for our line list. 

There are other approaches for the atomic data calibration in the literature, such as adjusting the $\log gf$ values from the theoretical EW values to match with the EW measured from the observed 
spectra \citep[e.g.][]{sousa2008, boeche2016, Andreasen2016}. However, measuring the EW in high precision for blended lines of Arcturus can be quite challenging. 

The fact that this line list produces reliable results compared to the benchmark values (see next Sections) is a strong indication that this calibration works. Moreover, 
the improvement of the fit of the synthetic spectra for both atlases using the calibrated data shows that our atomic data are refined (see an example of the calibration results in 
Fig.~\ref{calibration}). The $\log gf$ calibration procedure is provided with \textit{FASMA}.

\section{Normalization}\label{normalization}

In order to match the synthetic spectrum with the observed, the continuum points have to be well defined. \textit{FASMA} performs local 
normalization for the adopted intervals. 
Before the normalization process, we exclude cosmic rays that appear above the continuum in an automatic way by filtering points higher than 3\,$\sigma$ values from the median flux of the 
interval. 
Since the intervals for this work are in length of a few \AA{}ngstr\"{o}ms, a linear normalization is sufficient. 
We select 10 points with the highest flux values in each interval and fit them with a line. 
Then, we divide all points to this line. However, in cases of noisy spectra, a linear fit to the 
maximum values of the flux, leads to an overestimation of the continuum. We apply a correction to the flux depending on the noise level defined by the signal-to-noise 
ratio\footnote{Signal-to-noise is calculated per pixel throughout this paper.} which is calculated automatically from selected spectral regions\footnote{For this task, we use the 
PyAstronomy function, \texttt{estimateSNR}, where each region for the S/N calculation is divided into subsections and is fitted using a second degree polynomial. 
The S/N is then computed by the fit divided by the reduced $\chi^{2}$ of each subsection.}. 
The final flux is adjusted to the following correction: 
\begin{equation}
 \mathrm{flux\_corrected} = \mathrm{flux\_normalized} + \varepsilon \cdot \mathrm{noise},  
\end{equation}
where flux\_$\mathrm{normalized}$ is the linearly normalized flux from the maximum flux points, noise corresponds to $\tfrac{1}{S/N}$, and $\varepsilon$ is a scaling factor of 0.5,\,1.0,\,1.5,\,2.0 
depending on the S/N value. The $\varepsilon$ value is estimated empirically by visual inspection of spectra of S/N values from 20 to 500.

\begin{figure}
  \centering
   \includegraphics[width=1.05\linewidth]{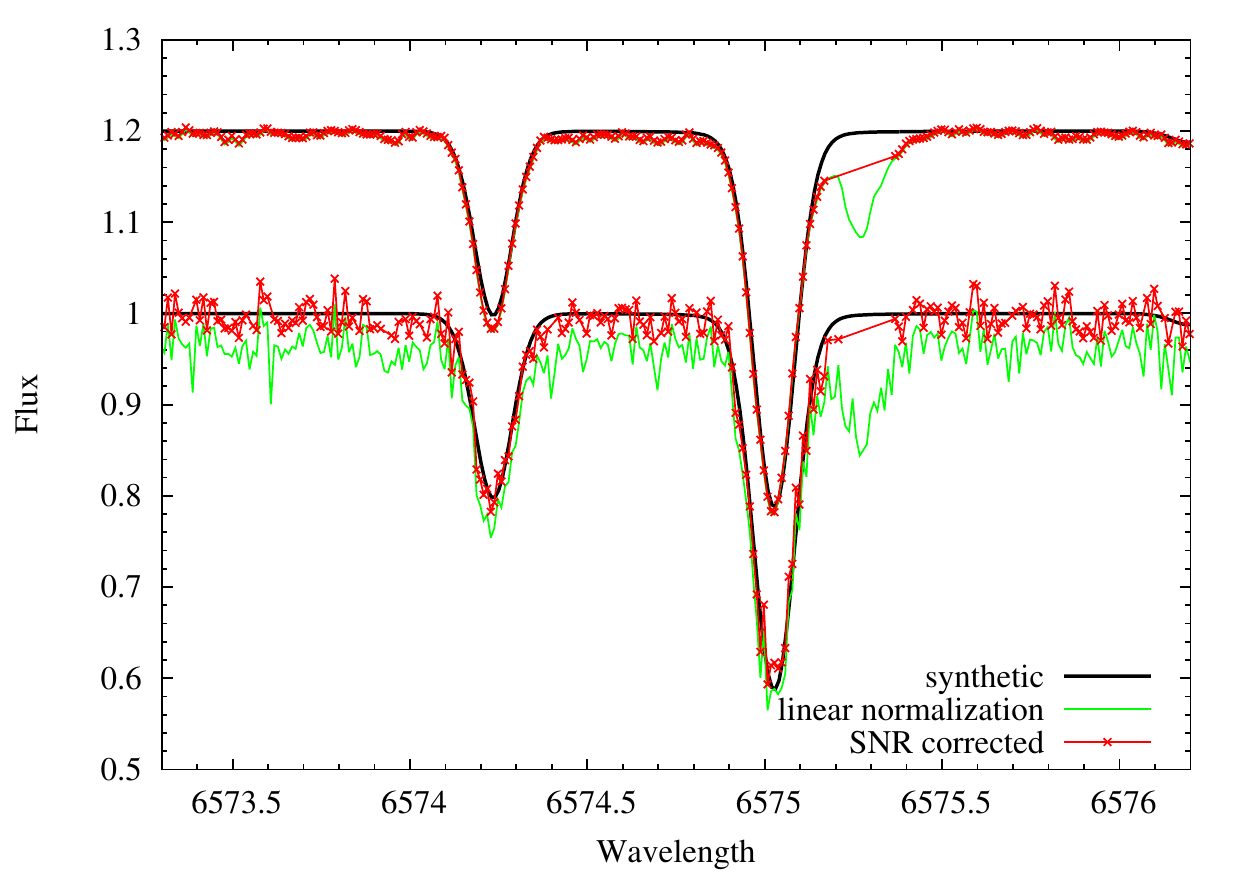}
  \caption{Normalization of the solar spectrum for one of our wavelength intervals. The upper spectrum has a S/N of $\sim$200 and the bottom $\sim$50. 
  We plot for both cases the solar synthetic spectrum (black line), the observed spectrum with a linear normalization (green line) and the observed spectrum 
  with a continuum correction and excluding bad points (red line points). The $\epsilon$ factor for the upper case is 1 and the latter is 3.}
  \label{snr}
  \end{figure}

Moreover, we apply a rough filter to exclude ``bad lines'', such as telluric or unidentified lines. Even though the optical range is not severely contaminated by telluric lines, 
we expect some weak lines to appear. 
This filter works only during the minimization process (with the \textit{refine} option on) using a synthetic spectrum to define the continuum points. 
The synthetic spectrum is obtained with the best-fit parameters of a first run of the minimization procedure to be realistically close to the observations. 
We discard the points of our observations which show 3\% difference from the synthetic ones and at the same time, the corresponding model points have to be close to the maximum flux 
($>$ 0.98) to ensure that we are not excluding any line points (see example in Fig.~\ref{snr}). Then, the minimization routine starts over. 

\section{Minimization}\label{minimization}

\textit{FASMA} includes the parameter optimization procedure based on the Levenberg-Marquardt algorithm \citep{Marquardt63} to solve the nonlinear least-squares problem, 
yielding the parameters that minimize the $\chi^{2}$:

\begin{equation}\label{x2} 
\chi^{2} = \sum_{i=1}^{N} \frac{(Obs _{i} - Synth _{i})^{2}}{\sigma_{i}^{2}}
\end{equation}
where $Obs$ refers to the observed flux points, $Synth$ to the synthetic ones, $\sigma$ to the error on the observed flux, and $N$ to the number of flux points within our defined intervals. 
The Levenberg-Marquardt technique combines the gradient descent method for searches that approach the minimum from far and the expansion method as the search converges.
Far from the solution, the parameters are updated by steps of scaled negative gradient towards the steepest-descent direction and close to the minimum, we assume the least squares function 
is approximately quadratic and calculate the minimum there. Convergence is achieved when at least one of the following criteria is satisfied: i) the relative error in the sum of squares is less 
than 10$^{-5}$, or ii) the relative error in the parameters between two consecutive iterations is less than 10$^{-4}$. These values are set empirically. 
Smaller values indicate better precision but require more computational time. For instance, a decrease of a factor of ten to these values, in the case of the Sun, would only change 
$\log g$ to the third decimal.

The user can set the following free parameters: $T_{\mathrm{eff}}$, $\log g$, $[M/H]$, $\upsilon_{mic}$, $\upsilon_{mac}$, and $\upsilon\sin i$. 
Metallicity in this work is defined as the average abundance of all elements with atomic number higher than two, producing absorption in our spectral regions. 
We could assume that $[M/H]$ is an approximation to $[Fe/H]$ because the dominant lines in our regions are the iron lines. 
However, for metal-poor stars, the overall metallicity can be enhanced by other elements (relative to iron), and in that case the previous assumption does 
not hold \citep[e.g.][]{adibekyan12}. 
Once the atmospheric parameters are derived, \textit{FASMA} is able to calculate the iron abundance for a given star. Having the parameters fixed, 
we calculate the synthetic spectrum and compare with the observed by changing only the iron abundance through a $\chi^{2}$ minimization to obtain the best-fit iron abundance. 
In future releases we intend to expand this analysis to other elements present in our chosen intervals.

\subsection{Methodology}\label{methodology}

\textit{FASMA} offers the option to the user either to provide initial guesses for the parameters or set the spectral type and luminosity class which are translated 
to a rough estimation of $T_{\mathrm{eff}}$ and $\log g$. However, in many cases there is little prior information on the atmospheric parameters of the star which means the 
starting point of the minimization procedure could be far from reality. This will affect the computation time since more iterations are required to reach the final solution 
but could also impact the final solution in case the minimum is a local one and not global. 

Our goal is to create a procedure which will be as independent as possible from initial conditions and for this reason, we performed all tests in this work assuming a general 
case where we have no prior information on the parameters of our sample. Therefore, all initial conditions in this work are set to solar values.

Moreover, some parameters are tied together making the problem degenerate. For example, macroturbulence and rotation are difficult to be 
distinguished using the standard minimization procedure for low rotational velocities (approximately below 5\,km\,s$^{-1}$). 
For this reason, we set $\upsilon_{mac}$ as a fixed parameter and initially set to solar value. We do the same for microturbulence as this parameter varies in a small range (0 -- 2\,km\,s$^{-1}$) 
for our sample. After a first minimization run, we obtain the best-fit values for $T_{\mathrm{eff}}$, $\log g$, $[M/H]$, and $\upsilon \sin i$ and using solar values for $\upsilon_{mac}$ 
and $\upsilon_{mic}$. We then refine our results with a second minimization process starting from the previous best-derived parameters and with new, updated $\upsilon_{mac}$ and $\upsilon_{mic}$ 
values calculated from empirical relations described below. During the second minimization run, $\upsilon_{mac}$ and $\upsilon_{mic}$ are changed at each iteration according to these 
empirical relations. 

Microturbulence is shown observationally to correlate mainly with $T_{\mathrm{eff}}$ and $\log g$ for FGK-type stars \citep[e.g.][]{nissen, adib2, ramirez13}. 
We set $\upsilon_{mic}$ according to the empirical correlation of our previous work for dwarf stars \citep{tsantaki13} and for the giant stars we use the calibration of \cite{Adibekyan2015}. 
Macroturbulence is a broadening mechanism which also correlates with atmospheric parameters (mainly $T_{\mathrm{eff}}$). 
We set $\upsilon_{mac}$ in our analysis following the relation of \cite{Doyle2014} for dwarf stars and of \cite{Hekker2007} for giants. 
The whole procedure is automatic when the user has the \textit{refine} option on. Otherwise, the final parameters are calculated with solar $\upsilon_{mac}$ and $\upsilon_{mic}$ values. 
The differences between the first and second run depend how far from solar values $\upsilon_{mac}$ and $\upsilon_{mic}$ are and can reach up to 200\,K for $T_{\mathrm{eff}}$, 0.1\,dex for 
$\log g$, and 0.2\,dex for $[M/H]$. 

\section{Internal error analysis}\label{error_analysis}

A careful error analysis should include the numerical precision errors in the minimization, errors in the flux, 
errors in the model assumptions, any degeneracies between the parameters, imperfect atomic data, non-LTE effects, etc., and combining all can be quite complicated.

The elements of the covariance matrix quantify the statistical errors on the best-fit parameters arising from the statistical fluctuations of the data. 
\textit{FASMA} provides the statistical uncertainties, i.e. the variances, from the diagonal terms of the covariance matrix of the best-fit solution. 
These errors are highly dependent on the flux errors ($\sigma_{i}$ in Eq.~\ref{x2}). Since flux errors for each flux measurement are not usually provided by the spectrographs, 
we assign an arbitrary flux error. It is possible though, to extract information of the uncertainties from the fit. Assuming the fit is perfect, i.e. the reduced $\chi^{2}$ is equal to unity, 
the uncertainties on the flux become: 
\begin{equation}\label{scaled}
\sigma_{scaled} = \sqrt{\chi^{2}_{red}} \; \; \sigma_{covar} 
\end{equation}
where $\chi^{2}_{red}$ is the reduced $\chi^{2}$ and $\sigma_{covar}$ are the uncertainties from the covariance matrix. 
Even if we use a more realistic flux error, e.g. proportional to the S/N, the re-scaled errors will not change in Eq.~\ref{scaled}.
The errors from the covariance matrix however, are unrealistically small as they do not account for systematic errors. 

On the other hand, Monte Carlo approximations would give more reliable error estimations but they are computationally expensive when we are dealing with more than a 
handful of stars. We used a different approach to estimate the errors by changing each of the free parameters of a certain value and then calculating how the other parameters vary. 
In particular, we add to the final $T_{\mathrm{eff}}$ $\pm$50\,K and run the minimization again by setting free only the others parameters ($\log g$, $[M/H]$ and $\upsilon\sin i$). 
We change in turn each of the other parameters ($\log g$ $\pm$ 0.1\,dex, $[M/H]$ $\pm$ 0.05\,dex, and $\upsilon\sin i$ $\pm$ 0.5\,km\,s$^{-1}$) separately. 
These values are reported as the average precision errors in high resolution studies as for example for the sample of \cite{sousa2008} in Sect.~\ref{451_results}. 
This procedure is summed to eight minimization processes (two per parameter) and results in six values per parameter. 
We assign the error of each parameter by the standard deviation of the six values.

In the following section, we test \textit{FASMA} for a wide sample of synthetic spectra and explore how our parameters are affected by different characteristics, such as: 
i) the initial conditions, ii) the S/N, iii) spectral resolution, and iv) rotation. 

\subsection{Tests on synthetic spectra}\label{synthetic_test}

\begin{figure}
 \centering
 \includegraphics[width=1.1\linewidth]{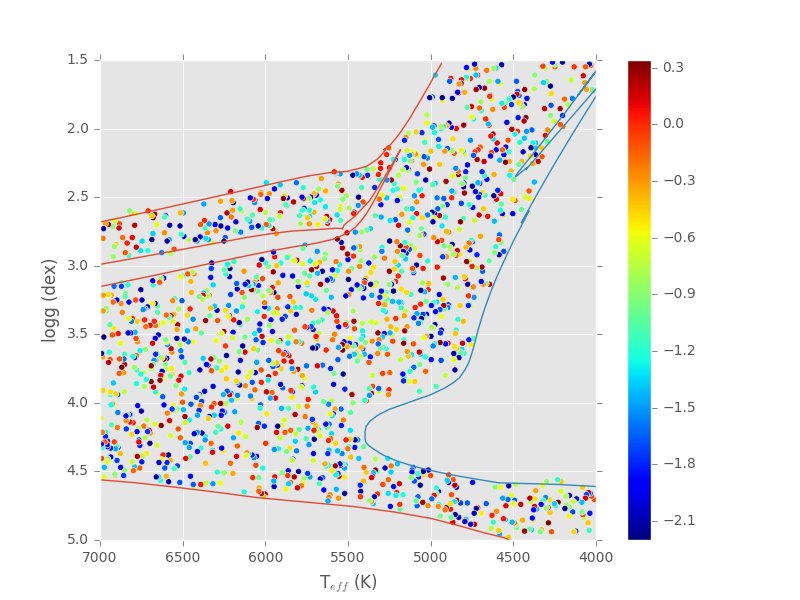}  
 \caption{The Hertzsprung-Russell diagram of the parameter space of our synthetic spectra. The isochrones correspond to Z\,=\,0.035 and 12.7\,Gyr (blue line), and Z\,=\,0.0001 and 1\,Gyr 
 (red line). The plot is color coded to metallicity.}
\label{fig_synth_sample}
\end{figure}

\begin{table*}
\caption{Mean differences ($\Delta$), median differences (med), standard deviations ($\sigma$), and mean absolute deviation (MAD) for the sample of 1700 synthetic spectra.}
\label{synthetic_table}
\scalebox{0.8}{
\begin{tabular}{llcccccccccccccccccccc}
\hline\hline
S/N & set-up &               \multicolumn{4}{c}{$T_{\mathrm{eff}}$ (K)}             & \multicolumn{4}{c}{$\log g$ (dex)}            & \multicolumn{4}{c}{$[M/H]$ (dex)}           & \multicolumn{4}{c}{$[a/Fe]$ (dex)}          & \multicolumn{4}{c}{$\upsilon\sin i$ (km\,s$^{-1}$)}      \\
    &        & $\Delta$ & med & $\sigma$ & MAD & $\Delta$ & med & $\sigma$ & MAD & $\Delta$ & med & $\sigma$  & MAD & $\Delta$ & med & $\sigma$ & MAD & $\Delta$ & med & $\sigma$  & MAD \\
\hline

\multirow{3}{*}{150} & High-res & 0  & 0  & 11  & 4   & 0.00 & 0.00 & 0.03 & 0.01 & 0.00  & 0.00  & 0.01 & 0.00 & 0.00 & 0.00 & 0.01 & 0.00 & -0.1 & 0.0 & 0.3 & 0.1 \\
                     & HR10     & -3 & -4 & 23  & 15  & 0.02 & 0.03 & 0.06 & 0.04 & -0.01 & -0.01 & 0.01 & 0.02 & 0.00 & 0.00 & 0.01 & 0.01 & 0.1  & 0.1 & 0.0 & 0.0 \\
                     & HR15n    & 6  & 3  & 103 & 46  & 0.01 & 0.01 & 0.21 & 0.08 & 0.00  & 0.00  & 0.09 & 0.04 & 0.00 & 0.00 & 0.03 & 0.02 & 0.6  & 0.8 & 1.2 & 1.0 \\
\hline
\multirow{3}{*}{100} & High-res & 1  & 0  & 7   & 16  & 0.00 & 0.00 & 0.03 & 0.01 & 0.00  & 0.00 & 0.01 & 0.01 & 0.00 & 0.00 & 0.01 & 0.01 & -0.1 & 0.0 & 0.4 & 0.2 \\
                     & HR10     & -5 & -4 & 27  & 18  & 0.02 & 0.03 & 0.07 & 0.05 & -0.01 & 0.00 & 0.01 & 0.02 & 0.01 & 0.00 & 0.02 & 0.01 & 0.1  & 0.1 & 0.0 & 0.0 \\
                     & HR15n    & 13 & 3  & 114 & 61  & 0.02 & 0.01 & 0.20 & 0.10 & 0.00  & 0.00 & 0.04 & 0.09 & 0.00 & 0.00 & 0.02 & 0.04 & 0.4  & 0.8 & 1.1 & 1.5 \\
\hline
\multirow{3}{*}{80}  & High-res & 0  & -1 & 14  & 7   & 0.00 & 0.00 & 0.03 & 0.02 & 0.00  & 0.00 & 0.01 & 0.01 & 0.00 & 0.00 & 0.01 & 0.00 & -0.1 & 0.0 & 0.4 & 0.2 \\
                     & HR10     & 11 & 11 & 35  & 24  & 0.06 & 0.06 & 0.09 & 0.06 & 0.01  & 0.01 & 0.02 & 0.02 & 0.00 & 0.00 & 0.02 & 0.01 & 0.1  & 0.1 & 0.0 & 0.0 \\
                     & HR15n    & 6  & 2  & 143 & 75  & 0.01 & 0.01 & 0.26 & 0.13 & 0.00  & 0.00 & 0.12 & 0.06 & 0.00 & 0.00 & 0.05 & 0.03 & 0.3  & 0.7 & 1.2 & 1.6 \\
\hline
\multirow{3}{*}{50}  & High-res & -1 & -1 & 17  & 10  & 0.00 & 0.00 & 0.04 & 0.02 & 0.00  & 0.00 & 0.01 & 0.01 & 0.00 & 0.00 & 0.01 & 0.01 & -0.1 & 0.0 & 0.2 & 0.4 \\
                     & HR10     & 12 & 11 & 44  & 31  & 0.05 & 0.06 & 0.11 & 0.07 & 0.01  & 0.01 & 0.03 & 0.02 & 0.00 & 0.00 & 0.03 & 0.02 & 0.1  & 0.1 & 0.0 & 0.0 \\
                     & HR15n    & 9  & 4  & 189 & 106 & 0.01 & 0.01 & 0.33 & 0.19 & 0.00  & 0.00 & 0.15 & 0.09 & 0.00 & 0.00 & 0.07 & 0.04 & 0.1  & 0.7 & 1.5 & 2.0 \\
\hline
\multirow{3}{*}{20}  & High-res & 0  & 0  & 36  & 22  & 0.00 & 0.00 & 0.05 & 0.03 & 0.00  & 0.00  & 0.02 & 0.03 & 0.01 & 0.00 & 0.01 & 0.02 & 0.0 & -0.1 & 0.7 & 0.4 \\
                     & HR10     & 12 & 13 & 102 & 67  & 0.05 & 0.06 & 0.22 & 0.14 & 0.01  & 0.01  & 0.07 & 0.05 & 0.00 & 0.00 & 0.06 & 0.04 & 0.1 & 0.0  & 0.0 & 0.0 \\
                     & HR15n    & -2 & 2  & 350 & 217 & 0.00 & 0.01 & 0.61 & 0.39 & -0.01 & -0.01 & 0.29 & 0.18 & 0.00 & 0.00 & 0.15 & 0.10 & -0.5 & 1.1 & 2.8 & 2.3 \\
\hline
\end{tabular}}
\end{table*}

Before analyzing real stars, we evaluate the performance of \textit{FASMA} by using synthetic spectra. This test will show possible correlations between the derived parameters, 
the efficiency of our minimization procedure, interpolation errors, and S/N dependences. We define a sample of ``synthetic stars'' covering the following parameters space: 
4000 $< T_{\mathrm{eff}} <$ 7000\,K, 1.5 $< \log g <$ 5.0\,dex, and --2.2 $< [M/H] <$ 0.5\,dex. We randomly select the parameters for the synthetic spectra to fall within two isochrones 
\cite[PARSEC isochrones;][]{bressen2012} corresponding to different ages and metallicities to recreate different stellar populations of our Galaxy. The total amount of parameters we selected 
are 1700 and the sample is shown in Fig.~\ref{fig_synth_sample}.

We enhance the abundances of alpha elements (O, Ne, Si, S, Ar, Ca, and Ti) with respect to iron according to the Galactic observations \cite[e.g.][]{Adibekyan2011}. 
In particular, we select a random $[\alpha/Fe]$ value from each bin below and for each metallicity range in a similar manner as in the work of \cite{Kordopatis2011}: 
\begin{itemize}
 \item $[\alpha/Fe]$ = 0.0\,dex     for $[M/H]$ > 0.0\,dex 
 \item 0.0 <$[\alpha/Fe]$< 0.1\,dex for \,--0.25 < $[M/H]$ < 0.0\,dex 
 \item 0.1 <$[\alpha/Fe]$< 0.2\,dex for \,--0.50 < $[M/H]$ < --0.25\,dex 
 \item 0.2 <$[\alpha/Fe]$< 0.3\,dex for \,--0.75 < $[M/H]$ < --0.50\,dex 
 \item 0.3 <$[\alpha/Fe]$< 0.4\,dex for \,--0.75 < $[M/H]$\,dex  
\end{itemize}

To create the synthetic spectra as realistic as possible, we include different microturbulence and macroturbulence velocities, depending on spectral type and luminosity class 
(see Sect.~\ref{methodology}). Rotation of 2\,km\,s$^{-1}$ is added to all spectra. The above set of synthetic spectra is convolved with three resolution kernels and are created with different 
parts of the line list corresponding to i) high resolution (R\,=\,78\,000) using the complete line list, ii) medium resolution (R\,=\,17\,000) using the line list which corresponds to the 
HR10 GIRAFFE set-up, and iii) medium resolution (R\,=19\,800) using the line list which corresponds to the HR15n GIRAFFE set-up. The high resolution spectra have wavelength step of 0.01\AA{} 
and the medium resolution spectra have 0.05\AA{}. For each of the three sets, we add Gaussian noise which corresponds to S/N values of 150, 100, 80, 50, and 20 (pixel$^{-1}$). 

\begin{figure*}%
 \centering
 \begin{minipage}{0.33\textwidth}
  \includegraphics[width=5.2cm, height=4.5cm]{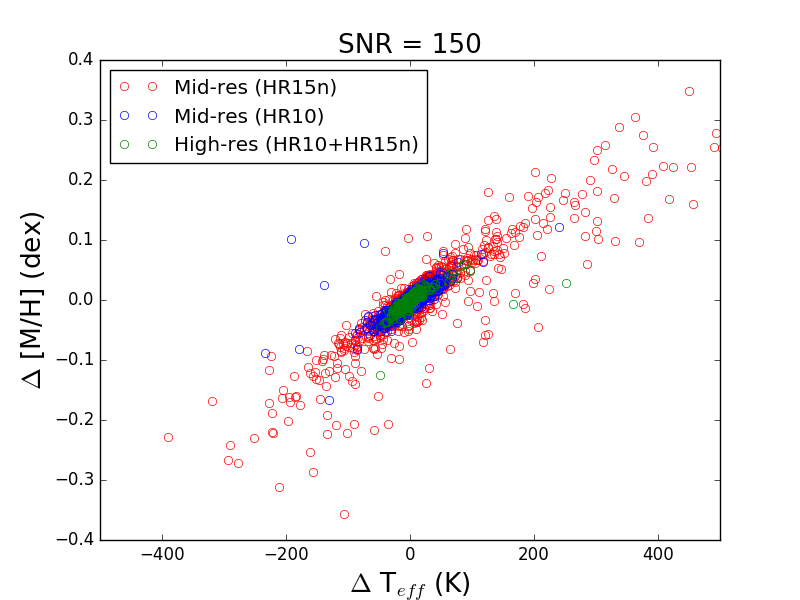} 
 \end{minipage}
 \begin{minipage}{0.33\textwidth}
  \includegraphics[width=5.2cm, height=4.5cm]{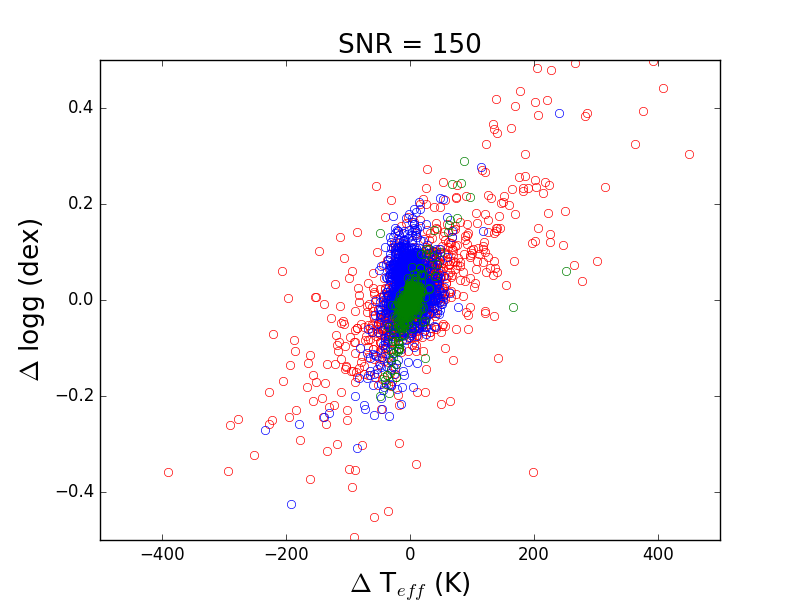} 
 \end{minipage}
 \begin{minipage}{0.33\textwidth}
  \includegraphics[width=5.2cm, height=4.5cm]{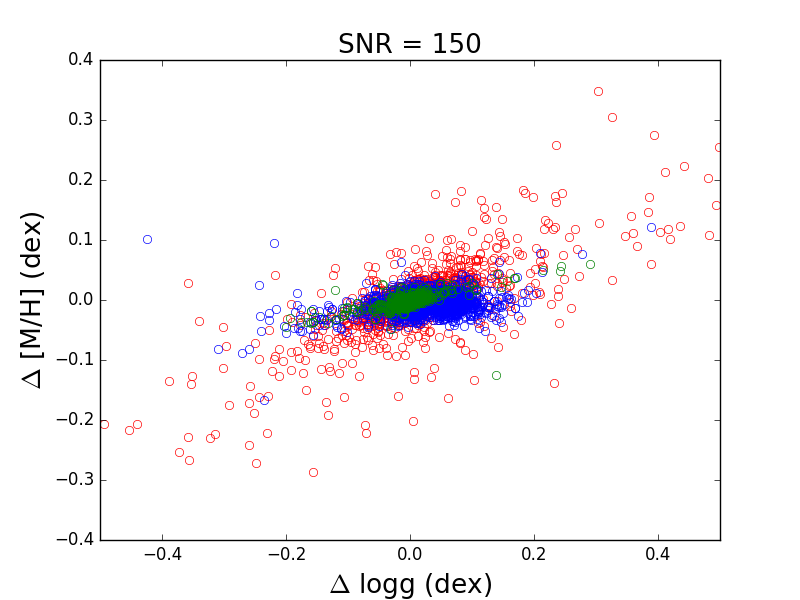} 
 \end{minipage}
 \centering
 \begin{minipage}{0.33\textwidth}
  \includegraphics[width=5.2cm, height=4.5cm]{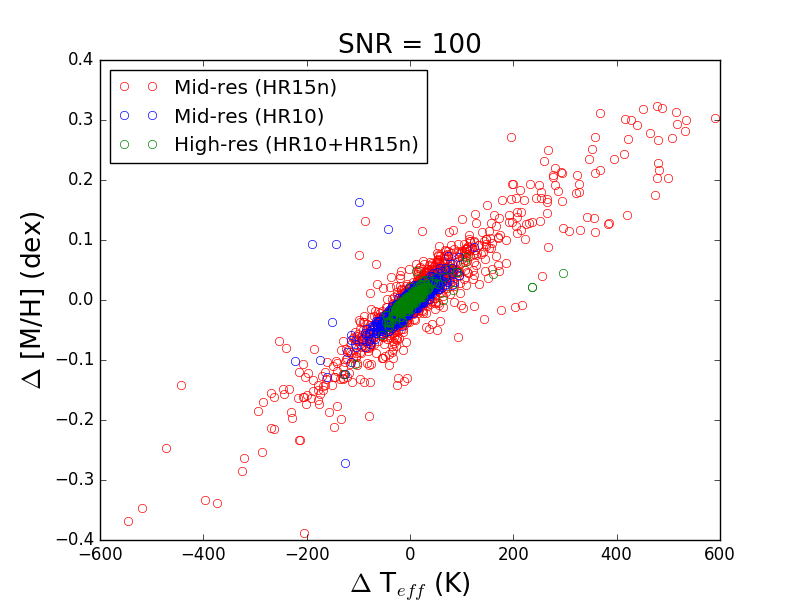} 
 \end{minipage}
 \begin{minipage}{0.33\textwidth}
  \includegraphics[width=5.2cm, height=4.5cm]{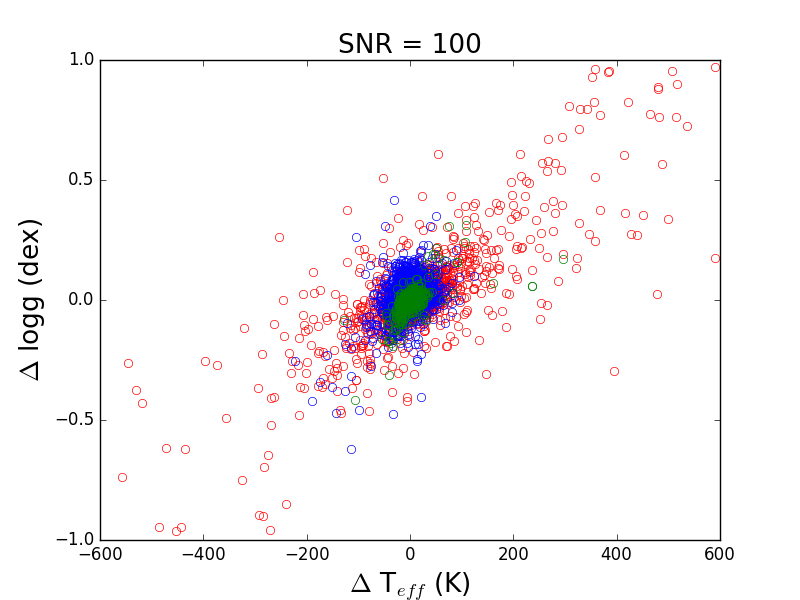} 
 \end{minipage}
 \begin{minipage}{0.33\textwidth}
  \includegraphics[width=5.2cm, height=4.5cm]{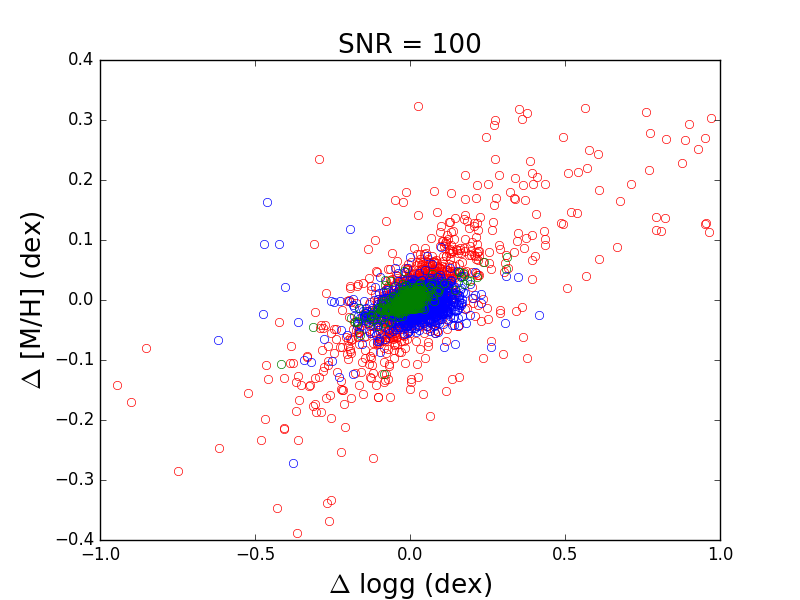} 
 \end{minipage}
 \centering
 \begin{minipage}{0.33\textwidth}
  \includegraphics[width=5.2cm, height=4.5cm]{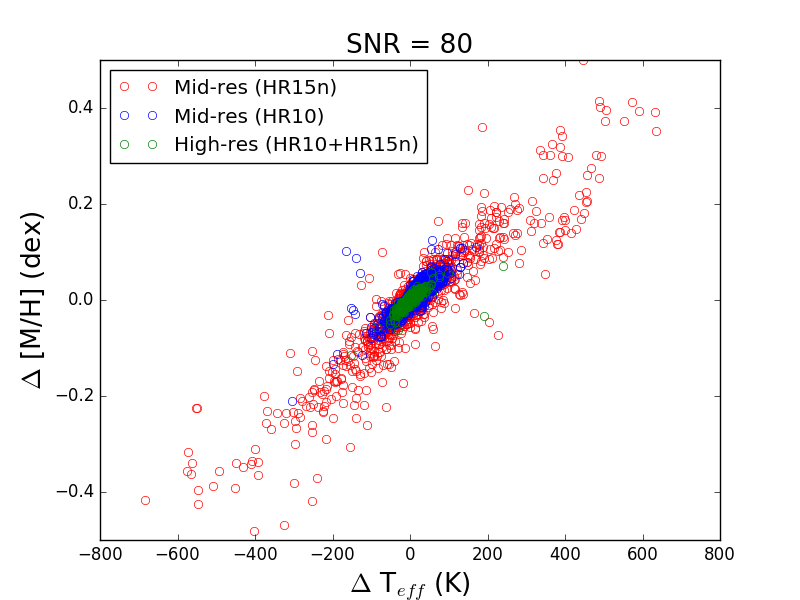} 
 \end{minipage}
 \begin{minipage}{0.33\textwidth}
  \includegraphics[width=5.2cm, height=4.5cm]{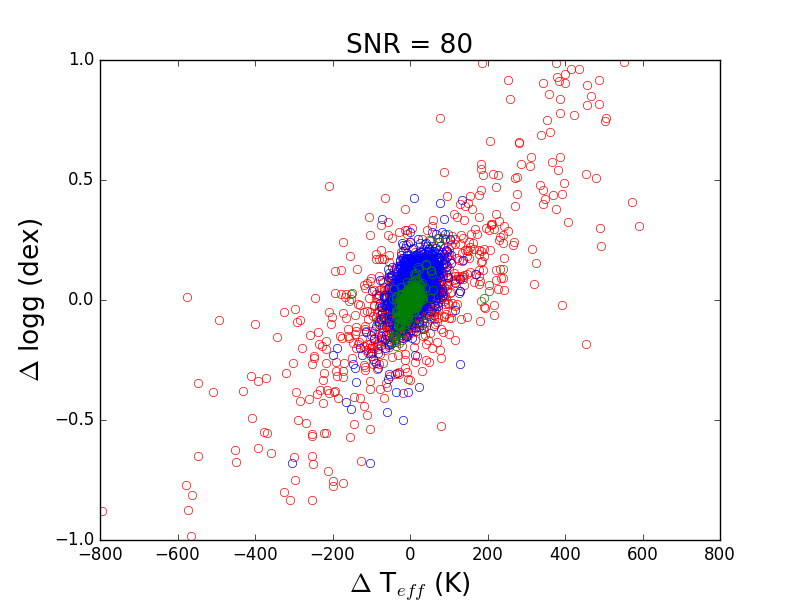} 
 \end{minipage}
 \begin{minipage}{0.33\textwidth}
  \includegraphics[width=5.2cm, height=4.5cm]{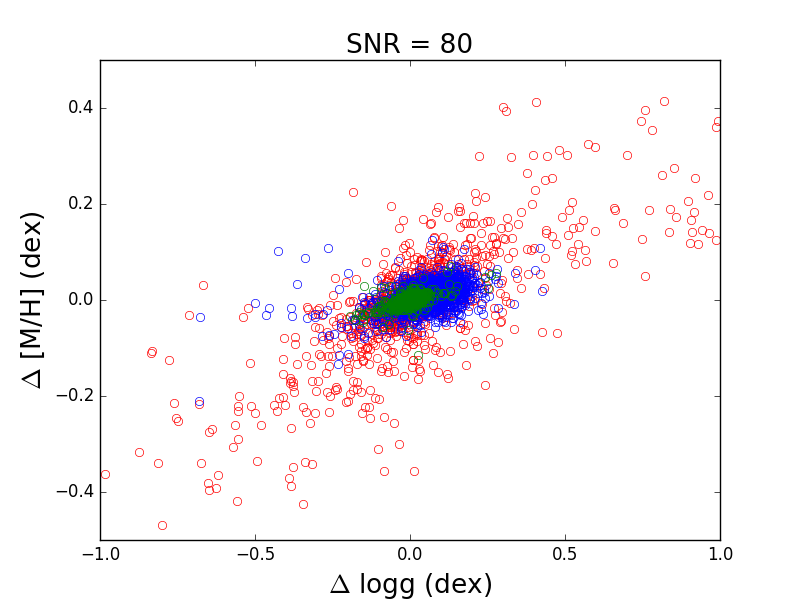} 
 \end{minipage}
 \centering
 \begin{minipage}{0.33\textwidth}
  \includegraphics[width=5.2cm, height=4.5cm]{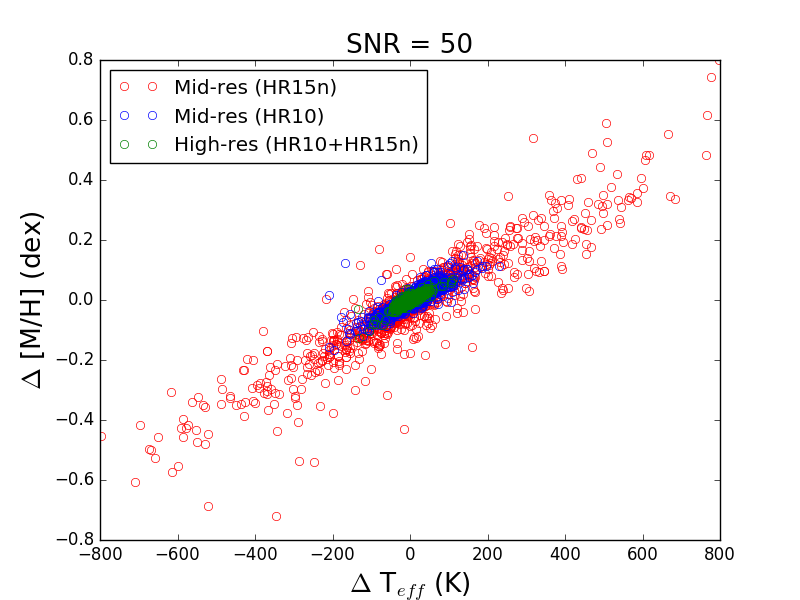} 
 \end{minipage}
 \begin{minipage}{0.33\textwidth}
  \includegraphics[width=5.2cm, height=4.5cm]{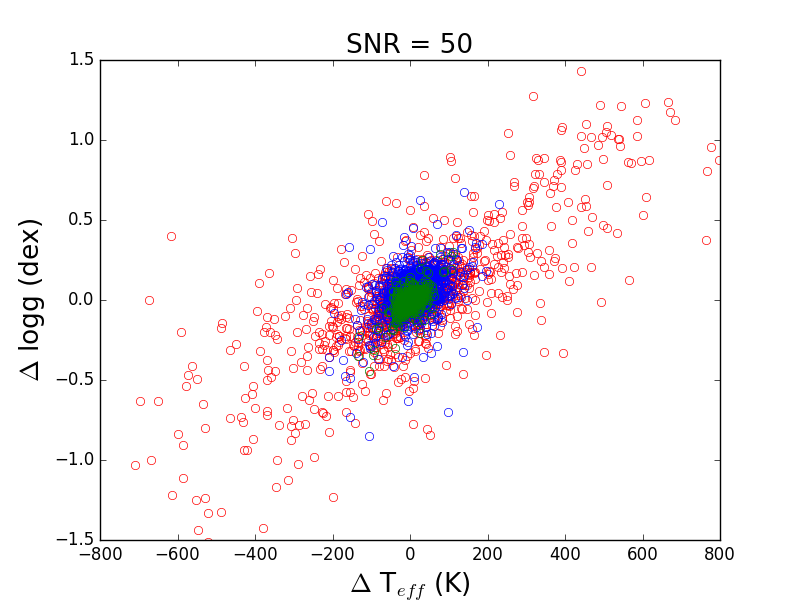} 
 \end{minipage}
 \begin{minipage}{0.33\textwidth}
  \includegraphics[width=5.2cm, height=4.5cm]{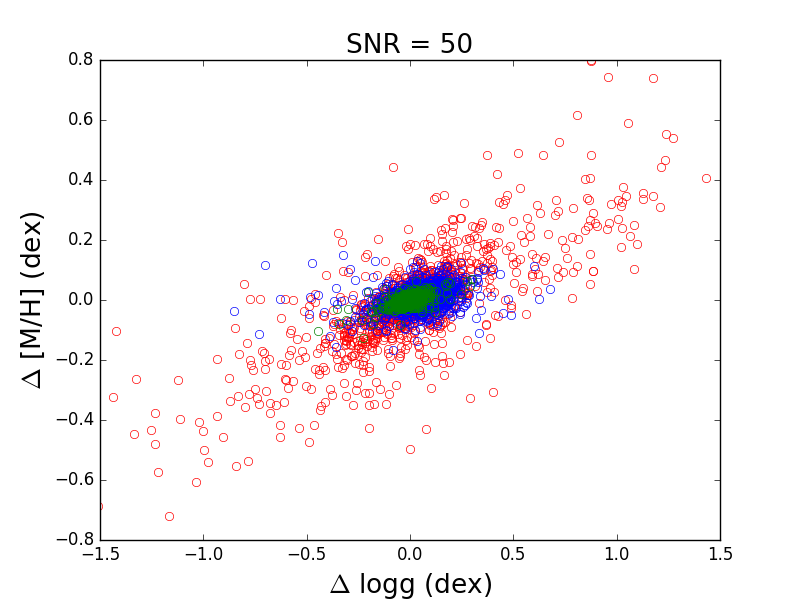} 
 \end{minipage}
 \centering
 \begin{minipage}{0.33\textwidth}
  \includegraphics[width=5.2cm, height=4.5cm]{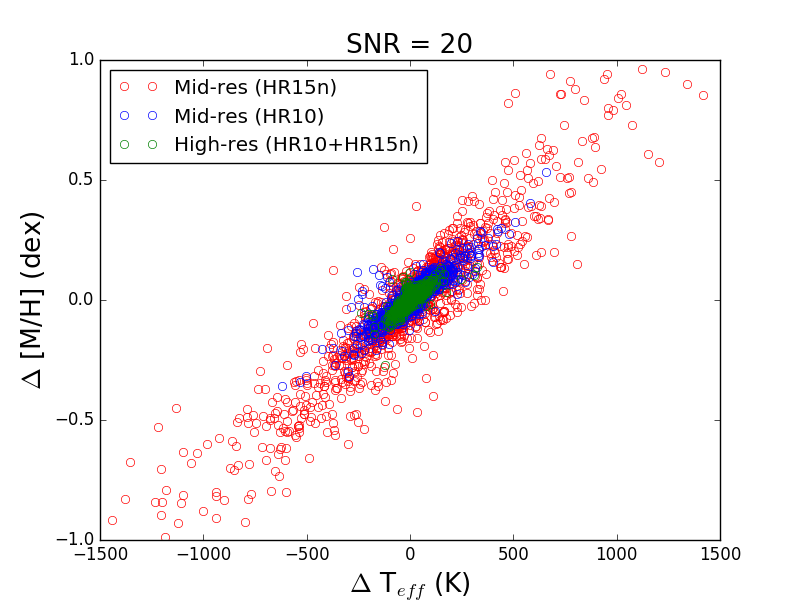} 
 \end{minipage}
 \begin{minipage}{0.33\textwidth}
  \includegraphics[width=5.2cm, height=4.5cm]{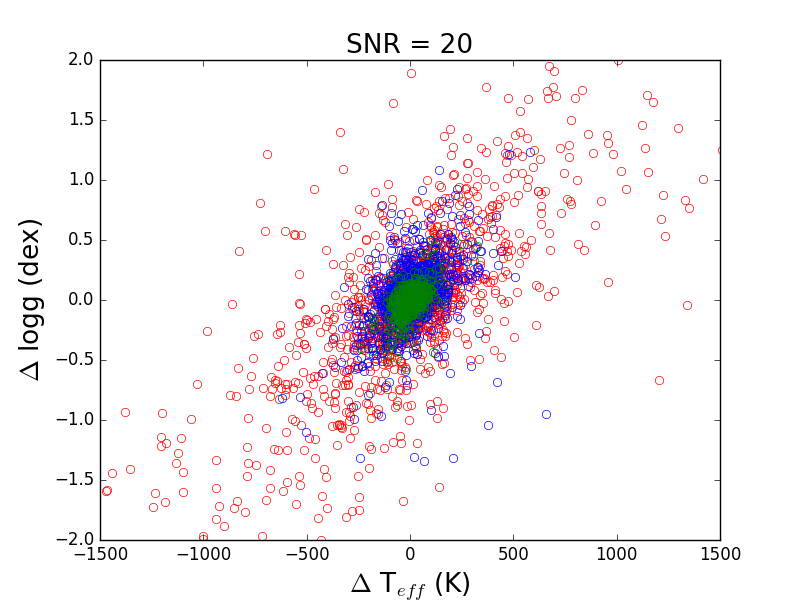} 
 \end{minipage}
 \begin{minipage}{0.33\textwidth}
  \includegraphics[width=5.2cm, height=4.5cm]{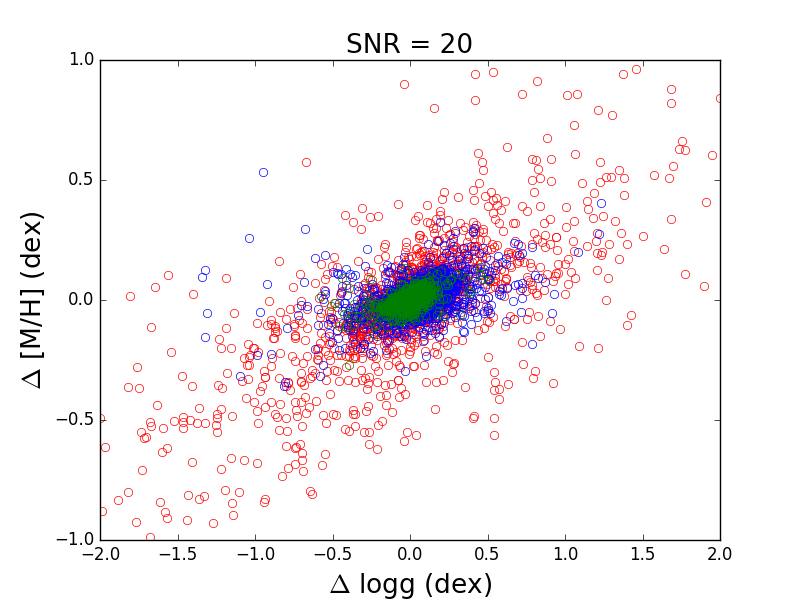} 
 \end{minipage}

 \caption{Correlations between the residuals of the main atmospheric parameters for the synthetic spectra for different resolutions depicted in different color. 
 Each row corresponds to different S/N values. The parameter space of the synthetic spectra covers FGK-type stars.}
 \label{snr_correlations}
\end{figure*}

Our results for the different resolutions and S/N are shown in the Table~\ref{synthetic_table}. The residual distributions are plotted in Fig.~\ref{distribution_synthetic} and 
the residual differences of the atmospheric parameters are presented in the Appendix~\ref{diff_synth}. 
The mean and median differences are very small for all parameters and resolution set-ups. The standard deviations and mean absolute deviations generally increase with decreasing S/N. 
The residual distributions are single peaked and the width of the distribution is very narrow for all parameters at the high resolution regime. 
The HR15n set-up performs worse from the three sets with almost 3 to 4 $\sigma$ higher values compared to HR10 because the wavelength intervals we selected contain less lines, 918 and 219 
lines for the HR10 and HR15n respectively. In Fig.~\ref{snr_correlations}, we report the correlations between the differences of the parameters for the different S/N values. 
We see a strong correlation between metallicity and effective temperature and a weaker between $\log g$ -- $T_{\mathrm{eff}}$. The correlation between $[M/H]$ -- $\log g$ prevails only 
for the lowest resolution set-up (HR15n). The differences in surface gravity are almost independent from the differences in metallicity for the high resolution and the HR10 set-up. 
The S/N does not affect the shape of these correlations but only their strength. 
\cite{Kordopatis2011} show similar correlations in their results of synthetic spectra with the strongest to be between $T_{\mathrm{eff}}$ -- $\log g$ and 
$T_{\mathrm{eff}}$ -- $[M/H]$ as in our case, whereas \cite{recio2016} show correlations between all parameters.

\begin{figure*}%
 \centering
 \begin{minipage}{0.21\textwidth}
  \includegraphics[width=4.5cm, height=4.2cm]{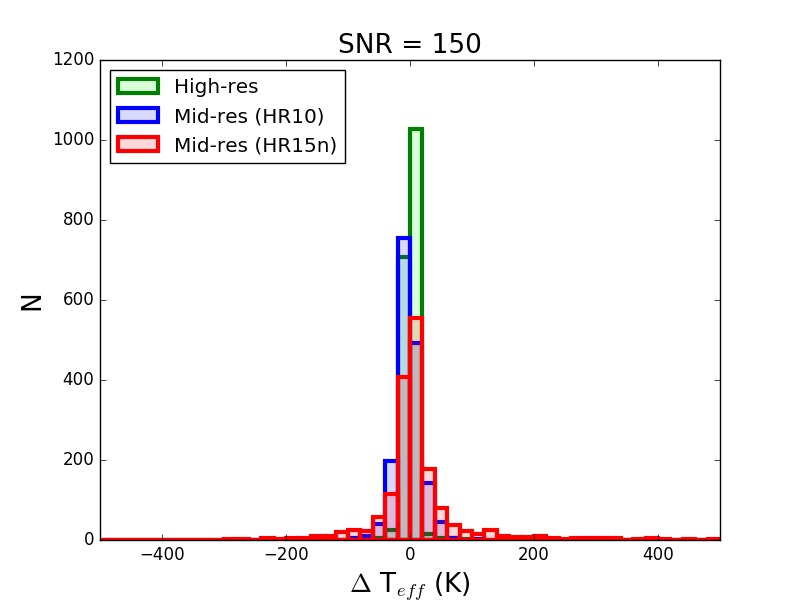} 
 \end{minipage}
\hspace{0.02\textwidth}%
 \begin{minipage}{0.21\textwidth}
  \includegraphics[width=4.5cm, height=4.2cm]{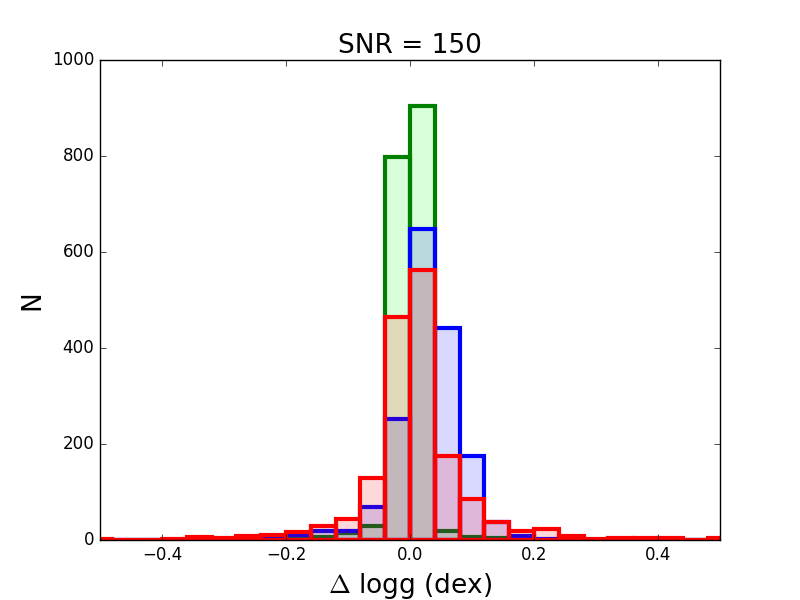} 
 \end{minipage}
\hspace{0.02\textwidth}%
 \begin{minipage}{0.21\textwidth}
  \includegraphics[width=4.5cm, height=4.2cm]{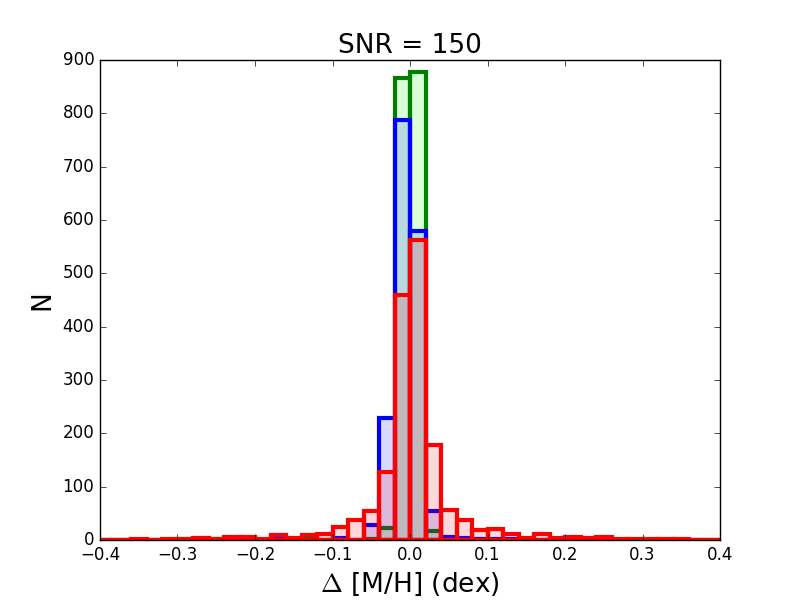} 
 \end{minipage}
\hspace{0.02\textwidth}%
 \begin{minipage}{0.21\textwidth}
  \includegraphics[width=4.5cm, height=4.2cm]{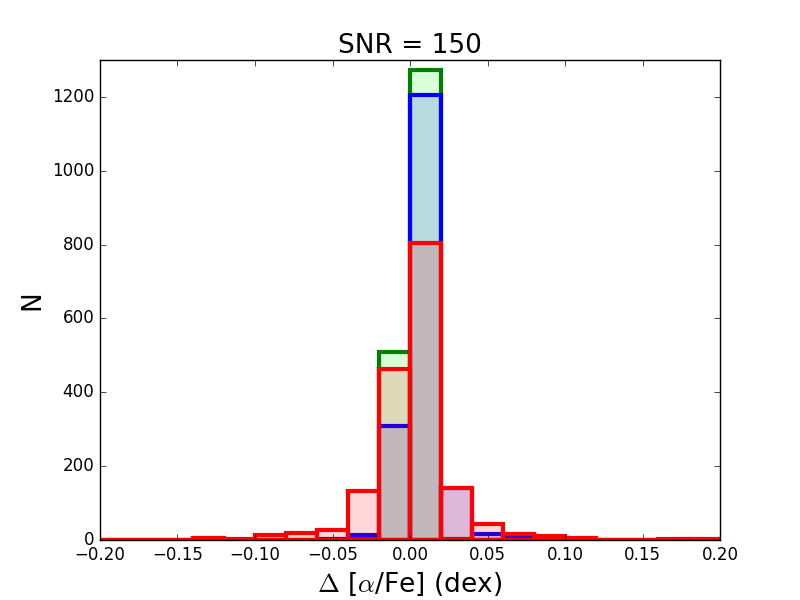} 
 \end{minipage}

 \centering
 \begin{minipage}{0.21\textwidth}
  \includegraphics[width=4.5cm, height=4.2cm]{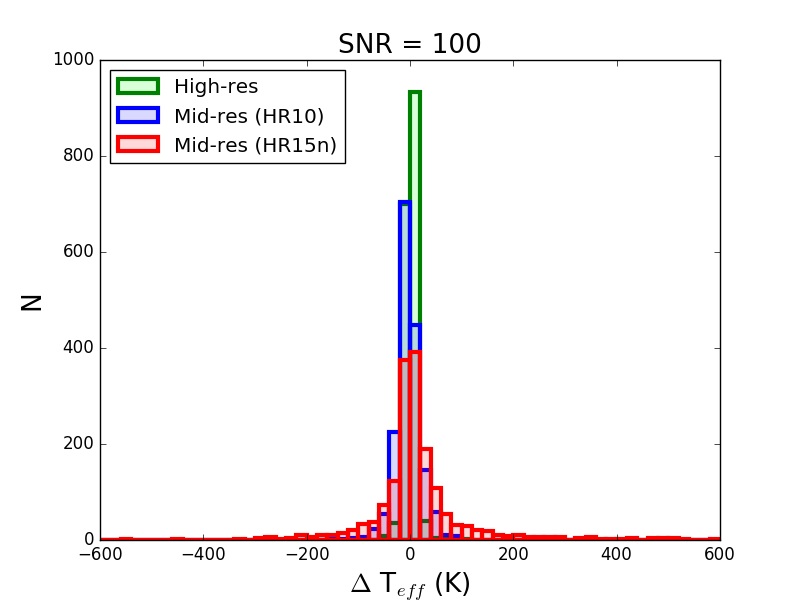} 
 \end{minipage}
\hspace{0.02\textwidth}%
 \begin{minipage}{0.21\textwidth}
  \includegraphics[width=4.5cm, height=4.2cm]{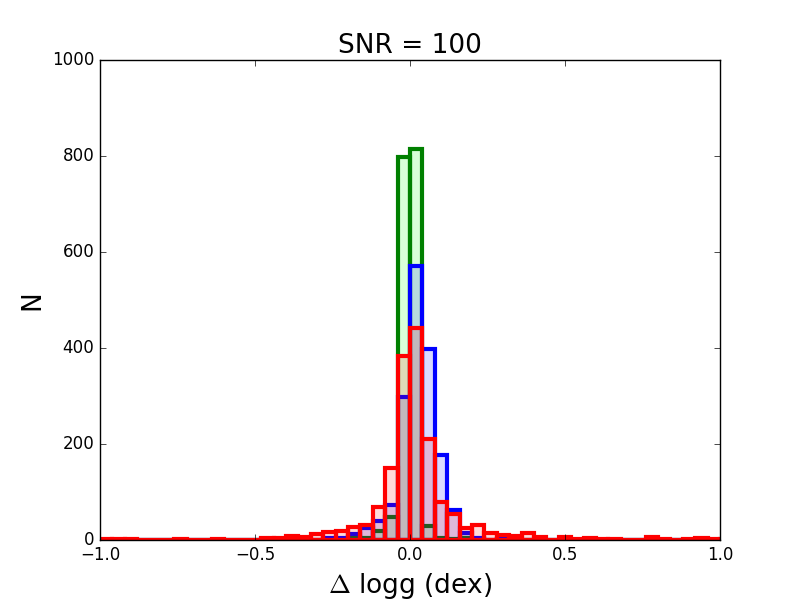} 
 \end{minipage}
\hspace{0.02\textwidth}%
 \begin{minipage}{0.21\textwidth}
  \includegraphics[width=4.5cm, height=4.2cm]{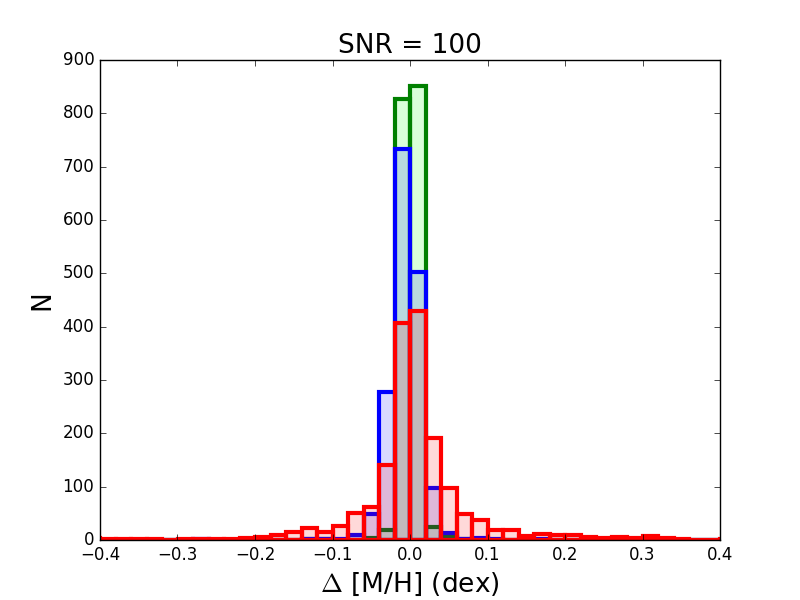} 
 \end{minipage}
\hspace{0.02\textwidth}%
 \begin{minipage}{0.21\textwidth}
  \includegraphics[width=4.5cm, height=4.2cm]{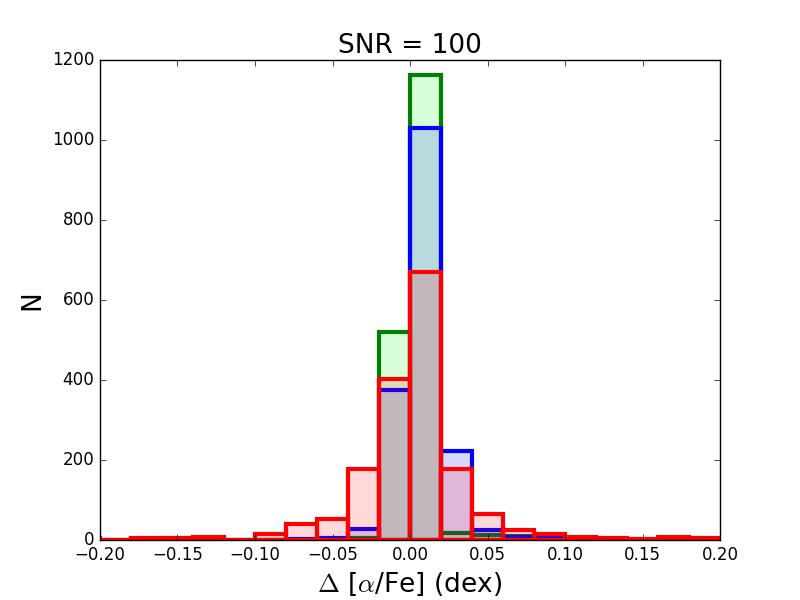} 
 \end{minipage}

 \centering
 \begin{minipage}{0.21\textwidth}
  \includegraphics[width=4.5cm, height=4.2cm]{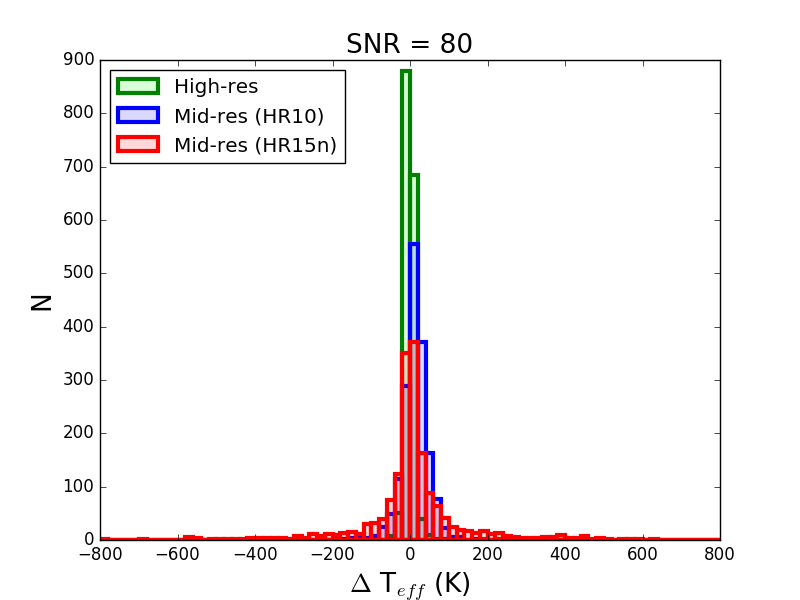} 
 \end{minipage}
\hspace{0.02\textwidth}%
 \begin{minipage}{0.21\textwidth}
  \includegraphics[width=4.5cm, height=4.2cm]{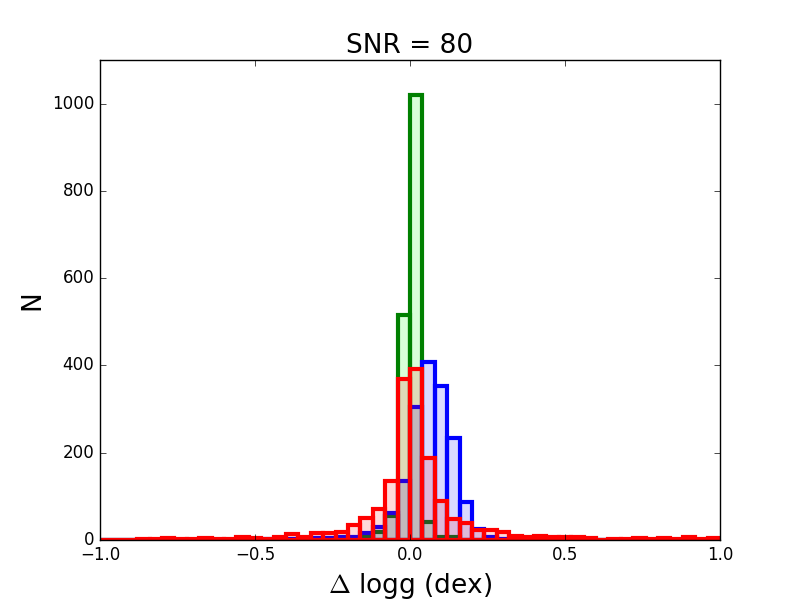} 
 \end{minipage}
\hspace{0.02\textwidth}%
 \begin{minipage}{0.21\textwidth}
  \includegraphics[width=4.5cm, height=4.2cm]{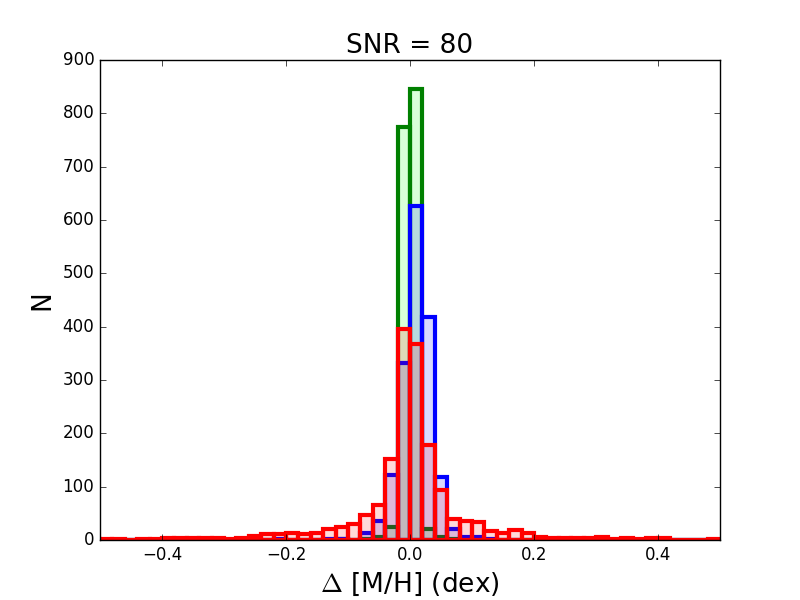} 
 \end{minipage}
\hspace{0.02\textwidth}%
 \begin{minipage}{0.21\textwidth}
  \includegraphics[width=4.5cm, height=4.2cm]{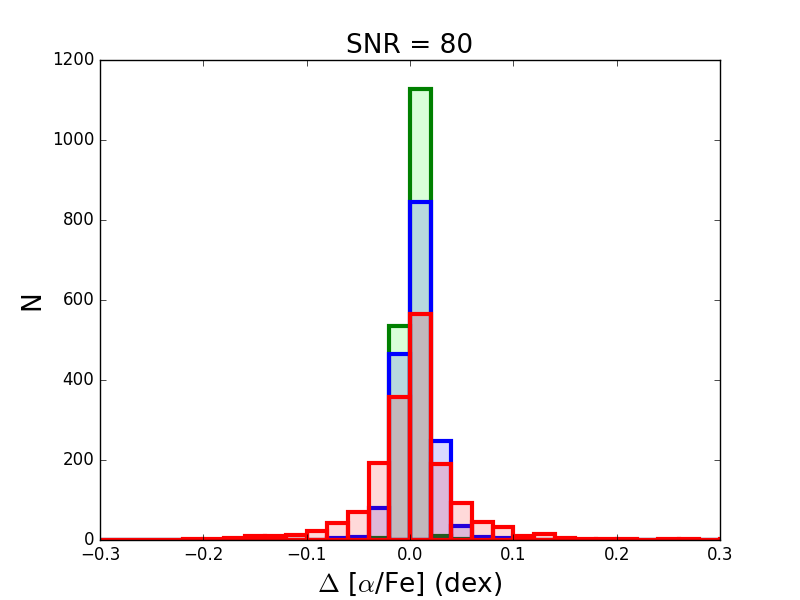} 
 \end{minipage}

 \centering
 \begin{minipage}{0.21\textwidth}
  \includegraphics[width=4.5cm, height=4.2cm]{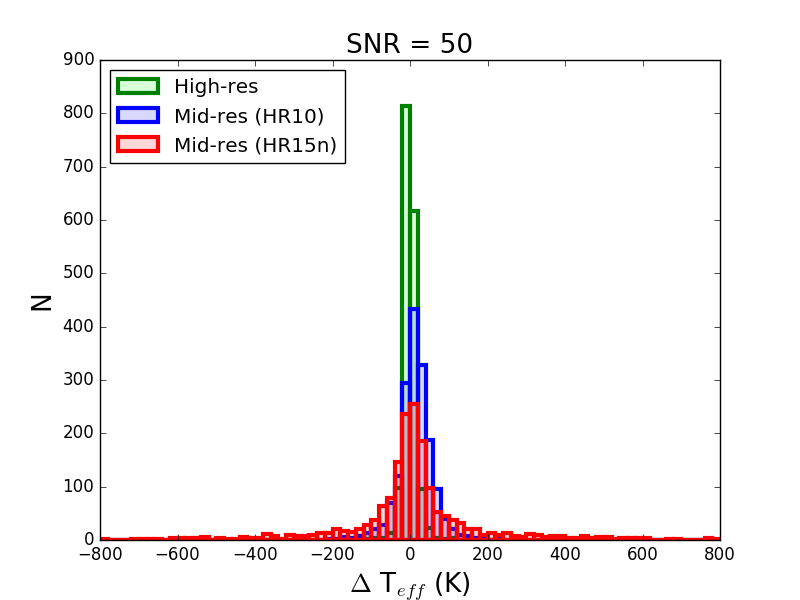} 
 \end{minipage}
\hspace{0.02\textwidth}%
 \begin{minipage}{0.21\textwidth}
  \includegraphics[width=4.5cm, height=4.2cm]{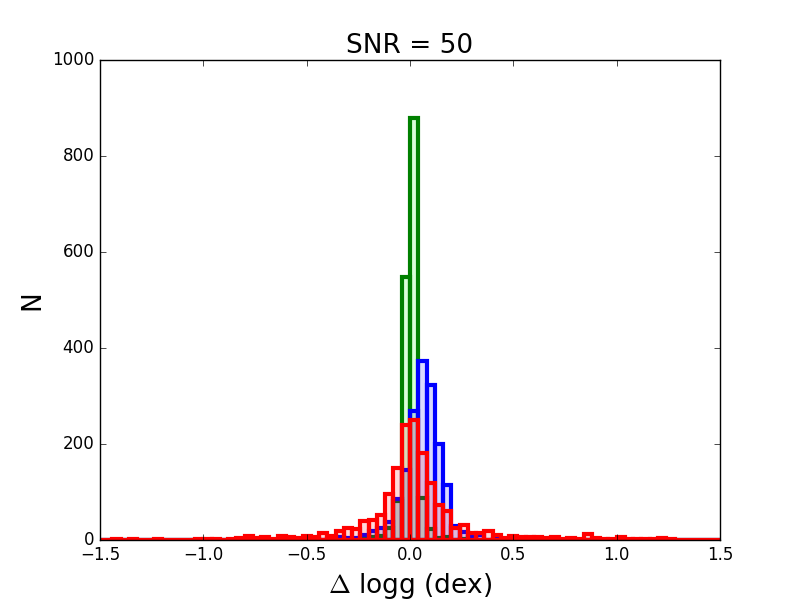} 
 \end{minipage}
\hspace{0.02\textwidth}%
 \begin{minipage}{0.21\textwidth}
  \includegraphics[width=4.5cm, height=4.2cm]{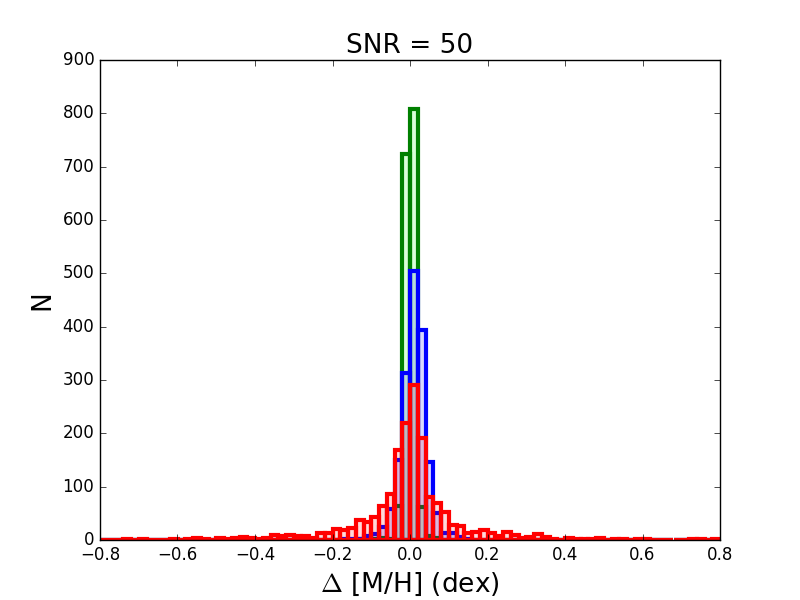} 
 \end{minipage}
\hspace{0.02\textwidth}%
 \begin{minipage}{0.21\textwidth}
  \includegraphics[width=4.5cm, height=4.2cm]{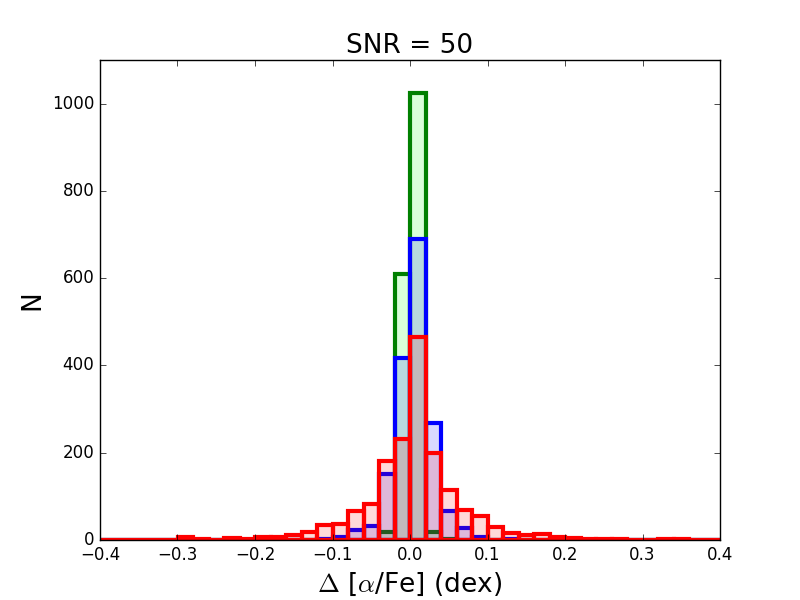} 
 \end{minipage}

  \centering
 \begin{minipage}{0.21\textwidth}
  \includegraphics[width=4.5cm, height=4.2cm]{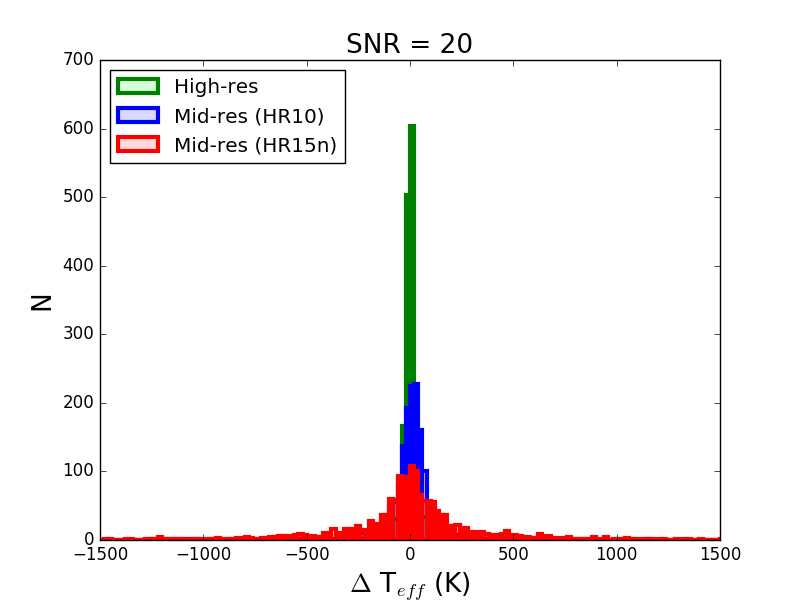} 
 \end{minipage}
\hspace{0.02\textwidth}%
 \begin{minipage}{0.21\textwidth}
  \includegraphics[width=4.5cm, height=4.2cm]{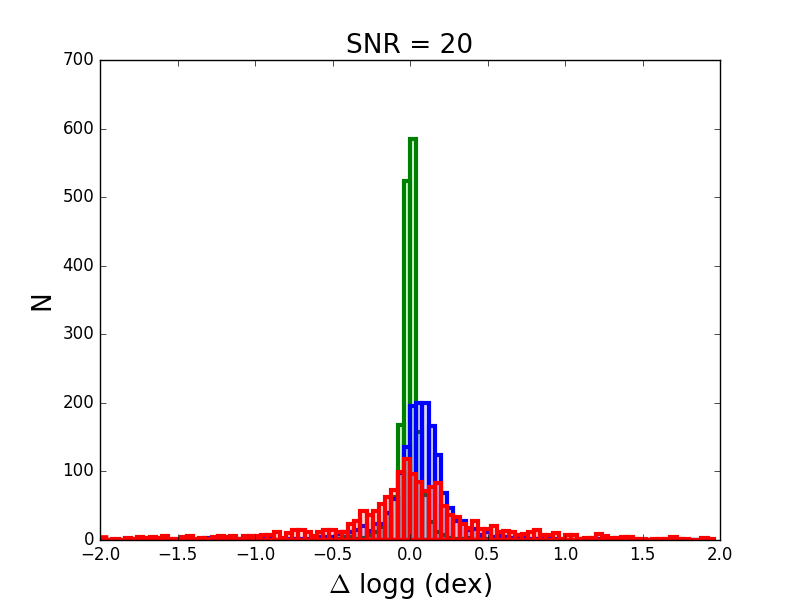} 
 \end{minipage}
\hspace{0.02\textwidth}%
 \begin{minipage}{0.21\textwidth}
  \includegraphics[width=4.5cm, height=4.2cm]{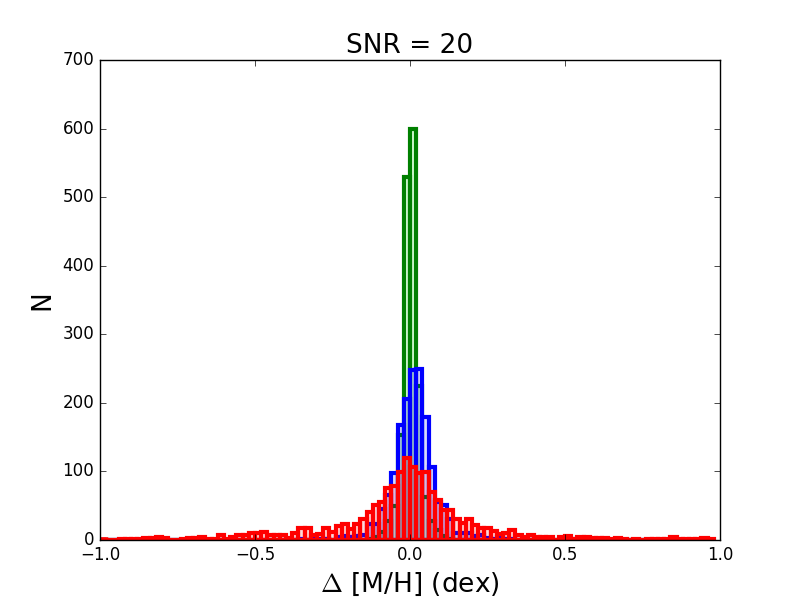} 
 \end{minipage}
\hspace{0.02\textwidth}%
 \begin{minipage}{0.21\textwidth}
  \includegraphics[width=4.5cm, height=4.2cm]{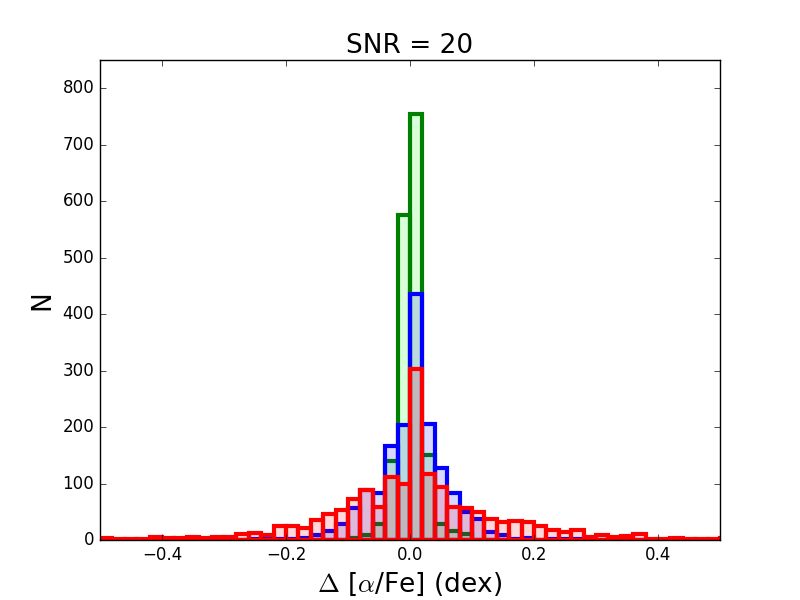} 
 \end{minipage}
  \caption{Distribution of the parameter residuals of the synthetic spectra for the different resolutions depicted in different colors. Each row corresponds to different S/N values. 
  The parameter space of the synthetic spectra covers FGK-type stars.}
  \label{distribution_synthetic}
 \end{figure*}

\subsection{Initial conditions}\label{initial_conditions}

In many cases we do not have any information on the stellar parameters and the initial guesses we assign can affect the convergence results. 
To check if the initial conditions affect the derived parameters, we perform a simple Monte Carlo test. We select three stars of different spectral types (F-type; Procyon, G-type; Sun, 
and K-type; del\,Eri), one giant (Arcturus), and one metal-poor star (HD\,201891) as references. We randomly select 500 initial parameters, the same for all stars, from a pool of parameters:  
4000 $< T_{\mathrm{eff}} <$ 7000\,K, 1.5 $< \log g <$ 5.0\,dex, --2.5 $< [M/H] <$ 0.4\,dex, and 0 $< \upsilon\sin i <$ 12.0\,km\,s$^{-1}$. We derive their parameters with the methodology 
described previously. We report the mean differences of the 500 values (e.g. for $T_{\mathrm{eff}}$ of del\,Eri: 
$\overline{\Delta T_{\mathrm{eff}}}$ = $\frac{\sum T_{\mathrm{eff}, i} - 5022}{500}$ , where 5022\,K is the $T_{\mathrm{eff}}$ using solar initial values). 

\begin{figure*}%
 \vspace{-0.2cm}
 \centering
 \begin{minipage}{0.21\textwidth}
  \includegraphics[width=4.5cm, height=4.1cm]{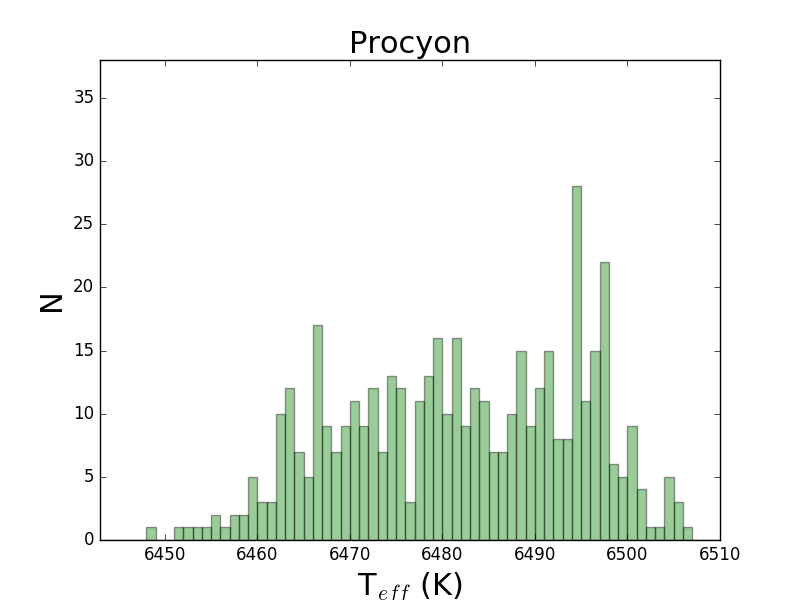} 
 \end{minipage}
\hspace{0.02\textwidth}%
 \begin{minipage}{0.21\textwidth}
  \includegraphics[width=4.5cm, height=4.1cm]{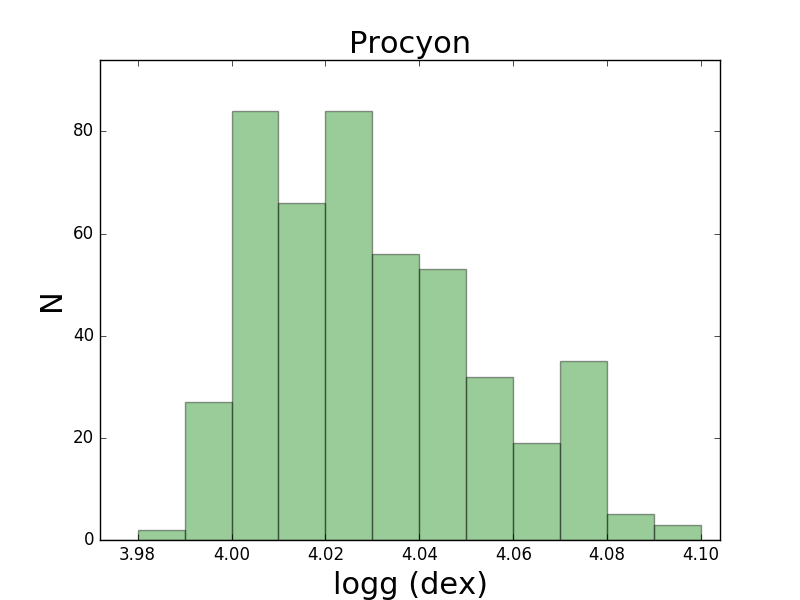} 
 \end{minipage}
\hspace{0.02\textwidth}%
 \begin{minipage}{0.21\textwidth}
  \includegraphics[width=4.5cm, height=4.1cm]{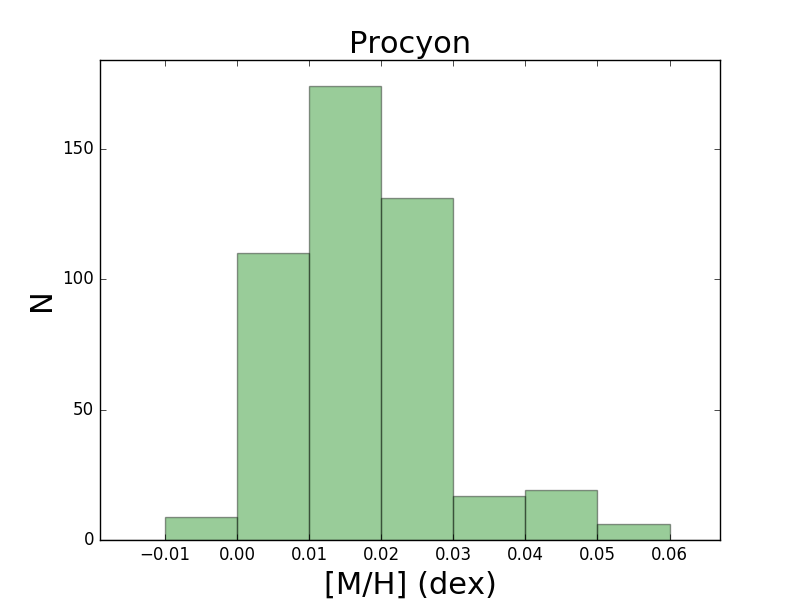} 
 \end{minipage}
\hspace{0.02\textwidth}%
 \begin{minipage}{0.21\textwidth}
  \includegraphics[width=4.5cm, height=4.1cm]{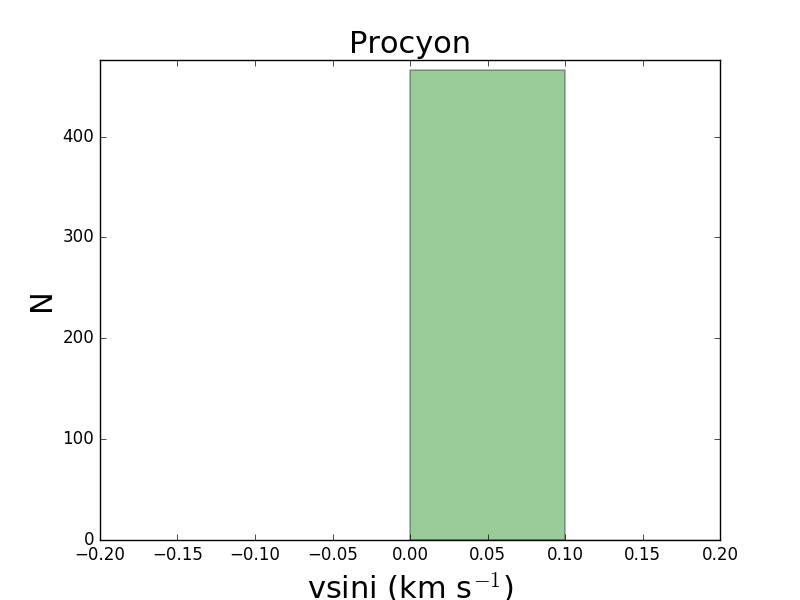} 
 \end{minipage}

  \centering
 \begin{minipage}{0.21\textwidth}
  \includegraphics[width=4.5cm, height=4.1cm]{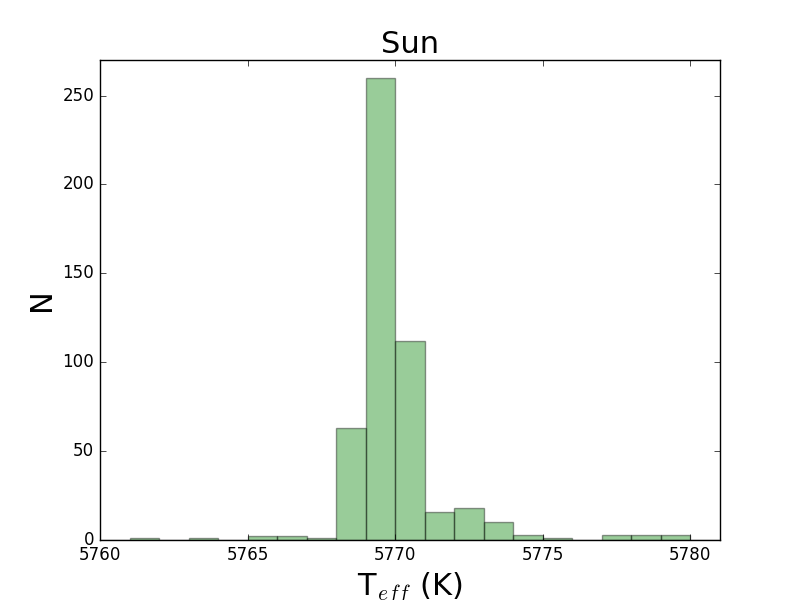} 
 \end{minipage}
\hspace{0.02\textwidth}%
 \begin{minipage}{0.21\textwidth}
  \includegraphics[width=4.5cm, height=4.1cm]{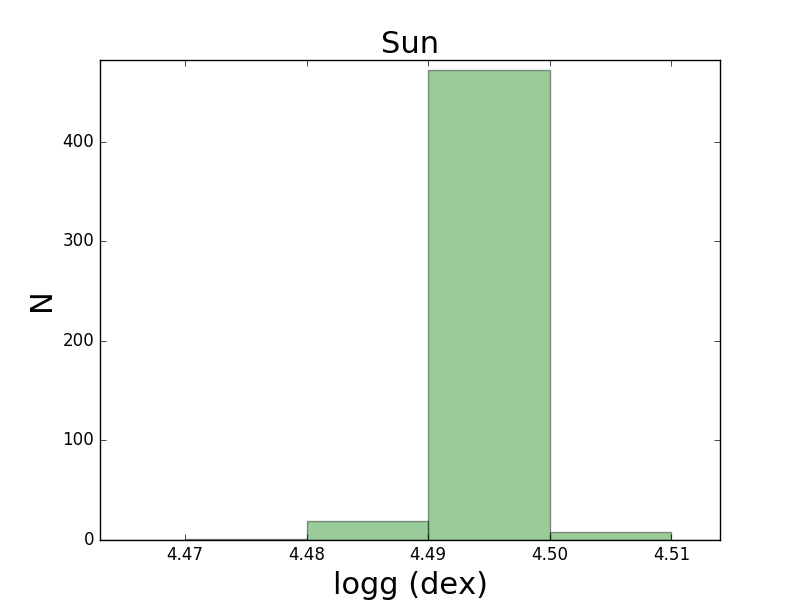} 
 \end{minipage}
\hspace{0.02\textwidth}%
 \begin{minipage}{0.21\textwidth}
  \includegraphics[width=4.5cm, height=4.1cm]{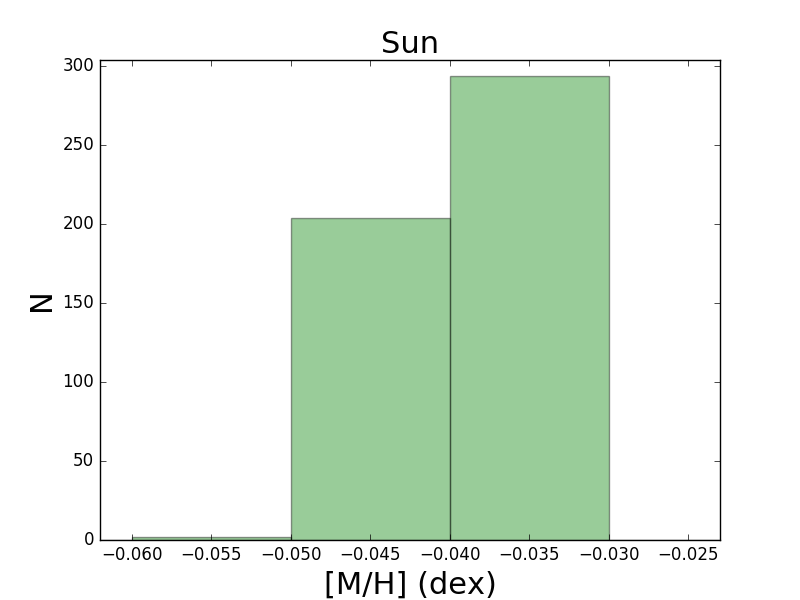} 
 \end{minipage}
\hspace{0.02\textwidth}%
 \begin{minipage}{0.21\textwidth}
  \includegraphics[width=4.5cm, height=4.1cm]{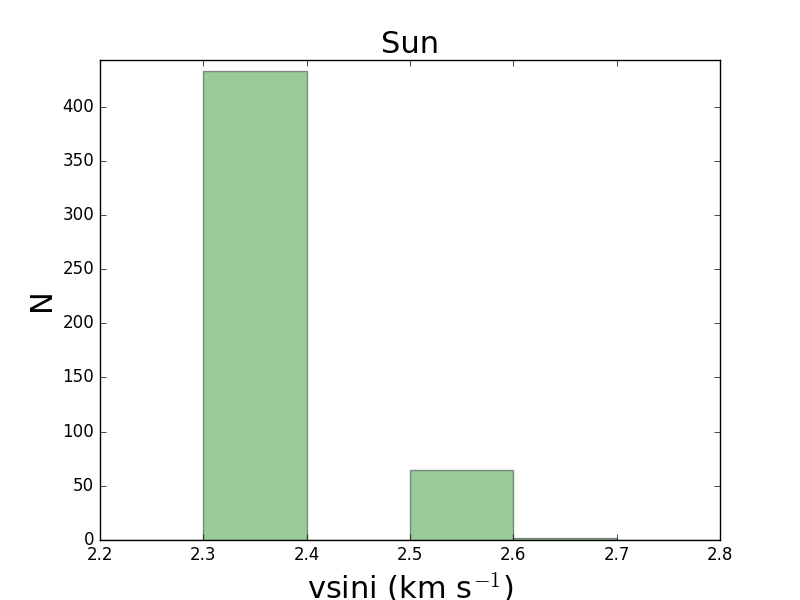} 
 \end{minipage}

  \centering
 \begin{minipage}{0.21\textwidth}
  \includegraphics[width=4.5cm, height=4.1cm]{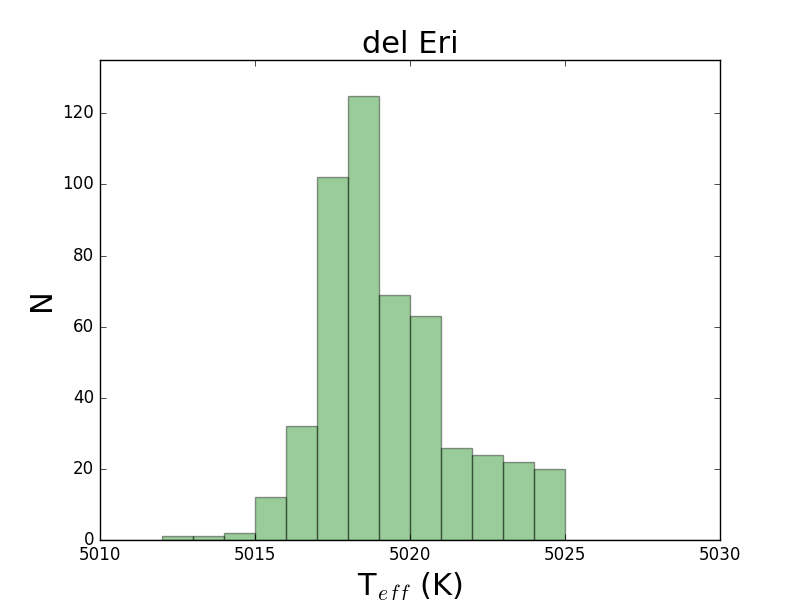} 
 \end{minipage}
\hspace{0.02\textwidth}%
 \begin{minipage}{0.21\textwidth}
  \includegraphics[width=4.5cm, height=4.1cm]{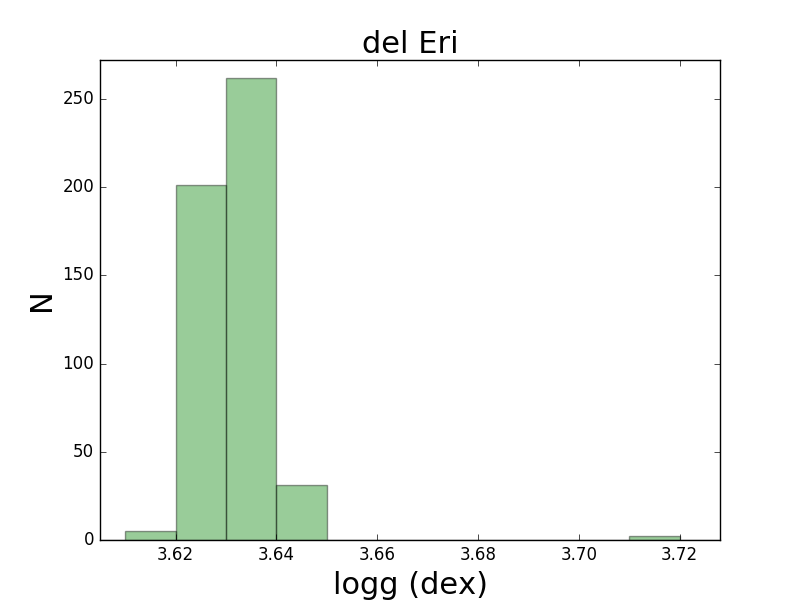} 
 \end{minipage}
\hspace{0.02\textwidth}%
 \begin{minipage}{0.21\textwidth}
  \includegraphics[width=4.5cm, height=4.1cm]{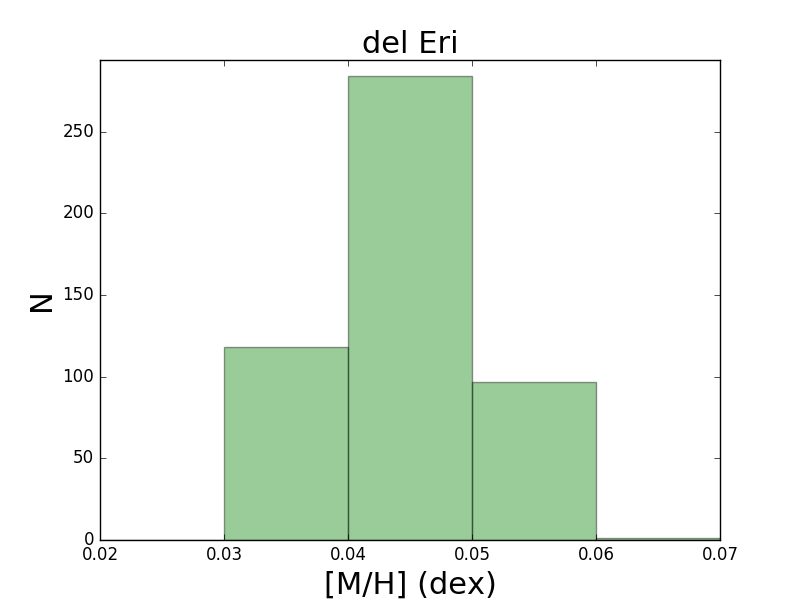} 
 \end{minipage}
\hspace{0.02\textwidth}%
 \begin{minipage}{0.21\textwidth}
  \includegraphics[width=4.5cm, height=4.1cm]{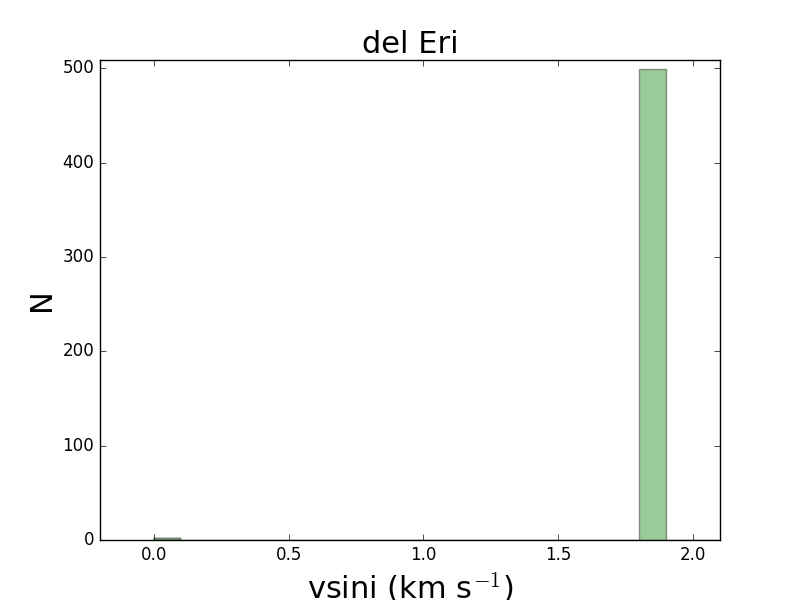} 
 \end{minipage}

  \centering
 \begin{minipage}{0.21\textwidth}
  \includegraphics[width=4.5cm, height=4.1cm]{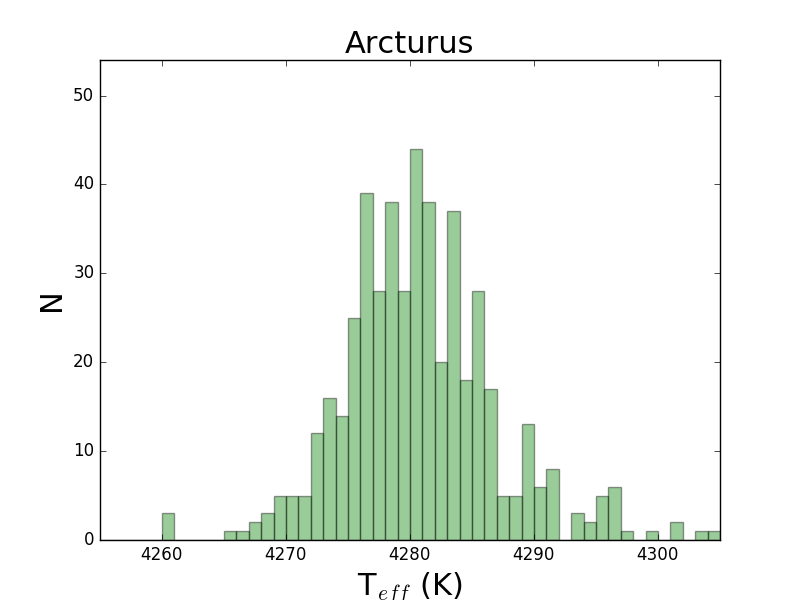} 
 \end{minipage}
\hspace{0.02\textwidth}%
 \begin{minipage}{0.21\textwidth}
  \includegraphics[width=4.5cm, height=4.1cm]{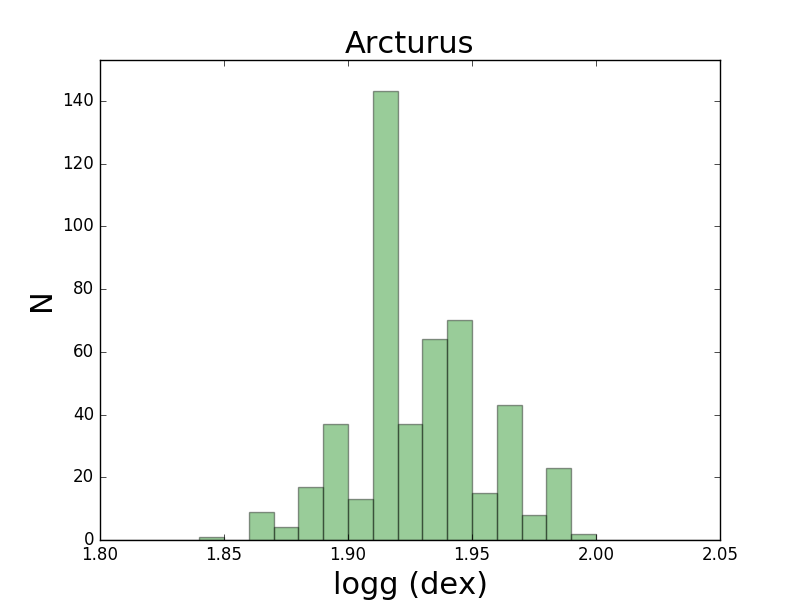} 
 \end{minipage}
\hspace{0.02\textwidth}%
 \begin{minipage}{0.21\textwidth}
  \includegraphics[width=4.5cm, height=4.1cm]{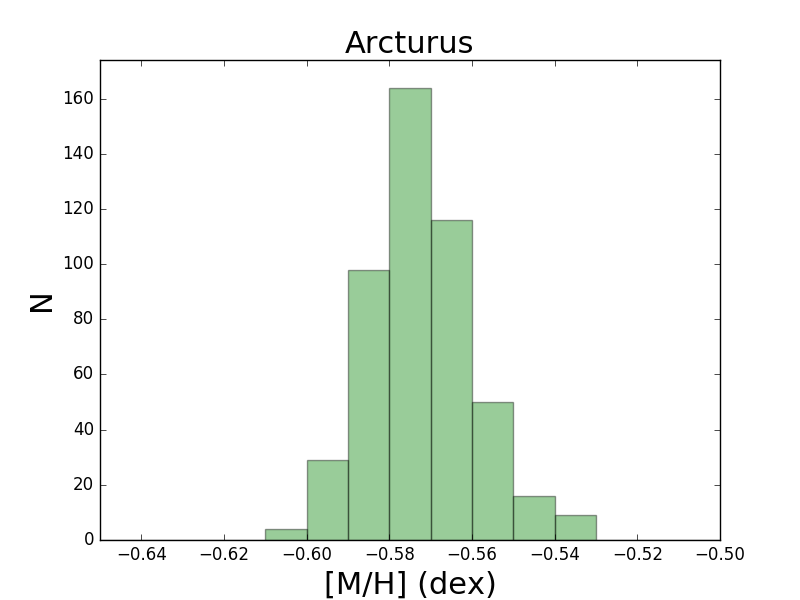} 
 \end{minipage}
\hspace{0.02\textwidth}%
 \begin{minipage}{0.21\textwidth}
  \includegraphics[width=4.5cm, height=4.1cm]{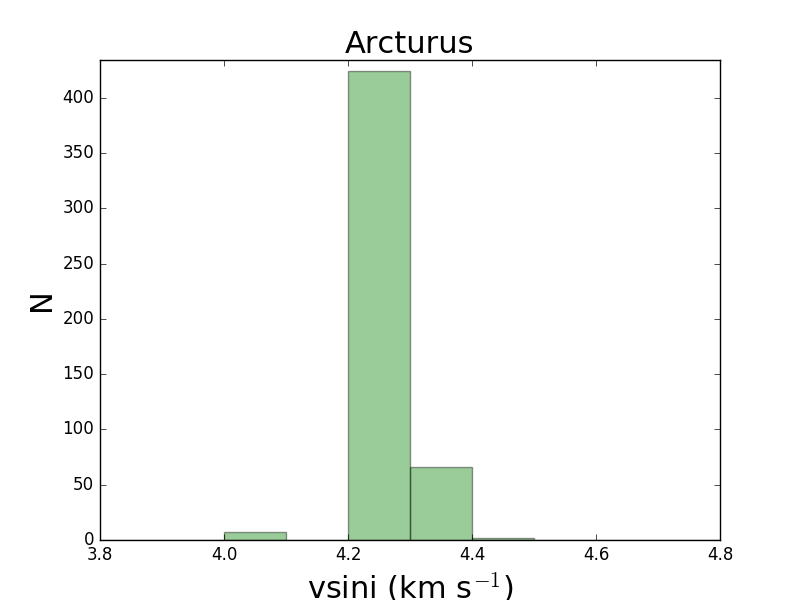} 
 \end{minipage}

 \centering
 \begin{minipage}{0.21\textwidth}
  \includegraphics[width=4.5cm, height=4.1cm]{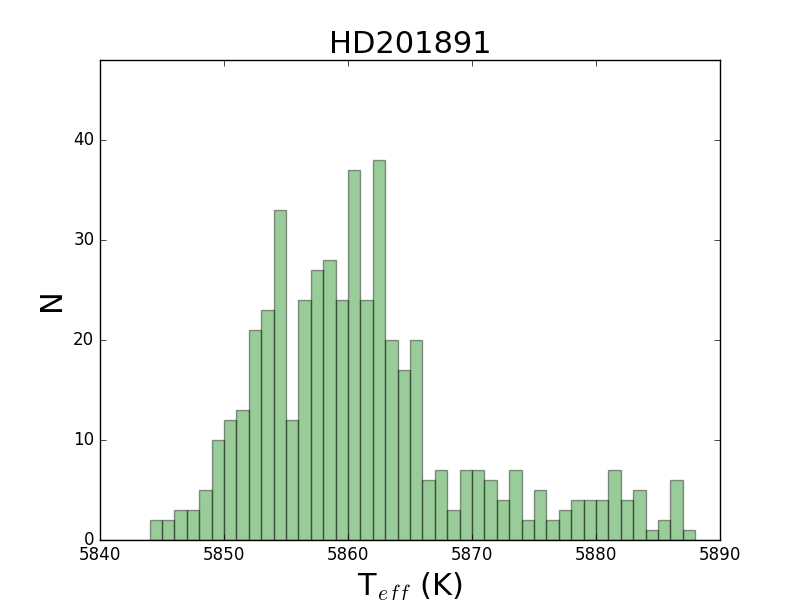} 
 \end{minipage}
\hspace{0.02\textwidth}%
 \begin{minipage}{0.21\textwidth}
  \includegraphics[width=4.5cm, height=4.1cm]{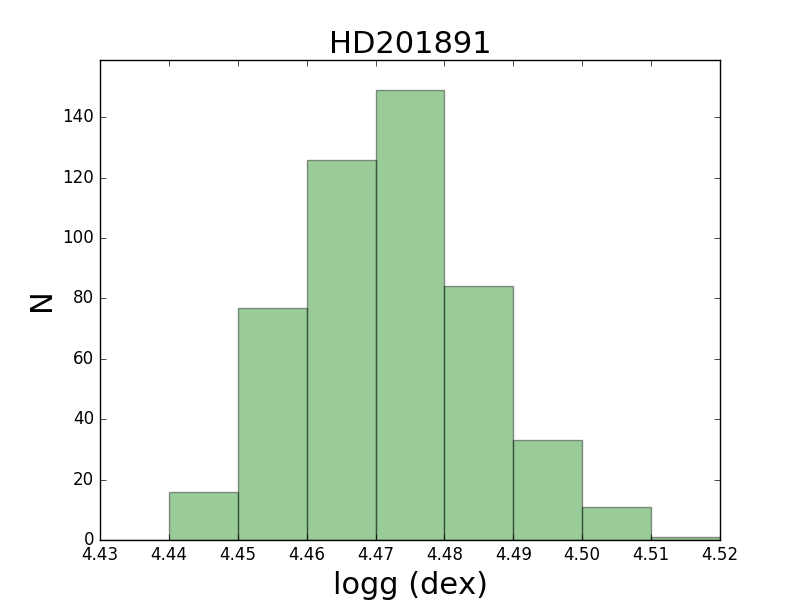} 
 \end{minipage}
\hspace{0.02\textwidth}%
 \begin{minipage}{0.21\textwidth}
  \includegraphics[width=4.5cm, height=4.1cm]{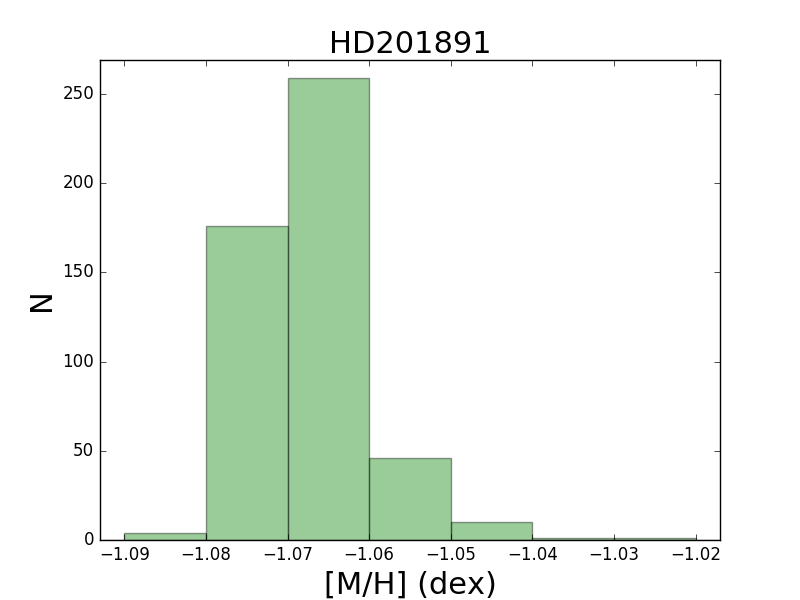} 
 \end{minipage}
\hspace{0.02\textwidth}%
 \begin{minipage}{0.21\textwidth}
  \includegraphics[width=4.5cm, height=4.1cm]{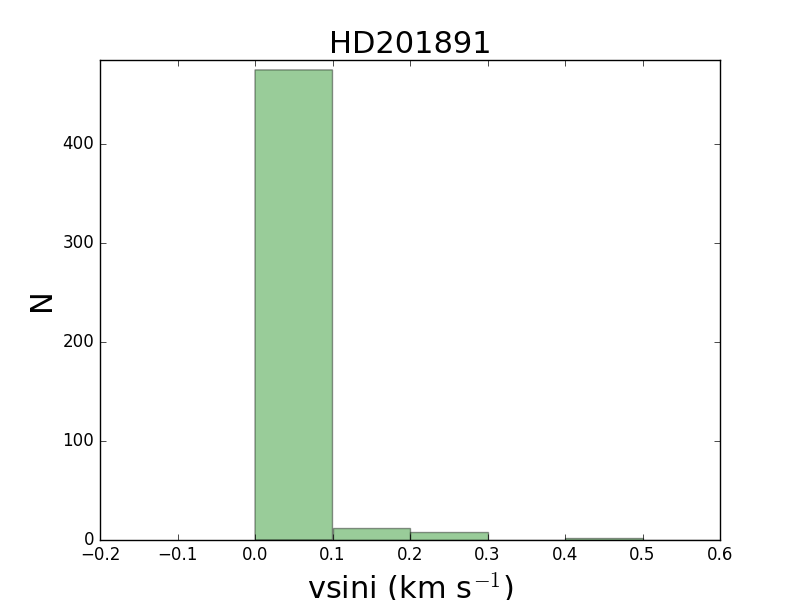} 
 \end{minipage}

 \caption{Distributions of stellar parameters starting the minimization from 500 different initial values, from top to bottom: Procyon (F-type), Sun (G-type), del\,Eri (K-type), 
 Arcturus (K-type giant), HD\,201891 (metal poor). The bin size for $T_{\mathrm{eff}}$ is 1\,K, for $\log g$ is 0.01\,dex, for $[M/H]$ is 0.01\,dex and for $\upsilon\sin i$ is 0.1 km\,s$^{-1}$.} 
 \label{initial_distributions}
\end{figure*}

We plot the distributions of the final results for the five stars in Fig.~\ref{initial_distributions}. The final distributions are not all Gaussian-like and do not all follow a similar pattern. 
From Table~\ref{initial_table} we show that the effective temperature of Procyon is mostly affected by the initial conditions while the differences for the rest of the parameters are close to 
zero. The dispersions for all parameters are the highest for Arcturus. 
We notice that the dispersions except for Arcturus are smaller than the ones derived previously for the synthetic spectra 
and we conclude that the choice of initial conditions does not significantly affect the precision of our results at least for stars with similar parameters to this example.

\begin{table*}
\caption{Mean differences ($\Delta$), median differences (med), standard deviations ($\sigma$), and mean absolute deviation (MAD) for the 500 different initial values.}
\label{initial_table}
\hspace*{-3em}
\scalebox{1.0}{\begin{tabular}{lcccccccccccccccc}
\hline\hline
Star & \multicolumn{4}{c}{$T_{\mathrm{eff}}$ (K)} & \multicolumn{4}{c}{$\log g$ (dex)} & \multicolumn{4}{c}{$[M/H]$ (dex)} & \multicolumn{4}{c}{$\upsilon\sin i$ (km\,s$^{-1}$)} \\
     & $\Delta$ & med & $\sigma$ & MAD            & $\Delta$ & med & $\sigma$ & MAD    & $\Delta$ & med & $\sigma$ & MAD   & $\Delta$ & med & $\sigma$ & MAD \\
\hline
del\,Eri   & 5  & 1  & 8  & 7  & 0.01 & 0.00 & 0.02 & 0.01 & 0.00 & 0.00 & 0.01 & 0.00 & 0.04 & 0.01 & 0.09 & 0.07 \\
Sun        & -1 & -1 & 2  & 1  & 0.00 & 0.00 & 0.00 & 0.00 & 0.00 & 0.00 & 0.00 & 0.00 & 0.01 & 0.00 & 0.03 & 0.02 \\
Procyon    & 15 & 16 & 13 & 11 & 0.03 & 0.03 & 0.02 & 0.01 & 0.00 & 0.00 & 0.01 & 0.01 & 0.00 & 0.00 & 0.00 & 0.00 \\
Arcturus   & 0  & 3  & 27 & 8  & 0.01 & 0.01 & 0.11 & 0.03 & 0.00 & 0.00 & 0.06 & 0.01 & 0.02 & 0.00 & 0.10 & 0.03 \\
HD\,201891 & 0  & -1 & 9  & 7  & 0.00 & 0.00 & 0.01 & 0.01 & 0.00 & 0.00 & 0.01 & 0.00 & 0.01 & 0.00 & 0.04 & 0.01 \\
\hline
\end{tabular}}
\end{table*}

\subsection{Signal-to-noise ratio}

\begin{figure*}
 \centering
 \begin{minipage}{0.5\textwidth}
   \includegraphics[width=1\linewidth]{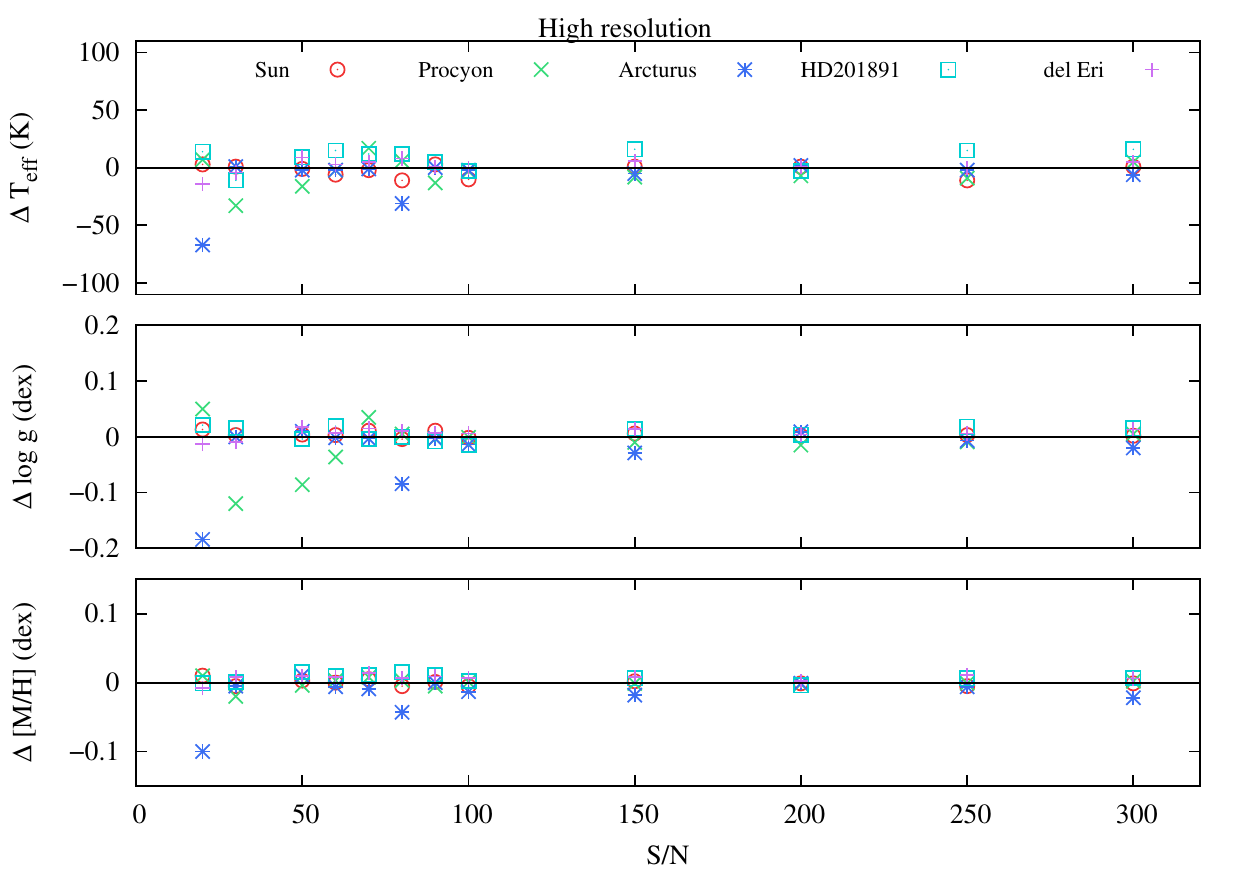}
 \end{minipage}
\hspace{-0.01\textwidth}%
 \begin{minipage}{0.5\textwidth}
   \includegraphics[width=1\linewidth]{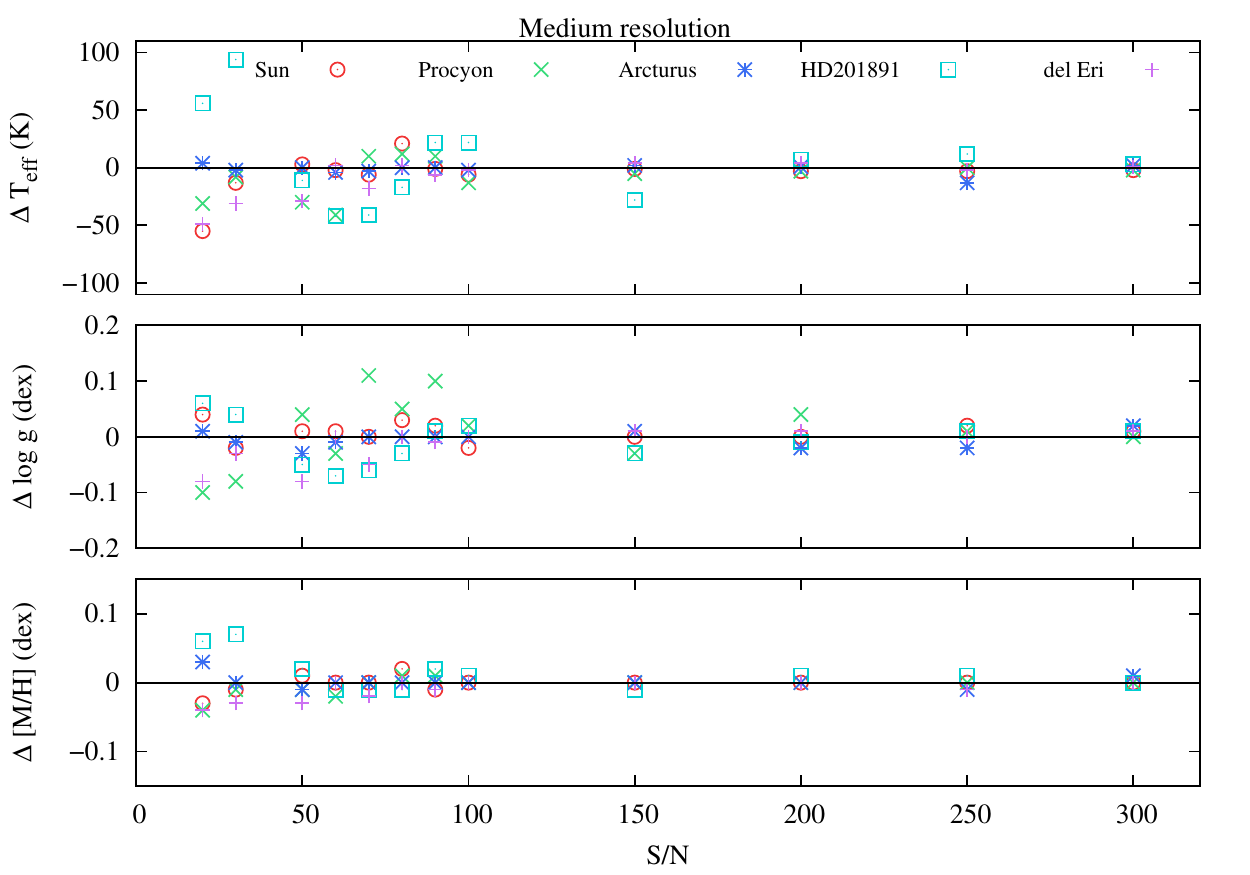} 
 \end{minipage}
  \caption{Change in parameters by adding different levels of noise (S/N\,=\,300, 250, 200, 150, 100, 90, 80, 70, 60, 50, 30, 20). The different symbols represent the different 
  type of stars. The upper panel corresponds to high resolution and the bottom to medium.}
  \label{fig_snr}
  \end{figure*}

The signal-to-noise ratio is a factor to indicate the quality of the data. In Sect.~\ref{synthetic_test}, we show how the parameters of synthetic spectra 
are affected for different S/N values and resolution. In this example, we explore how different S/N values measured per pixel, affect the precision of our results in the case of real spectra, 
in particular for the same five spectra as in Sect.~\ref{initial_conditions}. The spectra have S/N values between 550 and 1000 with the exception of HD\,201891 which has S/N 170 and adding 
higher S/N values is not meaningful for this star. Firstly, we normalize them and then we add different values of Gaussian noise\footnote{The mean and standard deviation 
of the normal distribution for the noise are: 0 and $\tfrac{1}{S/N}$ respectively.} (S/N = 300, 250, 200, 150, 100, 90, 80, 70, 60, 50, 30, 20). Moreover, we degrade 
the resolution of the spectra to R\,=\,17\,000 and adjust the wavelength spacing to 0.05\AA{} to mimic the GIRAFFE spectra. The final results are plotted in Fig.~\ref{fig_snr} for the two 
resolution regimes. 

The scatter in the parameters increases for lower signal-to-noise values and is stronger at medium resolution. The differences are very small even for the lowest S/N values similar in order of 
magnitude to the the correlation errors of Fig.~\ref{snr_correlations}. The reference stars of this example show a scatter but not obvious trends in the results suggesting the there are no 
significant systematic errors. 
We note however, that this test does not investigate the efficiency of our normalization method because the normalization was performed on the high 
S/N spectra before adding noise. Additionally, this test does not take into account correlated noise between adjacent pixels which is often the case due to re-sampling occuring for instance, 
when adding two or more spectra together. Therefore, we expect higher discrepancies which will possibly limit the full exploitation 
low S/N spectra of any survey. For instance, \cite{Pancino2017} show the distribution of the S/N for the individual GES spectra obtained so far with the majority of them to be 
around S/N$\sim$20. Nevertheless, many of them account for multiple observations and will be summed to obtain higher S/N and not all low S/N spectra will be used for abundance determinations.

\subsection{Spectral resolution}

The instrumental resolution could affect the precision of our parameters when our spectra are degraded enough. We select the stars in our sample (see Sect.~\ref{high}) observed in the 
highest resolution (HARPS; R$\sim$115\,000) and degrade their spectra to resolution values of typical spectrographs which operate in the optical: UVES (R$\sim$78\,000), UVES (R$\sim$45\,000 
for GES), HERMES (R$\sim$28\,000 for GALAH survey), GIRAFFE (R$\sim$19\,000 for GES-HR10), and GIRAFFE (R$\sim$17\,000 for GES-HR15n). To degrade the spectra, we use the same convolution function as in 
Sect.~\ref{synthesis}. For the medium resolution (R$\sim$19\,000 and $\sim$17\,000), we also change the numerical resolution (distance between two spectral elements) to 0.05\AA{} which is 
typical for GIRAFFE spectra in these settings whereas for the high resolution, we keep the same as for HARPS to 0.01\AA{}. 

We present the results for the stellar parameters for 24 stars in Table~\ref{table_resolution1}. 
When degrading the resolution, the standard deviations mainly for $T_{\mathrm{eff}}$, $\log g$, and $\upsilon  \sin i$ increase which means we lose some precision in our estimates. 
Metallicity is the parameter the least affected from resolution changes compared to changes in $T_{\mathrm{eff}}$ and $\log g$, showing very small scatter even for our lowest R. 
We note however, that the precision errors are larger when moving to lower resolution for all parameters including metallicity.

\begin{table}
\begin{center}
\caption{Mean differences and standard deviations ($\sigma$) for different resolution (R) regimes. The mean differences indicate the parameters from the degraded spectra minus the 
HARPS results for 24 stars.}
\label{table_resolution1}
\scalebox{0.9}{
\begin{tabular}{ccccccccc}
\hline\hline
R    & $\overline{\Delta T_{\mathrm{eff}}}$ & $\sigma$ & $\overline{\Delta \log g}$ & $\sigma$ & $\overline{\Delta [M/H]}$ & $\sigma$ &  $\overline{\Delta \upsilon  \sin i}$ & $\sigma$ \\
     & \multicolumn{2}{c}{(K)}          & \multicolumn{2}{c}{(dex)}  & \multicolumn{2}{c}{(dex)} & \multicolumn{2}{c}{(km\,s$^{-1}$)}  \\
\hline
17\,000 & 5  & 83 & 0.07  & 0.56 & 0.00  & 0.05 & -0.8 & 2.0 \\
19\,000 & -1 & 80 & 0.06  & 0.56 & -0.01 & 0.04 & -0.1 & 1.5 \\
28\,000 & 3  & 63 & -0.05 & 0.11 & 0.02  & 0.03 & 1.1  & 0.9 \\
45\,000 & 5  & 45 & -0.02 & 0.08 & 0.04  & 0.04 & 1.0  & 0.7 \\
78\,000 & 10 & 30 & 0.00  & 0.05 & 0.03  & 0.04 & 1.0  & 0.6 \\
\hline
\end{tabular} }
\end{center}
\end{table}

\subsection{Rotational velocities}\label{rotation}
 
\begin{figure*}
  \centering
   \includegraphics[width=0.7\linewidth]{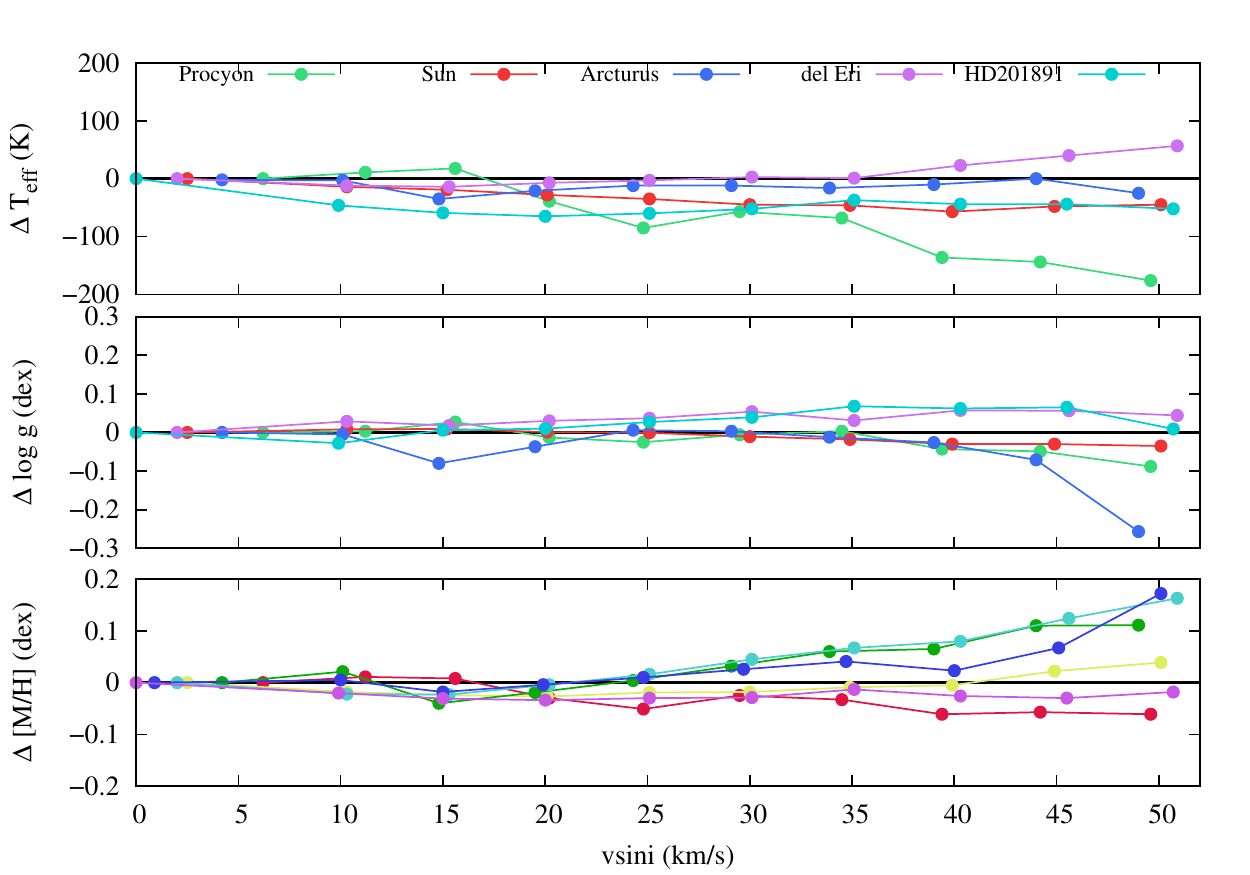}
  \caption{Change in stellar parameters by adding different rotational profiles with increment of 5\,km\,s$^{-1}$ for five different reference stars.}
  \label{fig_rotation}
  \end{figure*}

The spectral lines are affected by stellar rotation following a Doppler shift and changing their profile but preserving their EW. As rotational velocity increases, the 
spectral lines become shallower and blended with the neighboring ones. It is important to check the limitations of our method in a similar analysis as in our previous work
by adding different rotational profiles to stars of different spectral types, namely for the ones we used in Sect.~\ref{initial_conditions}. 
The rotational velocities are convolutions of $\upsilon  \sin i$ from 5 to 50\,kms$^{-1}$ in steps of 5\,km\,s$^{-1}$. In Fig.~\ref{fig_rotation} we plot the change in parameters with 
$\upsilon  \sin i$. We see that up to 35\,km\,s$^{-1}$ our parameters for the three spectral types are well constrained. We notice higher differences in temperature for Procyon. 
The giant and the metal poor star show higher deviations for metallicity and the giant star for surface gravity for the highest rotation profiles.

\section{Spectroscopic parameters for the 451 HARPS GTO sample}\label{451_results}
 
We implement \textit{FASMA} to a sample of 451 well-studied stars from the HARPS GTO planet search program \citep{mayor2003}. The sample is mainly comprised of FGK-type dwarfs with 
90\% of the combined spectra have S/N higher than 200. The stellar parameters of this sample were derived by imposing excitation and 
ionization equilibrium on weak iron lines using the ATLAS9 models in LTE. These stars were firstly analysed in terms of their parameters by \cite{sousa2008} 
and secondly by \cite{tsantaki13} with the same method but a shorter line list to apply corrections on the effective temperature for the cooler stars. Their parameters are in agreement with 
various spectroscopic and photometric works and thus, we consider these parameters reliable for a comparison sample.
 
\begin{figure*}
\centering
\begin{minipage}{0.5\textwidth}
  \centering
  \includegraphics[width=8.5cm, height=6.5cm]{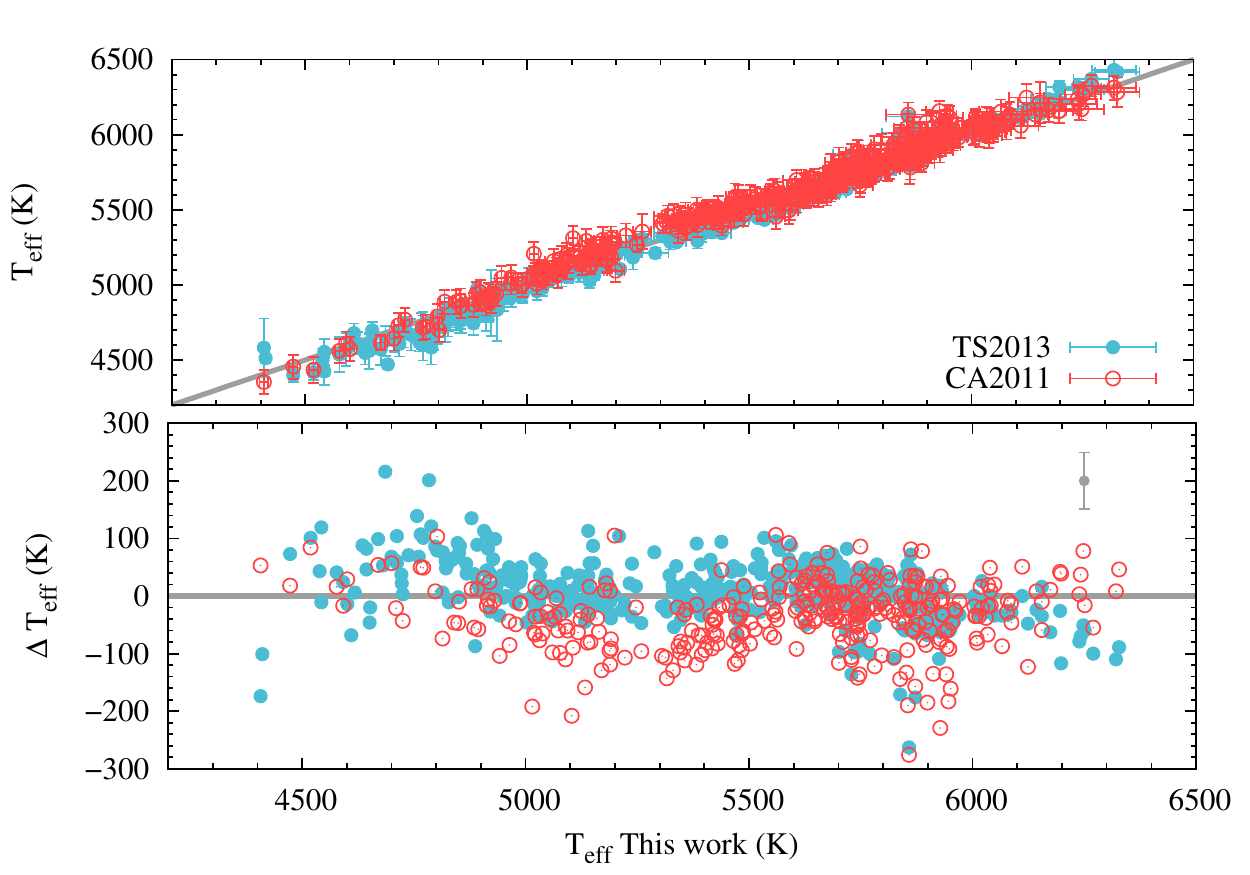} 
\end{minipage}
\hspace{-0.03\textwidth}%
\begin{minipage}{0.5\textwidth}
  \centering
  \includegraphics[width=8.5cm, height=6.5cm]{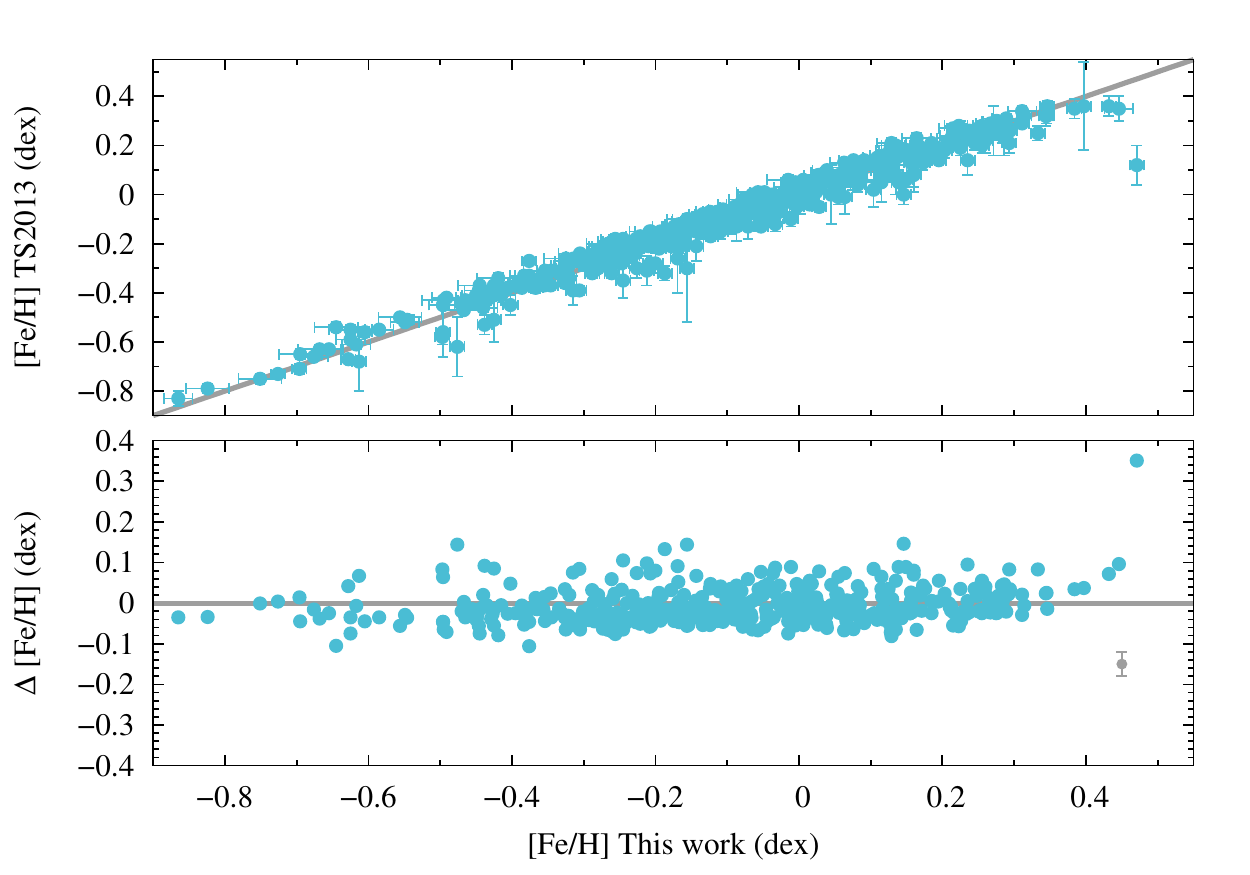} 
\end{minipage}
\begin{minipage}{0.5\textwidth}
  \centering
  \includegraphics[width=8.5cm, height=6.5cm]{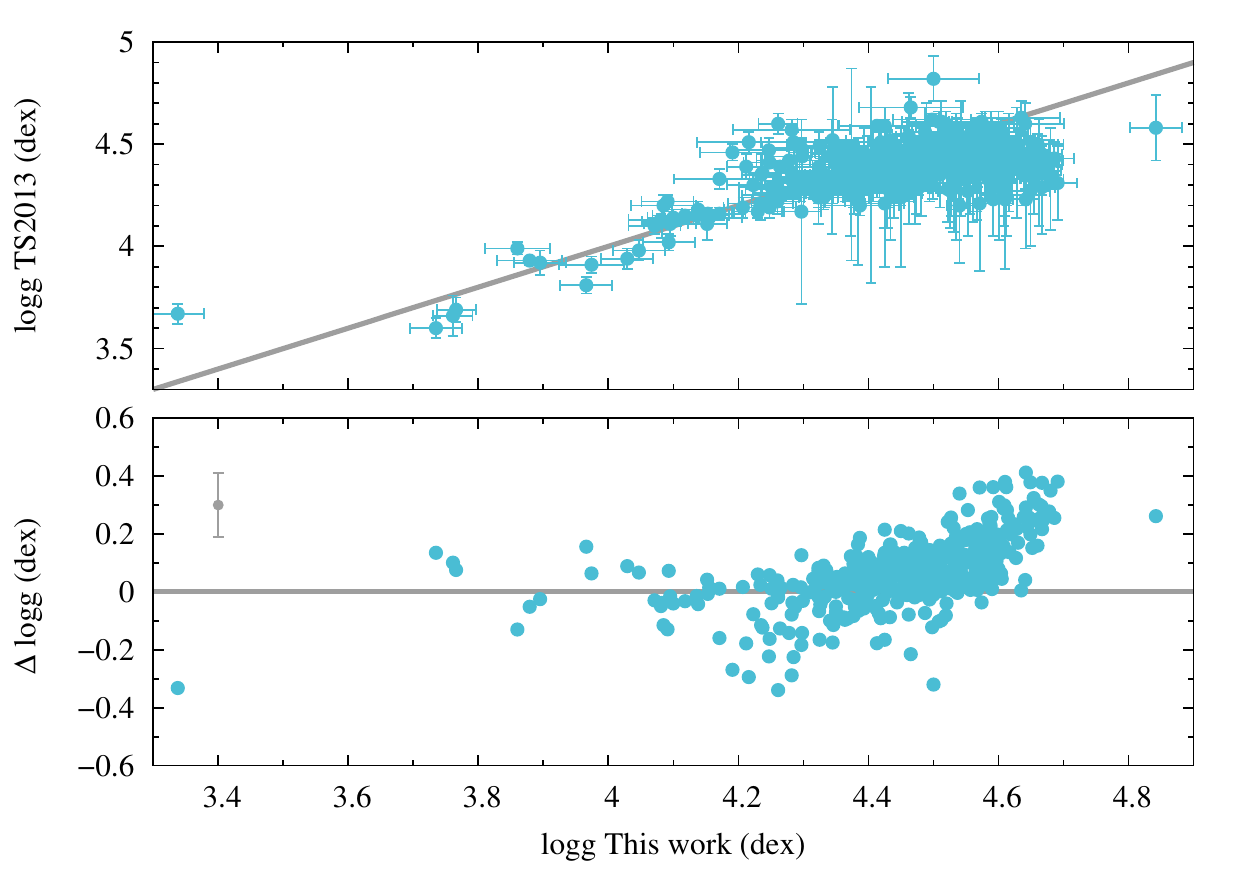} 
\end{minipage}
\hspace{-0.01\textwidth}%
\begin{minipage}{0.5\textwidth}
  \centering
  \includegraphics[width=8.5cm, height=6.5cm]{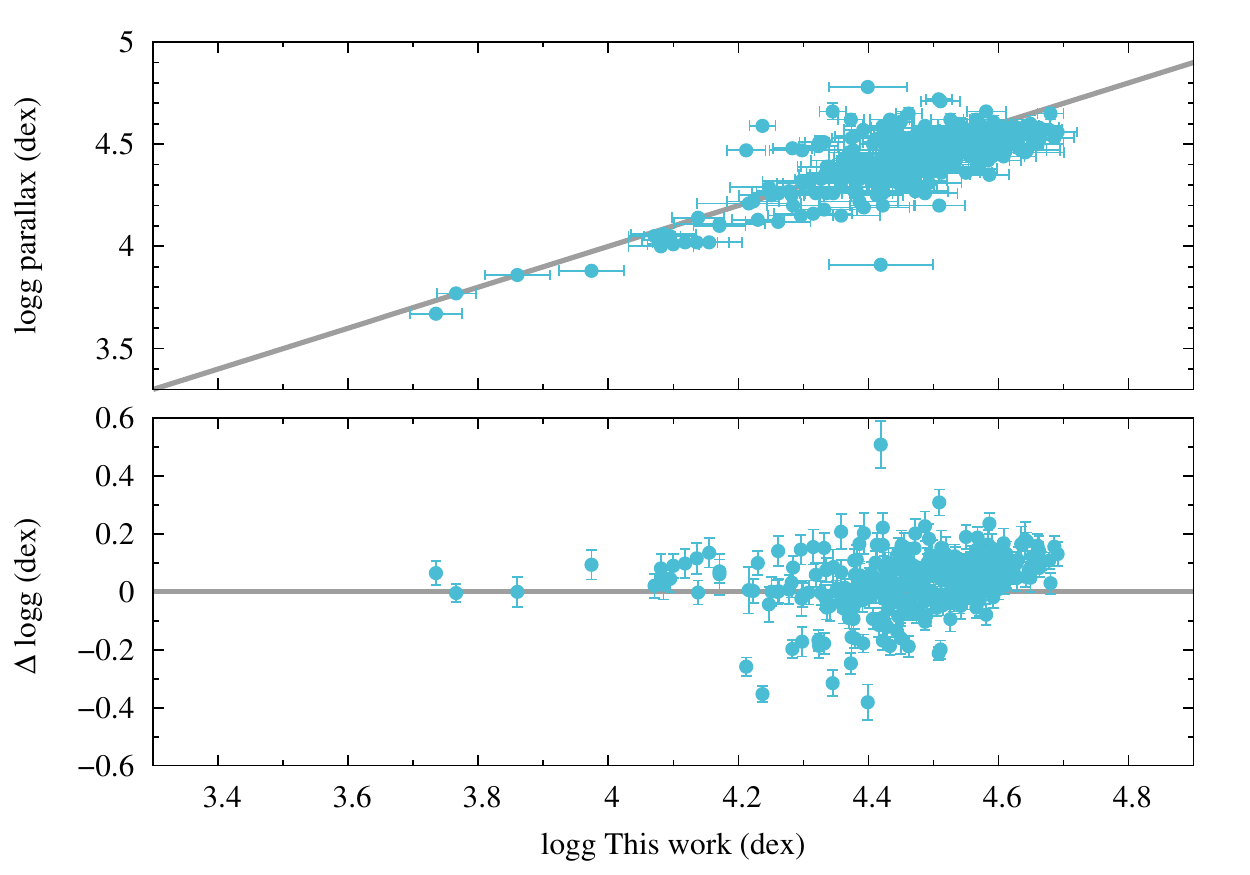} 
\end{minipage}
\caption{Differences in effective temperature, surface gravity and metallicity between the parameters of \citet[][TS2013]{tsantaki13} and this work for the HARPS sample. 
The differences correspond to this work minus others. For $T_{\mathrm{eff}}$, we also plot the comparison with the infrared flux method from \citet[][CA2011]{Casagrande2011} in red open circles.
The bottom right panel shows the difference between gravities derived using the \textit{Gaia} parallaxes. The gray point shows the average error.}
\label{451_fig}
\end{figure*}

We derived the stellar parameters with \textit{FASMA} and the methodology described before for the 451 stars and compared with the results of \cite{tsantaki13} using the same models. 
The iron metallicity for the comparison is derived as explained in Sect.~\ref{minimization}.
The results are depicted in Fig.~\ref{451_fig}. In the same plot, we show the comparison of our effective temperatures with a photometric method, namely the infrared flux method which is 
considered less model dependent, obtained from the Geneva-Copenhagen Survey \citep{Casagrande2011}. There is a very good agreement in all parameters for the comparison with \cite{tsantaki13} 
with the following mean differences: $\overline{\Delta T_{\mathrm{eff}}}$\,=\,--8\,K ($\sigma$\,=\,51\,K), $\overline{\Delta \log g}$\,=\,--0.07\,dex ($\sigma$\,=\,0.11\,dex), and 
$\overline{\Delta [Fe/H]}$\,=\,0.01 ($\sigma$\,=\,0.04\,dex). For the infrared flux method, the mean difference in temperature is: 
$\overline{\Delta T_{\mathrm{eff}}}$\,=\,--34\,K ($\sigma$\,=\,59\,K). We calculated surface gravities using the new parallaxes for the \textit{Gaia} DR1 \citep{Gaia2016} and masses 
from the PARAM 1.3 interface\footnote{\url{http://stev.oapd.inaf.it/cgi-bin/param}} \citep{dasilva06} for the stars with available measurements (342 out of 451). 
The comparison between the spectroscopic gravities and the ones derived from parallaxes (trigonometric gravities) are shown at the bottom left panel of Fig.~\ref{451_fig} with mean 
difference --0.03\,dex ($\sigma$\,=\,0.10\,dex). Our spectroscopic gravities appear to agree better with the trigonometric gravities for the highest $\log g$ values 
compared to the ones derived from the iron ionization balance. 

\section{Spectroscopic parameters for the \textit{Gaia}-ESO benchmark stars}\label{results}

In order to evaluate the performance of our code for a wider range of atmospheric parameters, we select the \textit{Gaia}-ESO benchmark stars as an ideal sample 
to check the accuracy of our method by measuring the discrepancies from the comparison. The effective temperature and surface gravity of this sample are derived 
independently from spectroscopic methods which depend on atmospheric models, using more fundamental relations (e.g. the Stefan-Boltzmann law and the Newton law of gravitation). 
Metallicities are derived using high resolution spectroscopy from various groups and the fundamental $T_{\mathrm{eff}}$ and $\log g$ values. The parameters of the 
benchmark stars are used in GES to homogenize the results of different analysis groups. The sample, excluding the M stars which require different treatment in 
terms of line list and models, contains 29 stars and is described in \citet{Heiter2015} where their $T_{\mathrm{eff}}$ and $\log g$ values are derived. Their metallicities are taken 
from \citet{jofre2014}. We added the latest metal-poor benchmark stars suggested by \citet{Hawkins2016}, reaching a total of 34 stars. 
Apart from their reliable parameters, the sample has spectra available in both high and medium resolution, both in high S/N. 

The spectral analysis in this work is based on the assumptions of LTE and 1D geometry in stellar atmospheres to reduce the complexity of the problem. 
However, there are conditions when the deviations of 1D static LTE models occur, especially for giant, hot, and metal-poor stars \citep{bergemann2012}. 
Non-LTE (NLTE) effects have an impact mainly on neutral iron lines and therefore on their abundance determinations. 
It is reported that stars sensitive to NLTE effects could have differences in their metallicity up to 0.10\,dex for the stars in this work \citep{jofre2014}. 
For consistency, we compare our results with iron metallicity for the benchmark stars not corrected for NLTE effects. 
For the metallicity comparison, we derive iron metallicity abundances for our sample. 

\subsection{``High'' resolution}\label{high}

\begin{figure}
  \centering
   \includegraphics[width=1.05\linewidth]{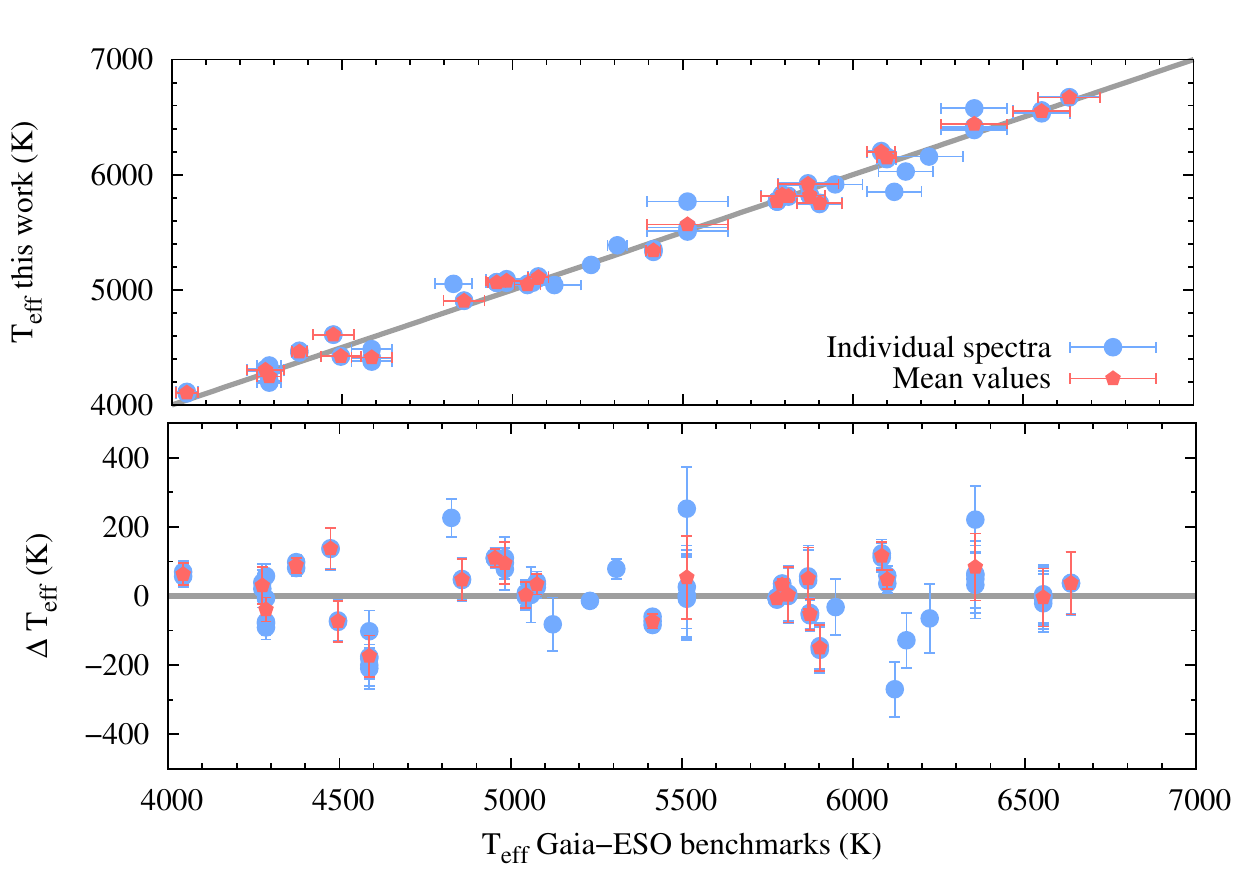} \\
   \includegraphics[width=1.05\linewidth]{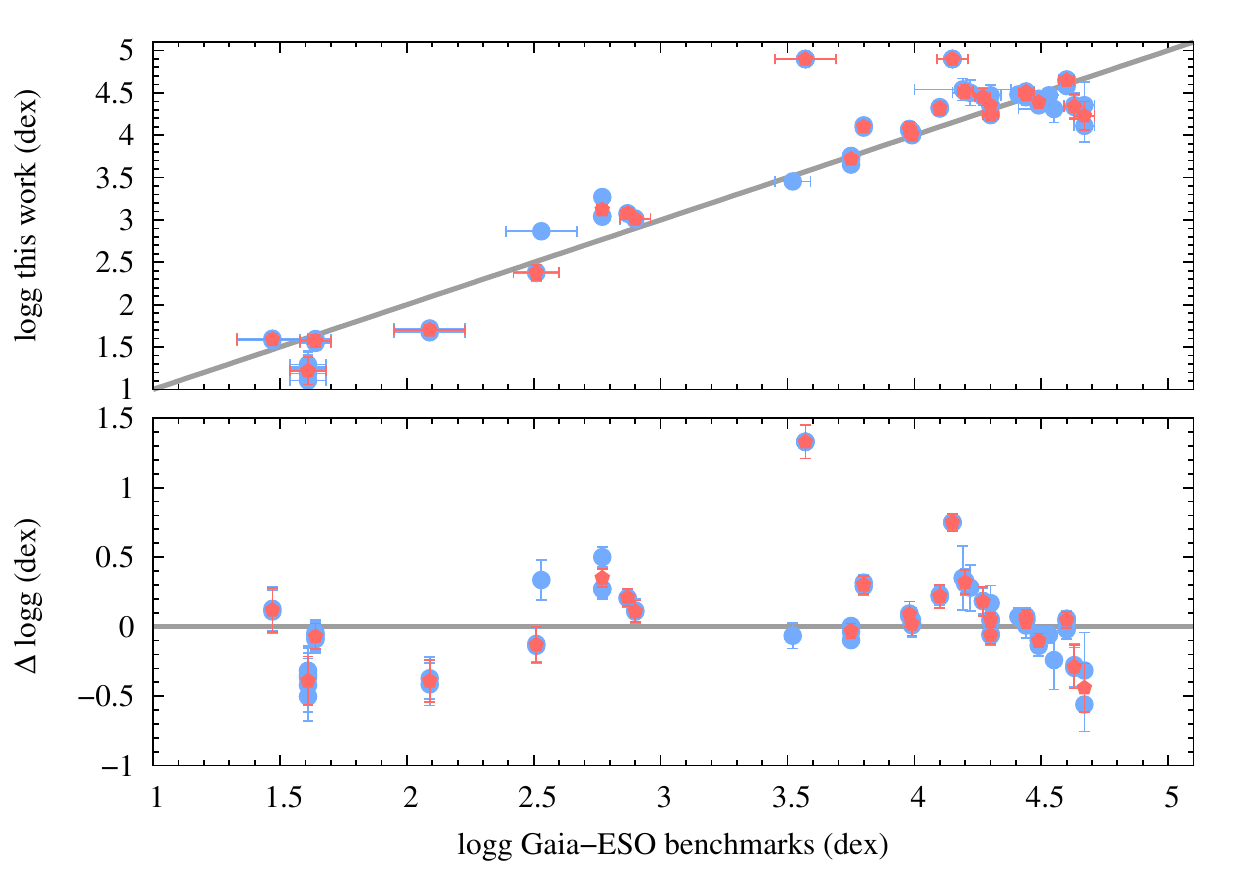} \\
   \includegraphics[width=1.05\linewidth]{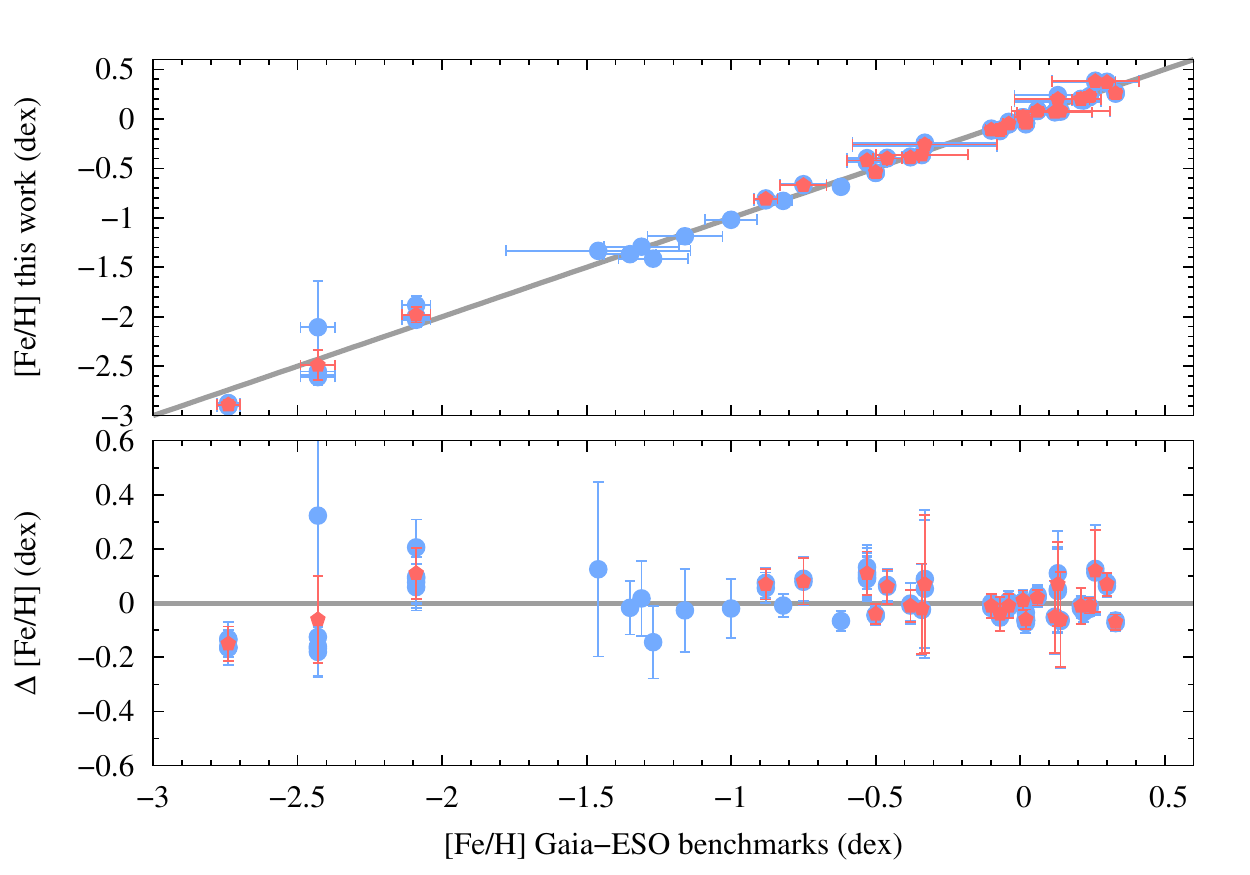}
  \caption{Differences in effective temperature, surface gravity and metallicity between the GES benchmark parameters and this work using the high resolution spectra and ATLAS9 models 
  (blue circles). For stars with multiple spectra, their mean values are plotted with red pentagons. }
  \label{high_res}
\end{figure}

The original spectra are taken from \citet{blanco2014b}\footnote{\url{http://www.blancocuaresma.com/s/benchmarkstars/}}. For most stars, there are more than one spectra available taken from 
different instruments (95 spectra in total). We derived the atmospheric parameters using the methodology described previously and the results are shown in Fig.~\ref{high_res}. 
This analysis is performed using ATLAS9 models. We also completed the same analysis using the public grid of ATLAS-APOGEE models (Fig.~\ref{high_res_apogee}) and the MARCS models 
(Fig.~\ref{high_res_marcs}). To better understand the discrepancies in our sample, we divide the stars into different luminosity classes. 
In Table~\ref{results_table_high}, we present the mean differences between this work and literature values for the 95 different spectra. 
We notice that all models show similar differences for all parameters and there is not evident choice on which grid delivers better results.

\begin{figure*}
  \centering
   \includegraphics[width=0.75\linewidth]{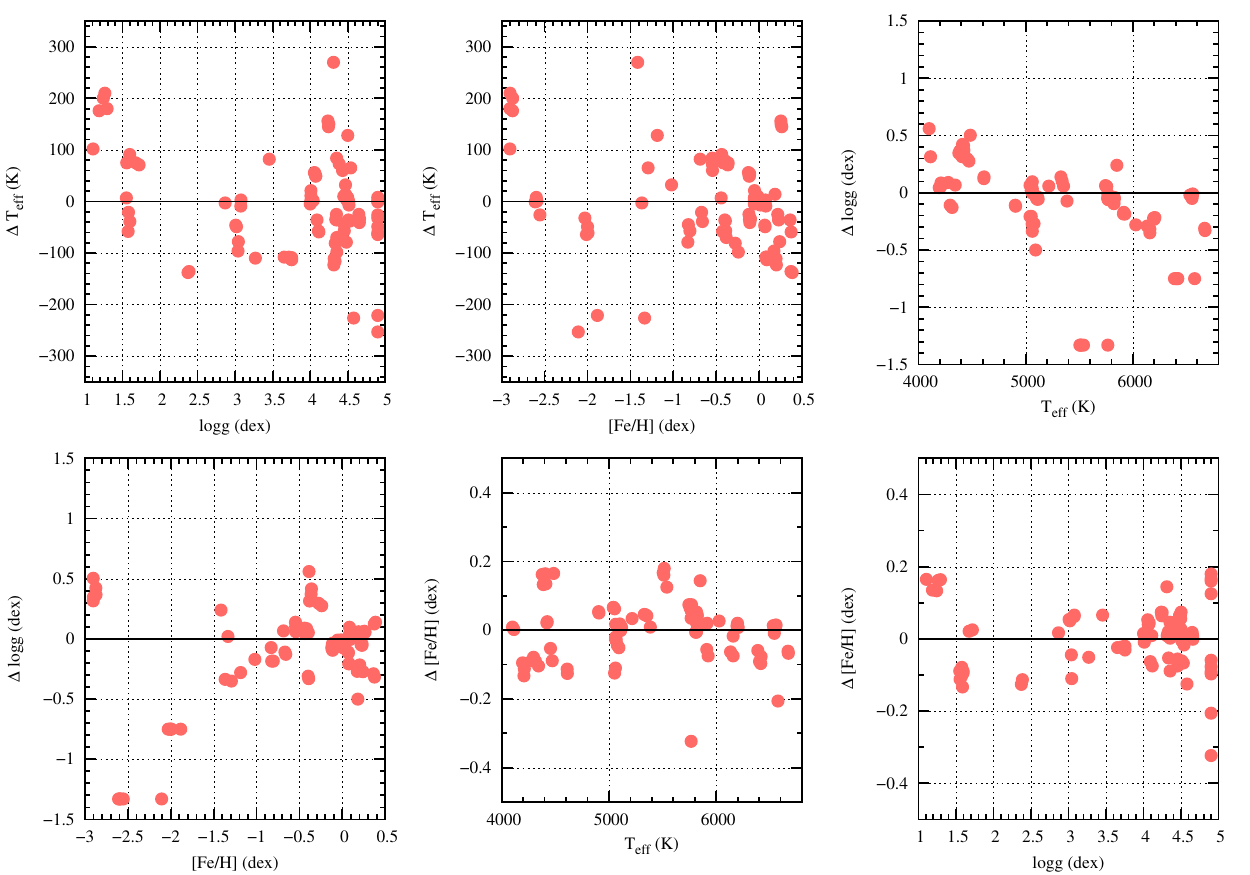}
  \caption{Correlations in parameters with differences with the benchmark values for the 95 spectra. The x-axis shows the values of this work.}
  \label{high_cor_kurucz}
\end{figure*}

We could not constrain the surface gravity of HD\,140283 (the outlier in the middle panel of Fig.~\ref{high_res}) possibly because there are very few iron lines due to its very low metallicity 
(--2.43\,dex) to indicate $\log g$ (second most metal-poor star in our sample). Due to this star we obtain the greatest differences for $\log g$ for the sub-giants 
because 5 out of 17 spectra belong to HD\,140283 (e.g. by excluding it we obtain $\overline{\Delta \log g}$ = 0.06\,dex for the ATLAS9 models). 
The highest differences for $T_{\mathrm{eff}}$ are observed for K-type stars. A possible explanation could be due to poor normalization because there are less continuum 
points at these temperatures or due to poor calibration of the atomic data. Metallicities are very well constrained for all luminosity classes.
Figure~\ref{high_cor_kurucz} shows the correlation of the residual differences with the parameters. We only see a correlation between $\log g$ and $T_{\mathrm{eff}}$ meaning 
that we underestimate gravities for the cooler stars and overestimate for the hotter stars.

\citet{blanco2014a} derived the atmospheric parameters for the same spectra as in this work but with different line lists and with their spectral synthesis package, iSpec. 
Their results (see their Table 6) are also in agreement with the GES benchmark values and consistent with ours as well. \citet{Smiljanic2014} provide parameters for the benchmark stars 
from different groups using EW methods and spectral synthesis techniques also derived from the same spectra as in this work and including new GES observations. 
Their differences from the benchmark parameters, namely for $T_{\mathrm{eff}}$ and $\log g$ (see their Table 4), are higher compared our results, possibly because some of the spectra they used 
have low S/N values. 

\begin{table*}
\begin{center}
\caption{Average difference and standard deviation between the synthetic spectral synthesis technique and the reference values for the sample in high resolution. N represents the number 
of spectra analysed in each group.}
\label{results_table_high}
\begin{tabular}{lccccccccc}
\hline\hline
     & $\overline{\Delta T_{\mathrm{eff}}}$ & $\sigma$ & $\overline{\Delta \log g}$ & $\sigma$ & $\overline{\Delta [Fe/H]}$ & $\sigma$ & $\overline{\Delta \upsilon\sin i}$ & $\sigma$ & N \\
     & \multicolumn{2}{c}{(K)} & \multicolumn{2}{c}{ (dex)} & \multicolumn{2}{c}{ (dex)} & \multicolumn{2}{c}{ (km\,s$^{-1}$)} &  \\
\hline
\multicolumn{10}{c}{ATLAS9} \\
Whole sample         & 7   & 91  & 0.12  & 0.38 & 0.00  & 0.08 & 0.5  & 1.8 & 95 \\
F-type dwarfs        & 0   & 102 & 0.31  & 0.34 & 0.03  & 0.08 & 0.8  & 2.3 & 16 \\
G-type dwarfs        & -5  & 70  & 0.05  & 0.09 & -0.03 & 0.04 & 0.5  & 1.4 & 29 \\
K-type dwarfs        & 74  & 62  & -0.14 & 0.23 & 0.02  & 0.05 & 0.7  & 2.3 & 9  \\
FGK-type sub-giants  & 36  & 88  & 0.44  & 0.61 & -0.02 & 0.12 & -0.4 & 2.1 & 17 \\
GK-type giants       & -20 & 105 & -0.05 & 0.27 & 0.00  & 0.10 & 1.0  & 1.4 & 24 \\
\hline
\multicolumn{10}{c}{ATLAS-APOGEE} \\
Whole sample         & -1  & 93  & 0.06  & 0.39 & -0.03 & 0.09 & 0.0  & 1.8 & 95 \\
F-type dwarfs        & 5   & 104 & 0.25  & 0.39 & -0.01 & 0.07 & 0.8  & 2.3 & 16 \\
G-type dwarfs        & -9  & 69  & -0.01 & 0.10 & 0.01  & 0.04 & 0.0  & 1.5 & 29 \\
K-type dwarfs        & 41  & 73  & -0.23 & 0.26 & -0.04 & 0.06 & 0.5  & 1.6 & 9  \\
FGK-type sub-giants  & 32  & 112 & 0.37  & 0.65 & -0.06 & 0.12 & -0.7 & 1.3 & 17 \\
GK-type giants       & -35 & 100 & -0.07 & 0.23 & -0.07 & 0.09 & -0.3 & 1.9 & 24 \\
\hline
\multicolumn{10}{c}{MARCS} \\
Whole sample         & -6  & 85  & 0.05  & 0.41 & -0.02 & 0.09 & 0.5  & 1.8 & 95 \\
F-type dwarfs        & -12 & 97  & 0.25  & 0.39 & -0.02 & 0.07 & 0.8  & 2.3 & 16 \\
G-type dwarfs        & -28 & 64  & -0.04 & 0.09 & 0.01  & 0.04 & 0.4  & 1.4 & 29 \\
K-type dwarfs        & 60  & 85  & -0.14 & 0.29 & -0.01 & 0.07 & 0.7  & 2.4 & 9  \\
FGK-type sub-giants  & -5  & 106 & 0.33  & 0.67 & -0.09 & 0.12 & -0.4 & 2.1 & 17 \\
GK-type giants       & 7   & 69  & 0.01  & 0.25 & -0.01 & 0.12 & 1.0  & 1.4 & 24 \\
\hline
\end{tabular}
\end{center}
\end{table*}

\subsection{``Medium'' resolution}

For medium resolution, we query the GES archive for spectra (ESO program: 188.B-3002(G)). 
The available spectra account for 25 out of 34 stars from our sample and are observed with the VLT-FLAMES multi-fiber facility fed to GIRAFFE spectrograph. 
The wavelength coverage of GIRAFFE is split in different set-ups, each with different resolution. 
In this work, we obtain spectra for each star from two set-up configurations of GIRAFFE: HR10 (5399--5619\,\AA{}) and HR15n (6470--6790\,\AA{}) with R$\sim$19\,000 and 17\,000 respectively. 
The spectra are firstly corrected for radial velocities shifts. Several spectra are co-added into one to increase the S/N. 

\begin{figure}
  \centering
   \includegraphics[width=1.05\linewidth]{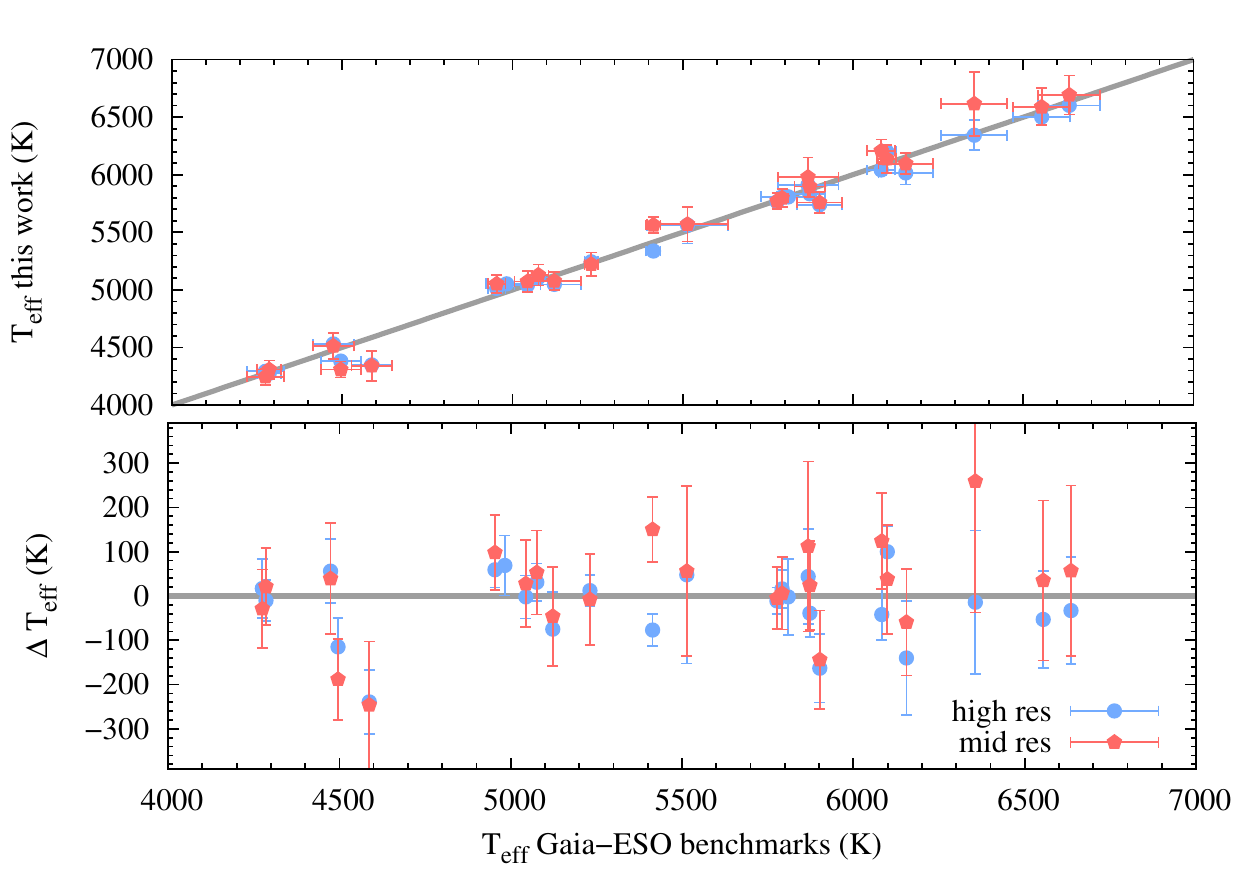} \\
   \includegraphics[width=1.05\linewidth]{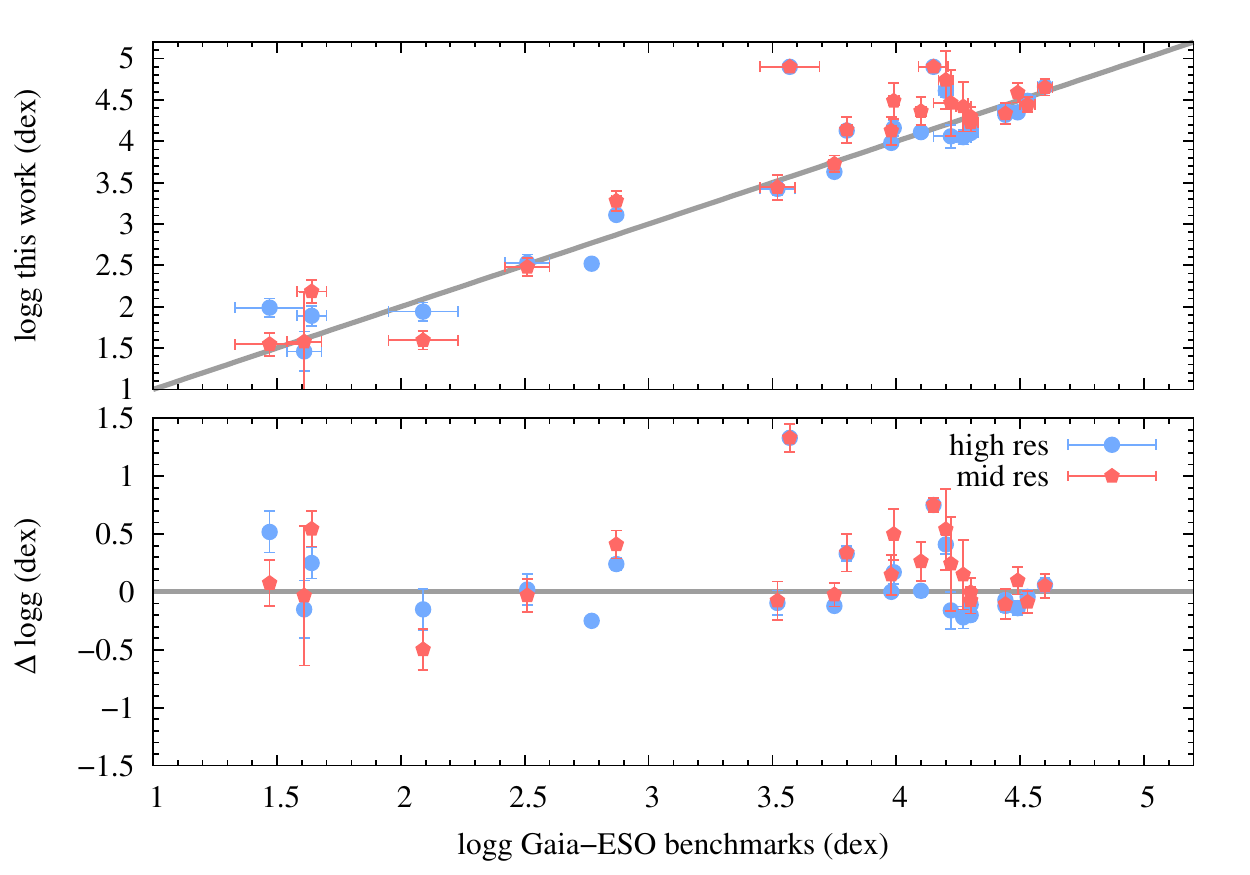} \\
   \includegraphics[width=1.05\linewidth]{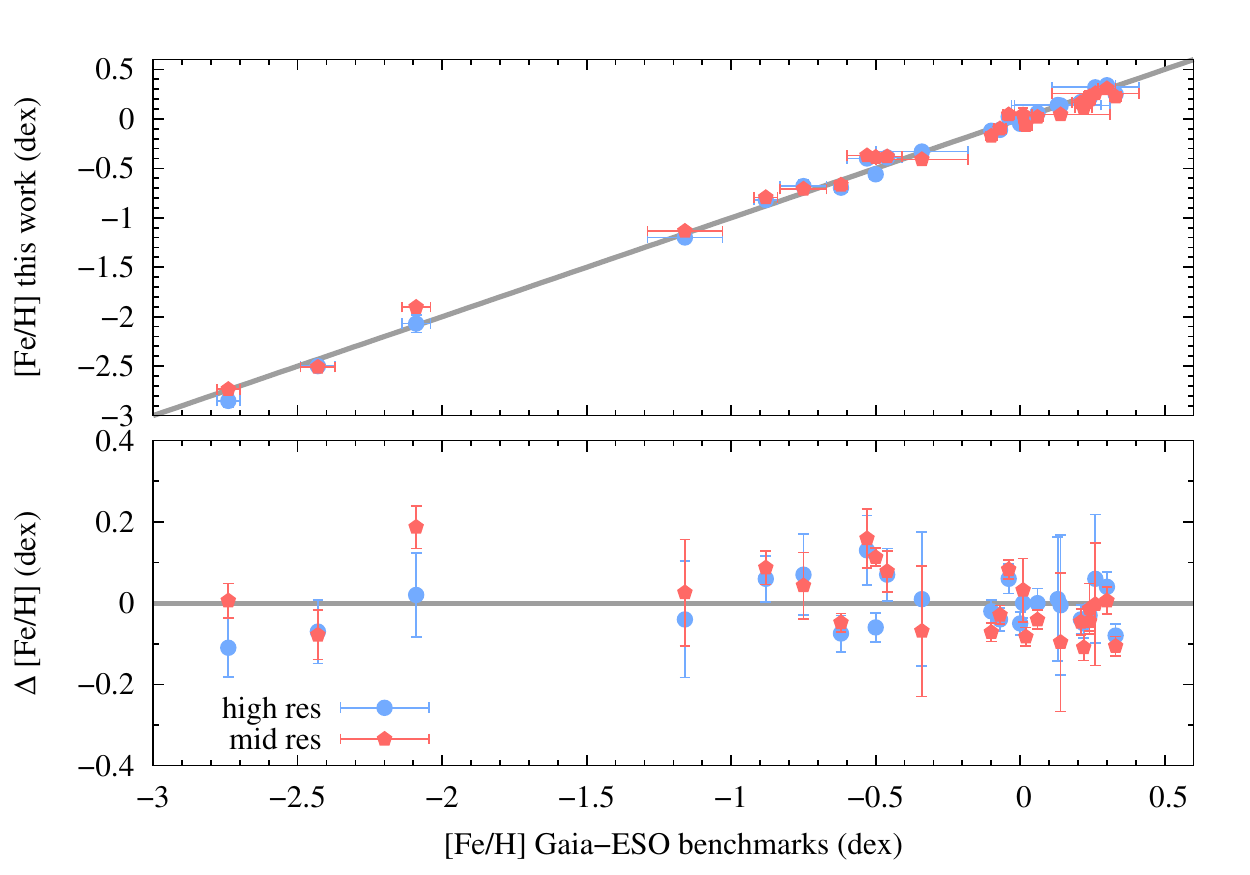}
  \caption{Differences in effective temperature, surface gravity and metallicity between the GES benchmark parameters and this work using the HR10 GIRAFFE spectra and ATLAS9 models (red pentagons). 
  For comparison we plot the mean parameters of high resolution spectra using the same line list (blue circles).}
  \label{mid_res_hr10}
\end{figure}

\begin{figure}
  \centering
   \includegraphics[width=1.05\linewidth]{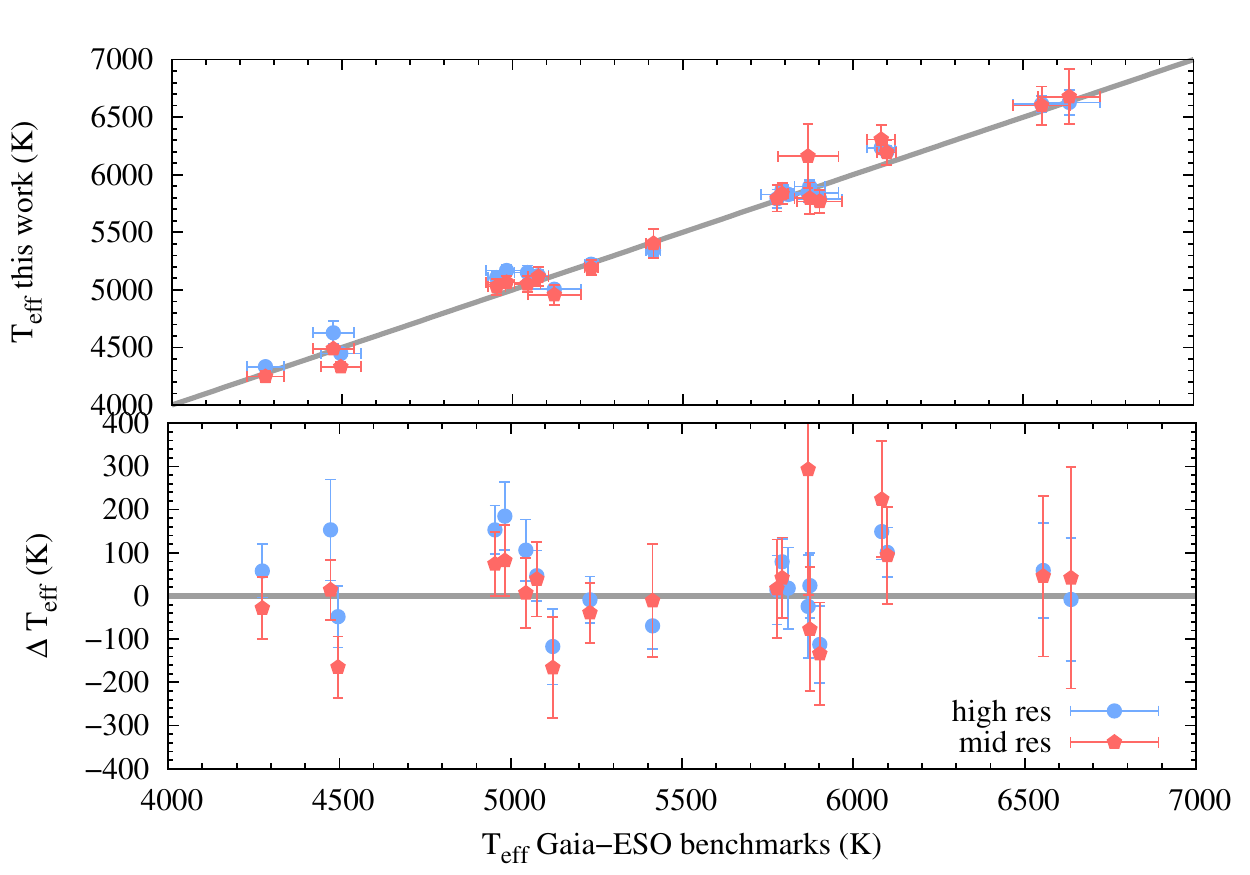} \\
   \includegraphics[width=1.05\linewidth]{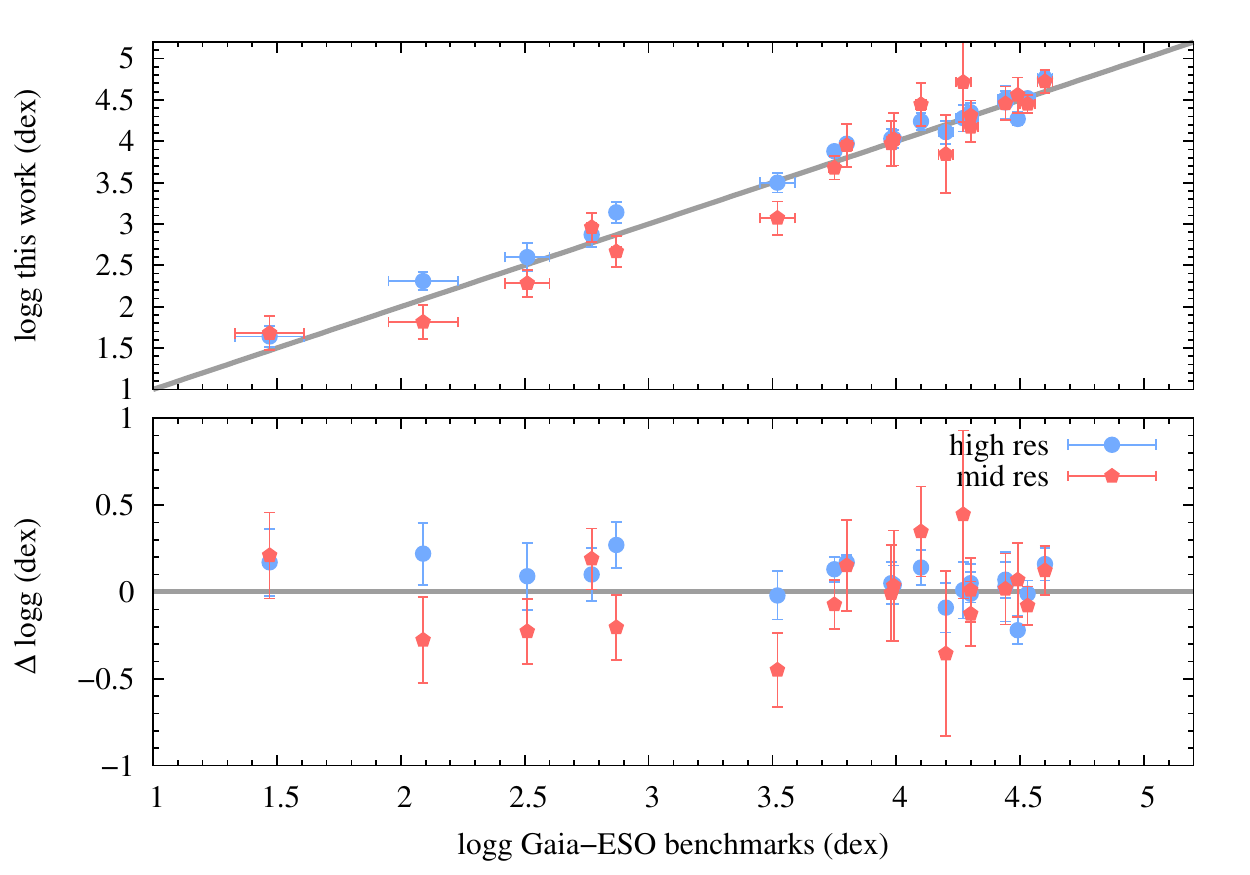} \\
   \includegraphics[width=1.05\linewidth]{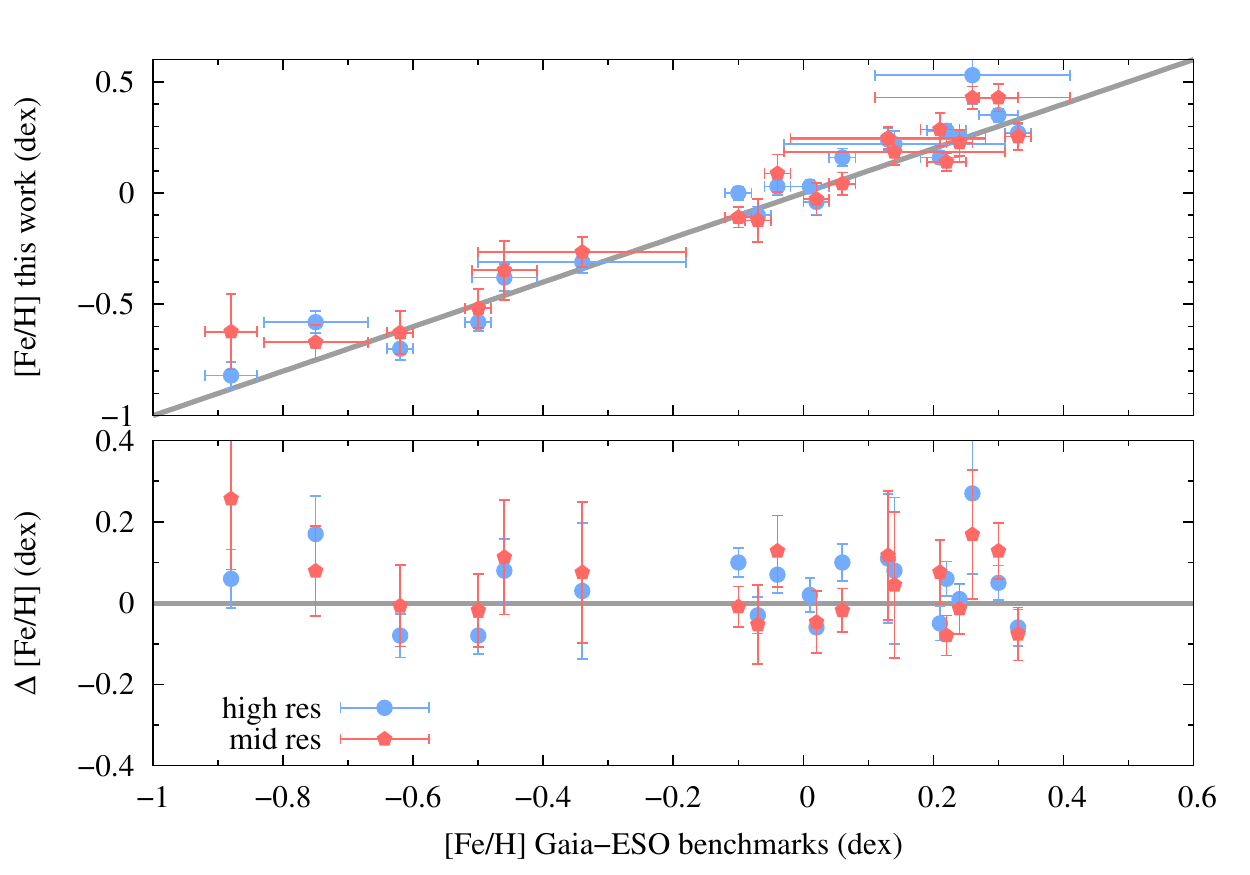}
  \caption{Same as in Fig.~\ref{mid_res_hr10} but for the HR15n set-up. }
  \label{mid_res_hr15n}
\end{figure}

We derive the stellar parameters for the two set-ups separately and the results are shown in Figs.~\ref{mid_res_hr10} and \ref{mid_res_hr15n} using the ATLAS9 models. 
For a homogeneous comparison, we derived the atmospheric parameters in high resolution using the spectra of Sect.~\ref{high} but we used the part of our line list which 
covers the two wavelength regions of the GIRAFFE spectrograph. The line lists used for HR10 and HR15n are smaller, 918 and 219 lines respectively, because the spectral regions 
are limited. We also provide results using the ATLAS-APOGEE models in  Figs.~\ref{mid_res_hr10_apogee}--\ref{mid_res_hr15n_apogee} and for the MARCS models in 
Figs.~\ref{mid_res_hr10_marcs}--\ref{mid_res_hr15n_marcs}. 

For the HR15n set-up, we could only constrain metallicity for the very metal poor stars ($[Fe/H]<$--2.0\,dex) by fixing the rest of the parameters to the literature values so we do not 
present them in the corresponding plots. In Table~\ref{results_table_medium}, we show the mean differences of our results with the GES values. We did not separate by luminosity class because 
of the small number of spectra. We point that the results from high resolution show smaller dispersion compared to the ones from medium resolution. 
The HR10 set-up has smaller dispersion compared to HR15n both in high and medium resolution for the effective temperature. The high dispersion in $\log g$ for high resolution is because of 
our outlier, HD\,140283. All models provide similar results. Metallicity is in very good agreement for both set-ups and resolutions. Yet, it appears to be an overestimation of temperature 
for HR15n in medium resolution which we should not neglect for the ATLAS9 models. Moreover, the error bars of the medium resolution are larger compared to high resolution.

 \begin{table*}
 \caption{Average difference and standard deviation between the synthetic spectral synthesis technique and the reference values. N represents the number of spectra analysed in each group.}
 \label{results_table_medium}
 \begin{tabular}{lccccccccc}
 \hline\hline
      & $\overline{\Delta T_{\mathrm{eff}}}$ & $\sigma$ & $\overline{\Delta \log g}$ & $\sigma$ & $\overline{\Delta [Fe/H]}$ & $\sigma$ & $\overline{\Delta \upsilon\sin i}$ & $\sigma$ & N\\
      & \multicolumn{2}{c}{(K)} & \multicolumn{2}{c}{ (dex)} & \multicolumn{2}{c}{(dex)} & \multicolumn{2}{c}{ (km\,s$^{-1}$)} & \\
 \hline
 \multicolumn{10}{c}{ATLAS9} \\

Medium resolution HR10     & 16  & 102 & 0.19  & 0.36 & 0.00  & 0.08 & -0.8 & 2.5 & 25 \\
High resolution HR10       & -22 & 79  & 0.09  & 0.36 & -0.01 & 0.06 & 0.4  & 1.8 & 81 \\
Medium resolution HR15n    & 19  & 112 & -0.01 & 0.23 & 0.04  & 0.09 & 2.9  & 3.3 & 21 \\
High resolution HR15n      & 32  & 87  & 0.06  & 0.11 & 0.03  & 0.09 & -0.1 & 1.5 & 68 \\
Medium resolution combined & 1   & 83  & 0.12  & 0.35 & 0.01  & 0.09 & -0.6 & 2.4 & 25 \\
\hline
\multicolumn{10}{c}{ATLAS-APOGEE} \\
Medium resolution HR10  & 1   & 130 & 0.16  & 0.35 & -0.05 & 0.10 & -1.1 & 2.6 & 25 \\
High resolution HR10    & -28 & 102 & 0.01  & 0.40 & -0.03 & 0.08 & 0.3  & 1.8 & 81 \\
Medium resolution HR15n & 16  & 161 & 0.01  & 0.29 & -0.01 & 0.10 & 2.6  & 3.3 & 21 \\
High resolution HR15n   & -23 & 82  & 0.02  & 0.15 & 0.05  & 0.09 & 0.8  & 1.7 & 68 \\
\hline
\multicolumn{10}{c}{MARCS} \\
Medium resolution HR10  & 1   & 130 & 0.16 & 0.35 & -0.05 & 0.10 & -1.1 & 2.6 & 25 \\
High resolution HR10    & -18 & 78  & 0.09 & 0.35 & -0.02 & 0.07 & 0.3  & 1.8 & 81 \\
Medium resolution HR15n & 45  & 160 & 0.04 & 0.29 & 0.02  & 0.12 & 2.5  & 3.3 & 21 \\
High resolution HR15n   & -12 & 88  & 0.07 & 0.28 & 0.04  & 0.11 & 0.8  & 1.7 & 68 \\
\hline
\end{tabular}
\end{table*}

In this work, we use spectra from the same spectrograph (GIRAFFE) but at different set-ups covering different wavelength regions of the spectrum and derive parameters for each of the two set-ups. 
In this case, we can increase the spectral information by combining, for a given star, the spectra from both set-ups to increase the number of lines, expecting to raise the precision of the method. 
Attention should be paid to convolve each synthetic spectrum with the corresponding resolution because it is slightly different for the two set-ups. We have calculated parameters for the 
combined spectra in Table~\ref{results_table_medium}. We note that the results are equally good to the results for medium resolution. Moreover, time consumption for the case of the combined 
spectra is less than calculating parameters separately for the two individuals. 

\section{Discussion}\label{discussion}

\subsection{Comparison with the EW method}

\textit{FASMA} also provides parameters based on the measurements of iron EW and by imposing ionization and excitation balance in a fully automatic way \citep[see details in][]{Andreasen2017}. 
An interesting test is to compare both methods to investigate any discrepancies in their results. For the analysis with the EW method, we use the iron line lists from \cite{sousa2008} and 
\cite{tsantaki13} where the latter used for the cooler stars. The EW values are measured automatically for the a given spectrum. 
The parameters are derived by satisfying the following criteria: 
i) the slope between iron abundance and excitation potential should be lower than 0.001, ii) the slope between iron abundance and reduced EW should be lower than 0.003, iii) the difference 
between the average abundances of \ion{Fe}{i} and \ion{Fe}{ii} should be less than 0.01 (these limits are defined in \cite{Andreasen2017}). The models used here are the ATLAS9.
We could not obtain viable solutions for 13 spectra out of the 95 with the EW method which were the most metal poor and coolest stars. 

\begin{figure}
  \centering
   \includegraphics[width=1.05\linewidth]{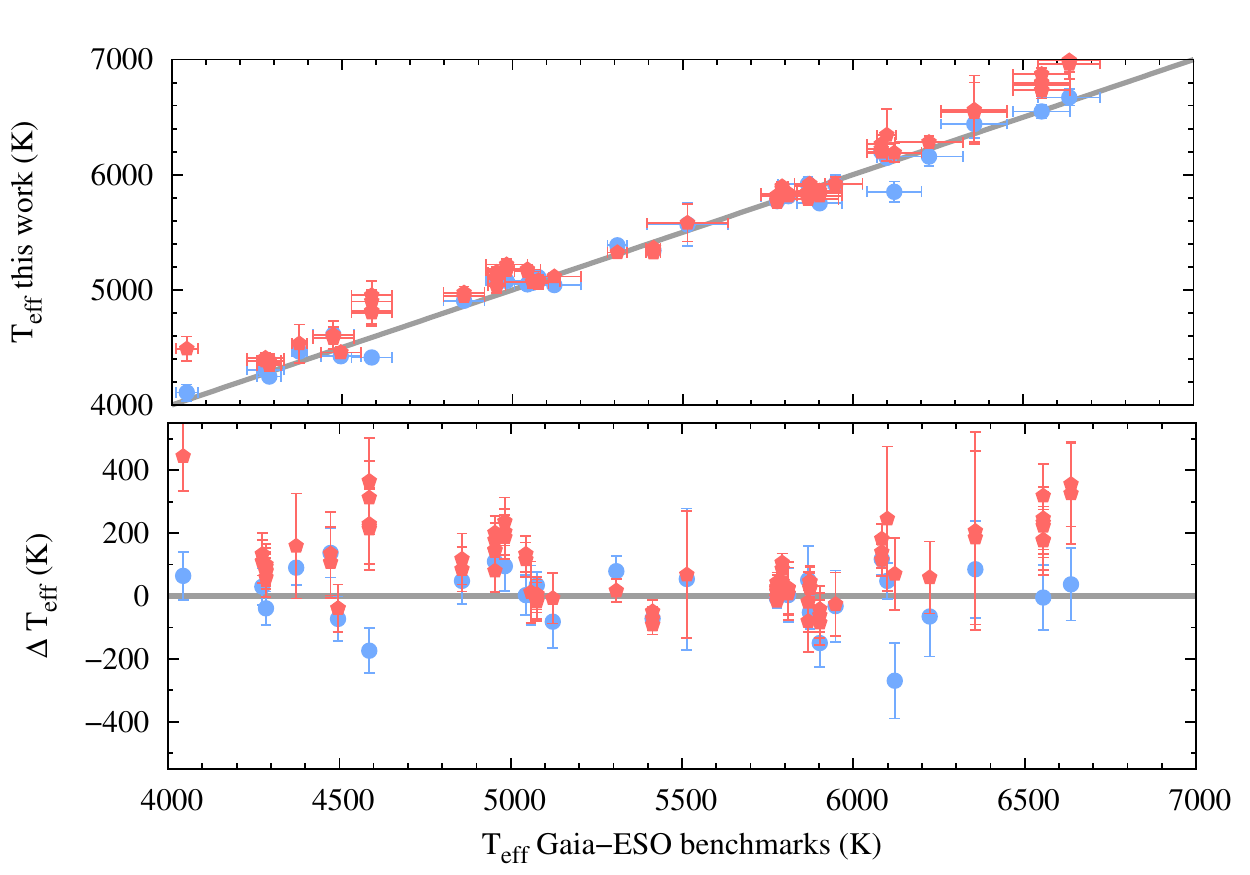} \\
   \includegraphics[width=1.05\linewidth]{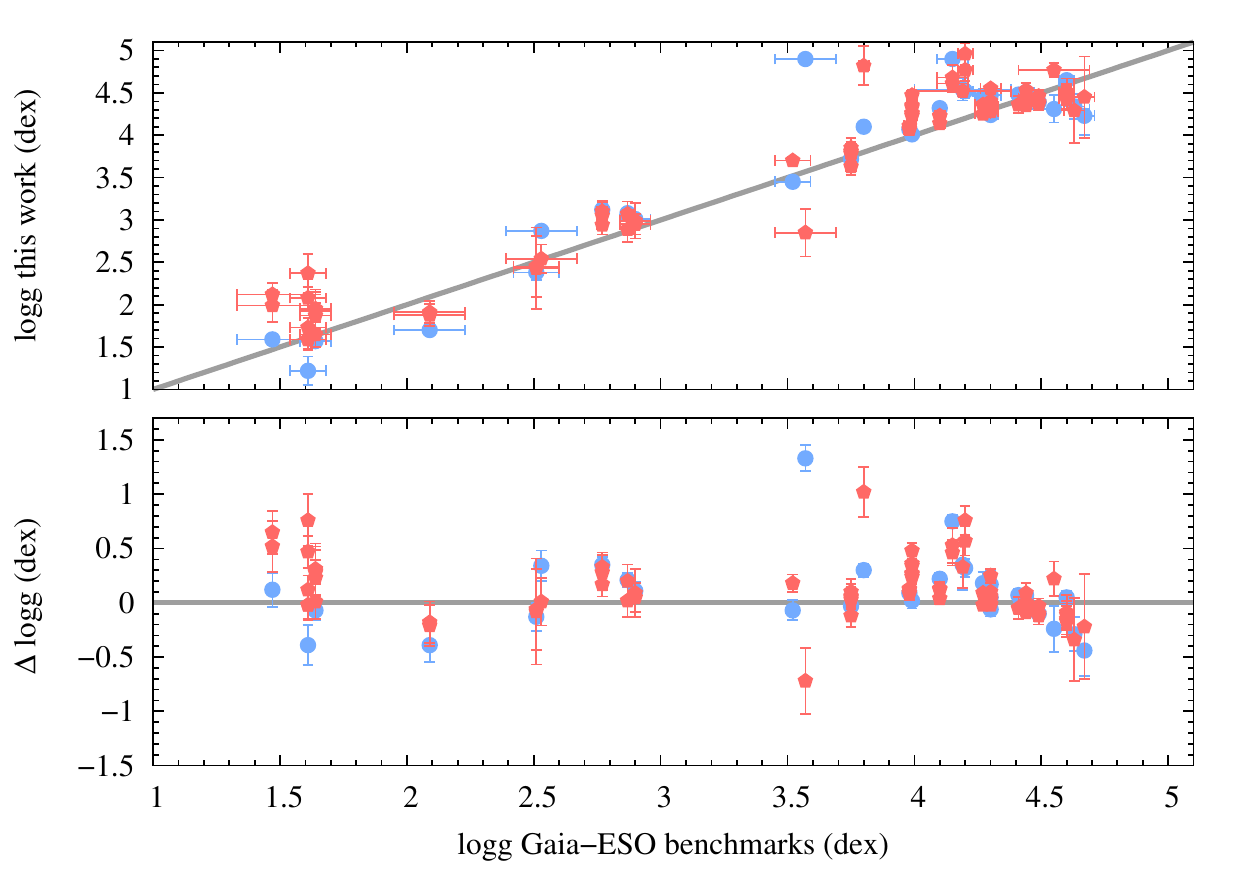} \\
   \includegraphics[width=1.05\linewidth]{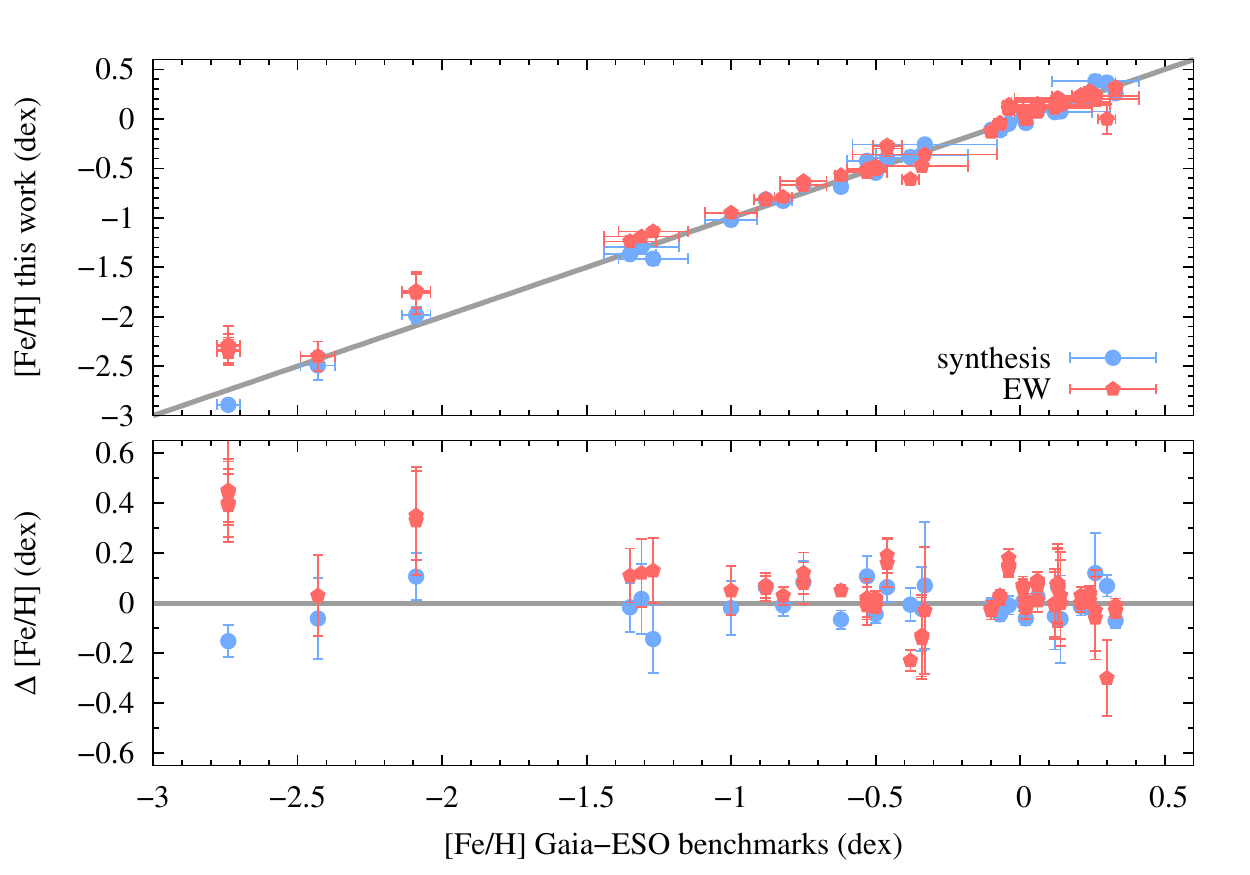}
  \caption{Differences in effective temperature, surface gravity and metallicity between the GES benchmark parameters and the synthesis results from this work (blue squares). The green circles 
  represent the differences with the EW results from this work.}
  \label{high_res_EW}
  \end{figure}

The statistics between the two methods for the common stars are presented in Table~\ref{results_table_high_EW} and the results are shown in Fig.~\ref{high_res_EW}. There is an 
overestimation of temperature for the EW method mainly because of the F-type stars ($T_{\mathrm{eff}}>$\,6000\,K). Surface gravities and metallicities on the other hand are well 
constrained for both methods. 

\begin{table}
\begin{center}
\caption{Average difference and standard deviation between the synthetic spectral synthesis technique and the EW method with the reference values in high resolution for 82 spectra.}
\label{results_table_high_EW}
\begin{tabular}{lcccccc}
\hline\hline
     & $\overline{\Delta T_{\mathrm{eff}}}$ & $\sigma$ & $\overline{\Delta \log g}$ & $\sigma$ & $\overline{\Delta [Fe/H]}$ & $\sigma$  \\
     & \multicolumn{2}{c}{(K)} & \multicolumn{2}{c}{ (dex)} & \multicolumn{2}{c}{ (dex)}  \\
\hline
EW method          & 97 & 115 & 0.11 & 0.26 & 0.06  & 0.13 \\
Spectral synthesis & 3  & 90  & 0.09 & 0.34 & 0.00  & 0.07 \\
\hline
\end{tabular}
\end{center}
\end{table}

\subsection{Correlated errors}

One important aspect to consider is potential degeneracies between the parameters. Surface gravity is the atmospheric parameter most difficult to constrain with spectral analysis 
methods, in particular with methods based on neutral or singly ionized iron lines and because of that, many works treat gravity as a fixed parameter to values from other methods 
\citep[e.g.][]{mortier13_gravity}. With the upcoming releases of \textit{Gaia} \citep{Gaia2016}, we will have precise parallaxes and therefore, distances for millions of stars. 
We can constrain surface gravity using the parallax information by calculating trigonometric gravities, assuming we can obtain accurate effective temperatures and masses.

\begin{figure*}
  \centering
   \includegraphics[width=0.8\linewidth]{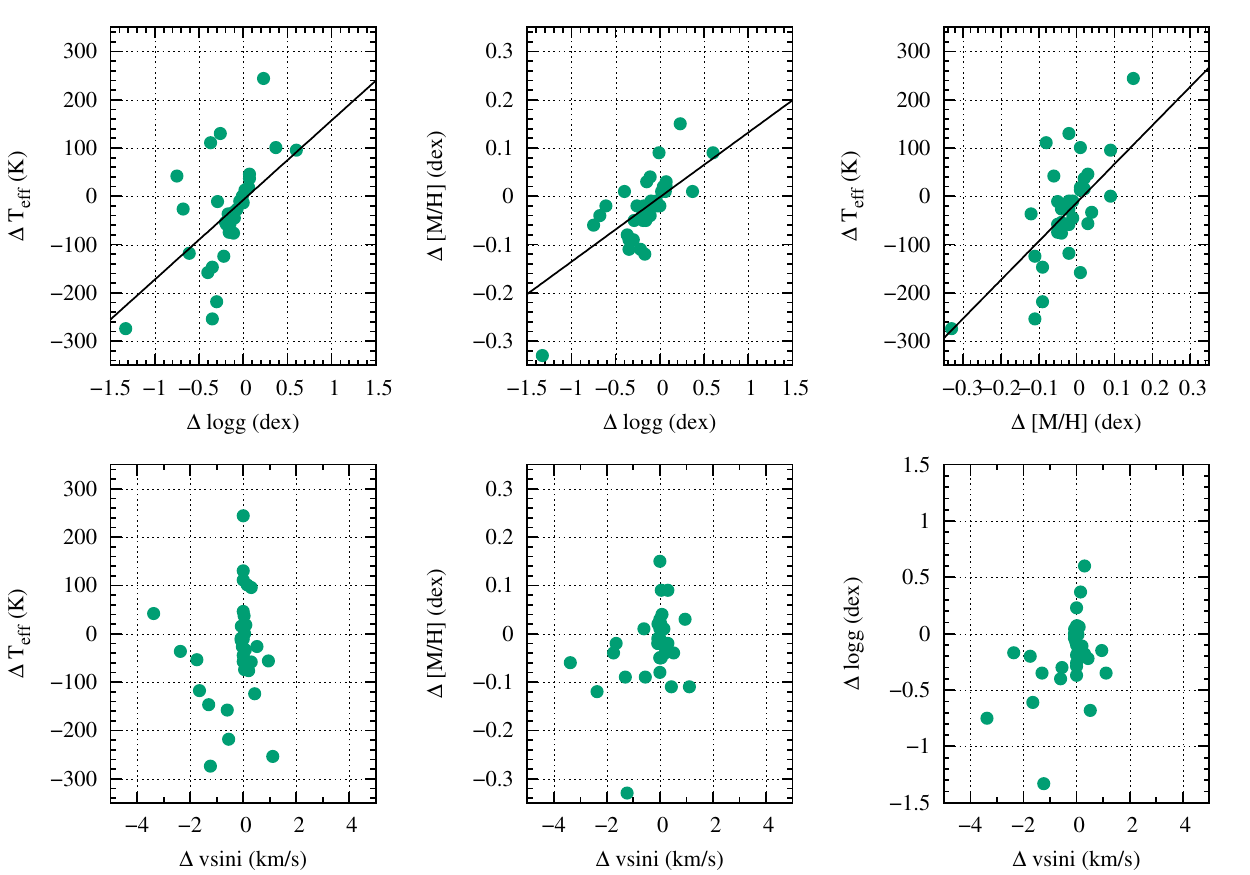}
  \caption{Differences in parameters between gravity constrained and unconstrained values (results from Sect.~\ref{high}). $\Delta \log g$ corresponds to $\log g$ from the benchmark values 
  minus $\log g$ from this work.}
  \label{loggfix}
\end{figure*}

We check any degeneracies in the parameters for the case we fix surface gravity to the benchmark values and let the other parameters vary. The differences between constrained and unconstrained 
(our standard methodology) values are presented in Fig.~\ref{loggfix}. We notice correlations between $T_{\mathrm{eff}}$ -- $[Fe/H]$, $[Fe/H]$ -- $\log g$, and 
$T_{\mathrm{eff}}$ -- $\log g$ but no obvious correlation with the $\upsilon\sin i$. The linear coefficients and the Pearson correlation coefficient are in Table~\ref{coefficients}. 
Similar correlations have been reported in several works that use synthesis on-the-fly but also in pattern matching techniques because of the interdependence of the parameters 
\citep{torres12, blanco2014a, Kordopatis2011}. On the other hand, it is worth mentioning that parameters based on the ionization and excitation balance show almost no dependence between 
them \citep{mortier13_gravity}. 

In Table~\ref{results_table_fix}, we show the mean differences between the constrained $ T_{\mathrm{eff}}$, $[Fe/H]$ and $\upsilon\sin i$ with the benchmark values for the high resolution 
sample divided per luminosity class. We notice that by fixing gravity, we obtain similar discrepancies in effective temperature and metallicity with the unconstrained values of 
Table~\ref{results_table_high} when comparing to the benchmark values. This test shows that fixing gravity to a more accurate estimation (e.g. from good parallax estimations) does not 
necessarily provide better parameter determinations for temperature and metallicity but it does reduces computation time because of a minus one free parameter.  

\begin{table}
\begin{center}
\caption{Average difference and standard deviation between the synthetic spectral synthesis technique with constrained gravity and the GES benchmark values in high resolution. 
N represents the number of spectra analysed in each group.}
\label{results_table_fix}
\scalebox{0.9}{
\begin{tabular}{lccccccc}
\hline\hline
     & $\overline{\Delta T_{\mathrm{eff}}}$ & $\sigma$ & $\overline{\Delta [Fe/H]}$ & $\sigma$ & $\overline{\Delta \upsilon\sin i}$ & $\sigma$ & N \\
     & \multicolumn{2}{c}{(K)} & \multicolumn{2}{c}{ (dex)} & \multicolumn{2}{c}{ (km\,s$^{-1}$)} &  \\
\hline
Whole sample        & -23  & 97  & -0.02 & 0.10 & -0.7 & 1.9 & 95 \\
F-type dwarfs       & -41  & 42  & -0.01 & 0.04 & -2.0 & 1.7 & 16 \\
G-type dwarfs       & -31  & 39  & -0.03 & 0.04 & 0.2  & 1.7 & 29 \\
K-type dwarfs       &  99  & 85  & 0.04  & 0.07 & 0.3  & 1.4 & 9  \\
FGK-type sub-giants & -82  & 159 & -0.12 & 0.23 & -0.3 & 1.8 & 17 \\
GK-type giants      & -62  & 92  & -0.01 & 0.13 & -0.7 & 2.0 & 24 \\
\hline
\end{tabular}}
\end{center}
\end{table}

\begin{table}
\begin{center}
\caption{Linear coefficients (y=bx+a) for the fits in Fig.~\ref{loggfix} and the Pearson correlation coefficient.}
\label{coefficients}
\begin{tabular}{lccc}
\hline\hline
	Relation 	       & a      & b      & Pearson \\
\hline
$T_{\mathrm{eff}}$ vs $\log g$ & -7.712 & 164.85 & 0.57 \\
$T_{\mathrm{eff}}$ vs $[M/H]$  & 0.1338 & 0.0018 & 0.61 \\
$[M/H]$ vs $\log g$            & -13.44 & 800.88 & 0.62 \\
\hline
\end{tabular}
\end{center}
\end{table}

\subsection{Presentation of the code} 

The code is written in python and the complete package is provided freely\footnote{\url{https://github.com/MariaTsantaki/fasma-synthesis}}. It is run either from the terminal or 
through a GUI interface for a more user-friendly approach. The spectral analysis in this work is relatively fast. The average time of our sample to achieve convergence with a standard 
computer, using the complete line list for each star in high resolution is $\sim$16~min whereas for medium resolution is less than 3~min (in this case the line list used is shorter).
The spectral package is already available online, however we plan to apply further improvements in the future. A list of upcoming updates will include: 
\begin{itemize}
 \item[-] abundances for other elements using the same analysis as deriving iron abundance,
 \item[-] expand our line list to cover the near infrared part of the spectrum to apply this work for GIRAFFE (HR21 set-up), APOGEE and \textit{Gaia} spectra, 
 \item[-] NLTE corrections. A way to correct for such effects is to use NLTE departure coefficients in order to create directly the synthetic spectrum \citep{Piskunov2016}.  
\end{itemize}

\section{Conclusions}

Precise and accurate determinations of the atmospheric stellar parameters are fundamental for deriving chemical composition, ages and evolutionary stages of stars. 
In this paper, we introduced a new package to determine the atmospheric stellar parameters for FGK-type stars based on the spectral synthesis technique. \textit{FASMA} contains 
all the necessary ingredients for a spectral synthesis analysis (line lists, models, minimization procedure) wrapped around MOOG and can be used directly for most optical surveys, 
such as for the \textit{Gaia}-ESO survey and also can be used for planet host characterizations. 

With our spectral package, we provide stellar parameters for a wide range of spectral types and luminosity classes and can be used to analyse large samples in a reasonable amount of time.
To test \textit{FASMA}, we use synthetic spectra which reveal correlations between the parameters with increasing strength towards lower S/N and lower resolutions. 
Our parameters show almost no dependence on the choice for initial parameters and are reliable for low S/N values. 
The effects of rotational velocities become visible after 35\,km\,s$^{-1}$. 

We compare our results with 451 stars from the HARPS sample and find very good agreement in all parameters. We also compare our results with the \textit{Gaia}-ESO benchmark stars 
using spectra both in high and medium resolution. Our results show very good agreement for metallicities in both high and medium resolution, even when using short line lists. 
The effecive temperatures in medium resolution show higher standard deviations compared to high resolution. When we have external better estimations of surface gravities, 
the effective temperature and metallicity determinations are not necessarily improved. 

We expect to improve this work by adding NLTE correction, wider coverage in wavelength and determinations of chemical abundances of other elements. 

\section*{Acknowledgments}

The authors thank the referee for the useful comments. MT and GB acknowledge support for this work from UNAM through grant PAPIIT IG100115.
G.D.C.T. was supported by the FCT/Portugal Ph.D. grant PD/BD/113478/2015. 
E.D.M., N.C.S., and S.G.S. acknowledge support from Funda\c{c}\~{a}o para a Ci\^{e}ncia e a Tecnologia (FCT) through national funds and by FEDER through COMPETE2020 by grants UID/FIS/04434/2013 \& 
POCI-01-0145-FEDER-007672, PTDC/FIS-AST/7073/2014 \& POCI-01-0145-FEDER-016880, and PTDC/FIS-AST/1526/2014 \& POCI-01-0145-FEDER-016886. N.C.S. and S.G.S. also acknowledge support 
from FCT through Investigador FCT contracts IF/00169/2012/CP0150/CT0002 e IF/00028/2014/CP1215/CT0002; and E.D.M. acknowledge support by the fellowship SFRH/BPD/76606/2011 
funded by FCT (Portugal) and POPH/FSE (EC). 
 
This research made use of the Vienna Atomic Line Database operated at Uppsala University, the Institute of Astronomy RAS in Moscow, and the University of 
Vienna. We thank the PyAstronomy and Astropy communities.


\bibliography{bibliography} 

\appendix
\clearpage
\newpage

\section{Line list}
 \begin{center}
 \begin{table}
 \caption{Line data used for the spectroscopic analysis.}\label{line_data}
\begin{tabular}{ccccc}
 \hline
 Wavelength & Element & $\chi_{ex}$ &  $\log gf$ & $\varGamma_{6}$ \\
   \AA{}    &         &  (eV)       &            &                 \\
 \hline
5339.188 &  $\ion{Ca}{ii}$  & 8.44 & -0.075 & -7.196 \\ 
5339.416 &  $\ion{Fe}{i}$  & 4.43 & -2.146 & -7.510 \\ 
5339.526 &  $\ion{Co}{i}$  & 4.23 & -0.377 & -7.510 \\ 
5339.937 &  $\ion{Fe}{i}$  & 3.27 & -0.609 & -7.221 \\ 
5340.189 &  $\ion{Fe}{i}$  & 4.29 & -2.229 & -7.520 \\ 
5340.447 &  $\ion{Cr}{i}$  & 3.44 & -0.724 & -7.262 \\
\end{tabular}
 \end{table}
 \end{center}
 
 \begin{center}
 \begin{table}
 \contcaption{Line data used for the spectroscopic analysis.}
 \begin{tabular}{ccccc}
 \hline
 Wavelength & Element & $\chi_{ex}$ &  $\log gf$ & $\varGamma_{6}$ \\
   \AA{}    &         &  (eV)       &            &                 \\
 \hline
5358.113 &  $\ion{Fe}{i}$  & 3.30 & -3.167 & -7.800 \\ 
5358.928 &  $\ion{Cr}{i}$  & 4.53 & -6.299 & -7.750 \\ 
5358.928 &  $\ion{Co}{i}$  & 4.15 & -0.563 & -7.510 \\ 
5359.199 &  $\ion{Co}{i}$  & 4.15 & 0.007 & -7.510 \\ 
5359.498 &  $\ion{K}{i}$  & 1.62 & -1.406 & 0.000 \\ 
5359.508 &  $\ion{Ce}{ii}$  & 1.78 & 0.032 & 0.000 \\
\end{tabular}
 \end{table}
 \end{center}
 
 \begin{center}
 \begin{table}
 \contcaption{Line data used for the spectroscopic analysis.}
 \begin{tabular}{ccccc}
 \hline
 Wavelength & Element & $\chi_{ex}$ &  $\log gf$ & $\varGamma_{6}$ \\
   \AA{}    &         &  (eV)       &            &                 \\
 \hline
5383.407 &  $\ion{Co}{i}$  & 4.15 & -3.226 & -7.510 \\ 
5383.408 &  $\ion{V}{i}$  & 2.58 & -0.113 & -7.520 \\ 
5384.156 &  $\ion{V}{i}$  & 3.87 & -5.883 & -7.270 \\ 
5384.630 &  $\ion{Ti}{i}$  & 0.83 & -2.666 & -7.639 \\ 
5384.870 &  $\ion{V}{ii}$  & 2.27 & -2.353 & -7.870 \\ 
5385.132 &  $\ion{Zr}{i}$  & 0.52 & -0.664 & 0.000 \\
\end{tabular}
 \end{table}
 \end{center}
 
 \begin{center}
 \begin{table}
 \contcaption{Line data used for the spectroscopic analysis.}
 \begin{tabular}{ccccc}
 \hline
 Wavelength & Element & $\chi_{ex}$ &  $\log gf$ & $\varGamma_{6}$ \\
   \AA{}    &         &  (eV)       &            &                 \\
 \hline
5399.465 &  $\ion{Mn}{i}$  & 3.85 & -8.236 & -6.350 \\ 
5399.465 &  $\ion{Mn}{i}$  & 3.85 & -5.951 & -6.350 \\ 
5399.478 &  $\ion{Mn}{i}$  & 3.85 & -7.504 & -6.350 \\ 
5399.485 &  $\ion{Mn}{i}$  & 3.85 & -1.337 & -6.350 \\ 
5399.494 &  $\ion{Mn}{i}$  & 3.85 & -0.828 & -6.350 \\ 
5399.497 &  $\ion{Mn}{i}$  & 3.85 & -0.982 & -6.350 \\
\end{tabular}
 \end{table}
 \end{center}
 
 \begin{center}
 \begin{table}
 \contcaption{Line data used for the spectroscopic analysis.}
 \begin{tabular}{ccccc}
 \hline
 Wavelength & Element & $\chi_{ex}$ &  $\log gf$ & $\varGamma_{6}$ \\
   \AA{}    &         &  (eV)       &            &                  \\
 \hline
5417.040 &  $\ion{Fe}{i}$  & 4.42 & -1.350 & -7.166 \\ 
5418.000 &  $\ion{Zr}{ii}$  & 1.76 & -1.247 & 0.000 \\ 
5418.120 &  $\ion{Si}{i}$  & 5.62 & -2.002 & -6.860 \\ 
5418.259 &  $\ion{Fe}{i}$  & 4.99 & -2.248 & -7.400 \\ 
5418.768 &  $\ion{Ti}{ii}$  & 1.58 & -2.002 & -7.850 \\ 
5418.856 &  $\ion{Ru}{i}$  & 1.12 & -1.017 & 0.000 \\
\end{tabular}
 \end{table}
 \end{center}

\begin{center}
 \begin{table}
 \contcaption{Line data used for the spectroscopic analysis.}
 \begin{tabular}{ccccc}
 \hline
 Wavelength & Element & $\chi_{ex}$ &  $\log gf$ & $\varGamma_{6}$ \\
   \AA{}    &         &  (eV)       &            &                  \\
 \hline
5447.675 &  $\ion{Ti}{i}$  & 3.57 & -4.733 & -7.660 \\ 
5447.917 &  $\ion{Ti}{i}$  & 2.49 & -1.472 & -7.770 \\ 
5453.987 &  $\ion{Fe}{i}$  & 4.15 & -2.434 & -7.470 \\ 
5454.099 &  $\ion{Ti}{ii}$  & 1.57 & -3.367 & -7.840 \\ 
5454.570 &  $\ion{Co}{i}$  & 4.07 & 0.065 & -7.510 \\ 
5455.103 &  $\ion{Fe}{i}$  & 3.25 & -4.445 & -7.820 \\
\end{tabular}
 \end{table}
 \end{center}

\begin{center}
 \begin{table}
 \contcaption{Line data used for the spectroscopic analysis.}
 \begin{tabular}{ccccc}
 \hline
 Wavelength & Element & $\chi_{ex}$ &  $\log gf$ & $\varGamma_{6}$ \\
   \AA{}    &         &  (eV)       &            &                  \\
 \hline
5487.915 &  $\ion{V}{i}$  & 2.37 & 0.197 & -7.627 \\ 
5488.120 &  $\ion{Co}{i}$  & 4.21 & -9.544 & -7.510 \\ 
5488.135 &  $\ion{Fe}{i}$  & 4.61 & -6.105 & -7.410 \\ 
5488.178 &  $\ion{Ti}{i}$  & 2.40 & -0.088 & -7.520 \\ 
5496.477 &  $\ion{Fe}{i}$  & 3.57 & -3.788 & -7.820 \\ 
5496.562 &  $\ion{Fe}{i}$  & 4.91 & -1.579 & -7.345 \\
\end{tabular}
 \end{table}
 \end{center}

\begin{center}
 \begin{table}
 \contcaption{Line data used for the spectroscopic analysis.}
 \begin{tabular}{ccccc}
 \hline
 Wavelength & Element & $\chi_{ex}$ &  $\log gf$ & $\varGamma_{6}$ \\
   \AA{}    &         &  (eV)       &            &                  \\
 \hline
5517.065 &  $\ion{Fe}{i}$  & 4.21 & -1.966 & -7.480 \\ 
5517.201 &  $\ion{V}{i}$  & 0.00 & -3.476 & -7.780 \\ 
5517.533 &  $\ion{Si}{i}$  & 5.08 & -2.392 & -7.320 \\ 
5518.070 &  $\ion{Ti}{i}$  & 2.41 & -1.397 & -7.520 \\ 
5518.107 &  $\ion{C}{2}$ & 0.69 & -0.044 & 0.000 \\ 
5518.359 &  $\ion{Ce}{ii}$  & 0.55 & -1.703 & 0.000 \\
\end{tabular}
 \end{table}
 \end{center}

\begin{center}
 \begin{table}
 \contcaption{Line data used for the spectroscopic analysis.}
 \begin{tabular}{ccccc}
 \hline
 Wavelength & Element & $\chi_{ex}$ &  $\log gf$ & $\varGamma_{6}$ \\
   \AA{}    &         &  (eV)       &            &                  \\
 \hline
5532.347 &  $\ion{C}{2}$ & 0.82 & 0.346 & 0.000 \\ 
5532.747 &  $\ion{Fe}{i}$  & 3.57 & -2.120 & -7.762 \\ 
5532.866 &  $\ion{Fe}{i}$  & 4.45 & -1.680 & -7.520 \\ 
5533.031 &  $\ion{Mo}{i}$  & 1.33 & 0.066 & 0.000 \\ 
5533.820 &  $\ion{Nd}{ii}$  & 0.56 & -0.994 & 0.000 \\ 
5534.261 &  $\ion{C}{2}$ & 0.80 & -0.265 & 0.000 \\
\end{tabular}
 \end{table}
 \end{center}

\begin{center}
 \begin{table}
 \contcaption{Line data used for the spectroscopic analysis.}
 \begin{tabular}{ccccc}
 \hline
 Wavelength & Element & $\chi_{ex}$ &  $\log gf$ & $\varGamma_{6}$ \\
   \AA{}    &         &  (eV)       &            &                  \\
 \hline
5559.893 &  $\ion{Fe}{i}$  & 4.64 & -2.565 & -7.310 \\ 
5560.211 &  $\ion{Fe}{i}$  & 4.43 & -0.947 & -7.189 \\ 
5560.547 &  $\ion{V}{i}$  & 0.04 & -3.590 & -7.761 \\ 
5560.675 &  $\ion{C}{2}$ & 0.50 & -0.368 & 0.000 \\ 
5561.445 &  $\ion{Ce}{ii}$  & 1.46 & 0.028 & 0.000 \\ 
5561.659 &  $\ion{C}{2}$ & 0.73 & 0.372 & 0.000 \\
\end{tabular}
 \end{table}
 \end{center}

\begin{center}
 \begin{table}
 \contcaption{Line data used for the spectroscopic analysis.}
 \begin{tabular}{ccccc}
 \hline
 Wavelength & Element & $\chi_{ex}$ &  $\log gf$ & $\varGamma_{6}$ \\
   \AA{}    &         &  (eV)       &            &                  \\
 \hline
5573.409 &  $\ion{Sm}{i}$  & 0.39 & -1.952 & 0.000 \\ 
5573.554 &  $\ion{C}{2}$ & 0.65 & 0.001 & 0.000 \\ 
5573.659 &  $\ion{Mn}{i}$  & 5.54 & 0.258 & -7.580 \\ 
5573.707 &  $\ion{Co}{i}$  & 4.23 & -7.617 & -7.510 \\ 
5573.724 &  $\ion{Mn}{i}$  & 4.36 & -1.668 & -7.770 \\ 
5573.734 &  $\ion{C}{2}$ & 0.65 & -4.652 & 0.000 \\
\end{tabular}
 \end{table}
 \end{center}

\begin{center}
 \begin{table}
 \contcaption{Line data used for the spectroscopic analysis.}
 \begin{tabular}{ccccc}
 \hline
 Wavelength & Element & $\chi_{ex}$ &  $\log gf$ & $\varGamma_{6}$ \\
   \AA{}    &         &  (eV)       &            &                  \\
 \hline
5597.805 &  $\ion{Cr}{i}$  & 3.84 & -1.508 & -7.340 \\ 
5598.295 &  $\ion{Fe}{i}$  & 4.65 & -0.306 & -7.085 \\ 
5598.425 &  $\ion{Fe}{i}$  & 3.63 & -7.385 & -7.750 \\ 
5598.466 &  $\ion{Fe}{i}$  & 2.42 & -8.272 & -7.770 \\ 
5598.480 &  $\ion{Ca}{i}$  & 2.52 & 0.010 & -7.541 \\ 
5598.825 &  $\ion{C}{2}$ & 0.62 & 0.047 & 0.000 \\
\end{tabular}
 \end{table}
 \end{center}

\begin{center}
 \begin{table}
 \contcaption{Line data used for the spectroscopic analysis.}
 \begin{tabular}{ccccc}
 \hline
 Wavelength & Element & $\chi_{ex}$ &  $\log gf$ & $\varGamma_{6}$ \\
   \AA{}    &         &  (eV)       &            &                  \\
 \hline
6533.110 &  $\ion{V}{ii}$  & 1.14 & -3.368 & 0.000 \\ 
6533.928 &  $\ion{Fe}{i}$  & 4.56 & -1.160 & -7.182 \\ 
6546.242 &  $\ion{Fe}{i}$  & 2.76 & -1.588 & -7.626 \\ 
6546.269 &  $\ion{Ti}{i}$  & 1.43 & -0.407 & -7.750 \\ 
6546.943 & $\ion{CN}{}$ & 0.72 & -1.467 & 0.000 \\ 
6568.914 &  $\ion{C}{2}$ & 1.88 & -8.375 & 0.000 \\
\end{tabular}
 \end{table}
 \end{center}

\begin{center}
 \begin{table}
 \contcaption{Line data used for the spectroscopic analysis.}
 \begin{tabular}{ccccc}
 \hline
 Wavelength & Element & $\chi_{ex}$ &  $\log gf$ & $\varGamma_{6}$ \\
   \AA{}    &         &  (eV)       &            &                 \\
 \hline
6640.480 & $\ion{CN}{}$ & 0.89 & -1.489 & 0.000 \\ 
6640.486 &  $\ion{C}{2}$ & 1.35 & -7.497 & 0.000 \\ 
6646.935 &  $\ion{Fe}{i}$  & 2.61 & -3.903 & -7.604 \\ 
6647.803 &  $\ion{Ni}{i}$  & 4.42 & -2.016 & -7.210 \\ 
6648.086 &  $\ion{Fe}{i}$  & 1.01 & -5.681 & -7.830 \\ 
6652.970 &  $\ion{Fe}{i}$  & 5.52 & -2.156 & 0.000 \\
\end{tabular}
 \end{table}
 \end{center}

\begin{center}
 \begin{table}
 \contcaption{Line data used for the spectroscopic analysis.}
 \begin{tabular}{ccccc}
 \hline
 Wavelength & Element & $\chi_{ex}$ &  $\log gf$ & $\varGamma_{6}$ \\
   \AA{}    &         &  (eV)       &            &                 \\
 \hline
6737.985 &  $\ion{Fe}{i}$  & 4.56 & -1.573 & -7.213 \\ 
6738.215 &  $\ion{Fe}{i}$  & 5.35 & -1.351 & -7.330 \\ 
6738.620 &  $\ion{Ti}{i}$  & 3.39 & -0.438 & -7.628 \\ 
6738.660 &  $\ion{Sc}{i}$  & 4.82 & -2.931 & -6.780 \\ 
6738.842 &  $\ion{Fe}{i}$  & 4.84 & -1.878 & -7.520 \\ 
6739.205 &  $\ion{Cu}{i}$  & 6.99 & 1.293 & -7.310 \\
\hline
 \end{tabular}
 \end{table}
 \end{center}

\clearpage

\section{Results for synthetic spectra}\label{diff_synth}

\begin{figure*}%
 \centering
  \begin{minipage}{0.33\textwidth}
  \includegraphics[width=6.5cm, height=5.5cm]{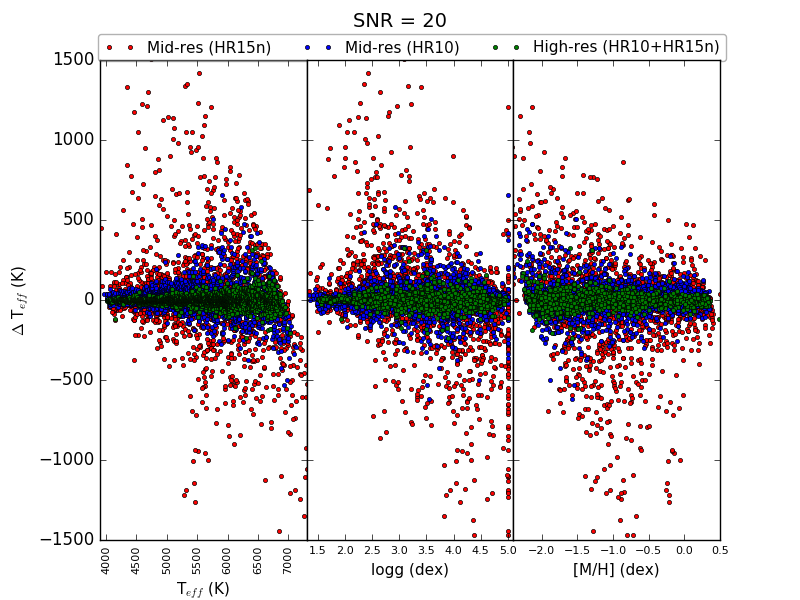} 
 \end{minipage}
 \begin{minipage}{0.33\textwidth}
  \includegraphics[width=6.5cm, height=5.5cm]{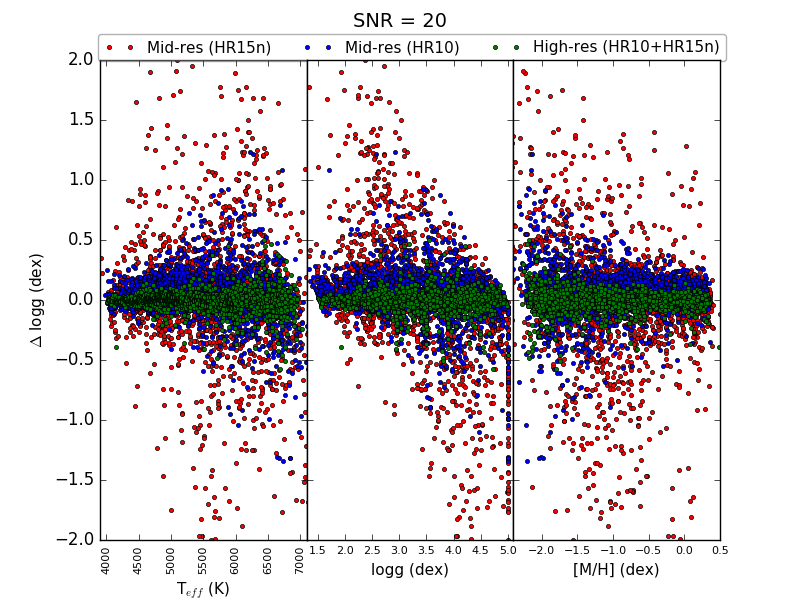} 
 \end{minipage}
 \begin{minipage}{0.33\textwidth}
  \includegraphics[width=6.5cm, height=5.5cm]{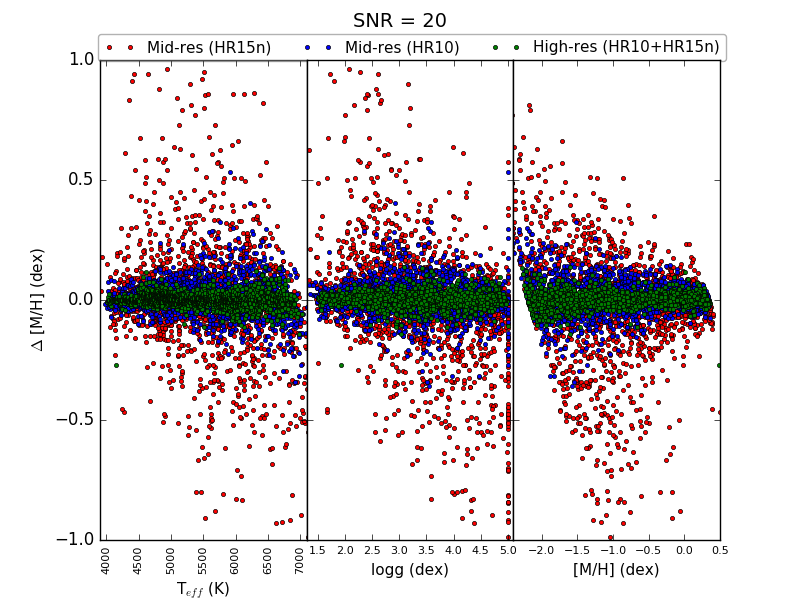} 
 \end{minipage}
 \caption{The residuals of the main atmospheric parameters for the synthetic spectra of FGK-type stars for different resolutions depicted in different color. 
 Each row corresponds to different S/N values.}
 \label{diff_params}
\end{figure*}

In Fig.~\ref{diff_params}, we plot the differences of the derived parameters with \textit{FASMA} for the synthetic spectra of Sect.~\ref{synthetic_test}. The atmospheric parameters of the 
sample cover the FGK-type stars. The steep differences for $\log g$ around 3.0 and 3.5\,dex appear because these are the limits where microturbulence and macroturbulence follow different 
treatments for the giant and dwarf stars according to the correlations mentioned in Sect.~\ref{synthetic_test}.

\begin{figure*}%
 \centering
 \begin{minipage}{0.33\textwidth}
  \includegraphics[width=6.5cm, height=5.5cm]{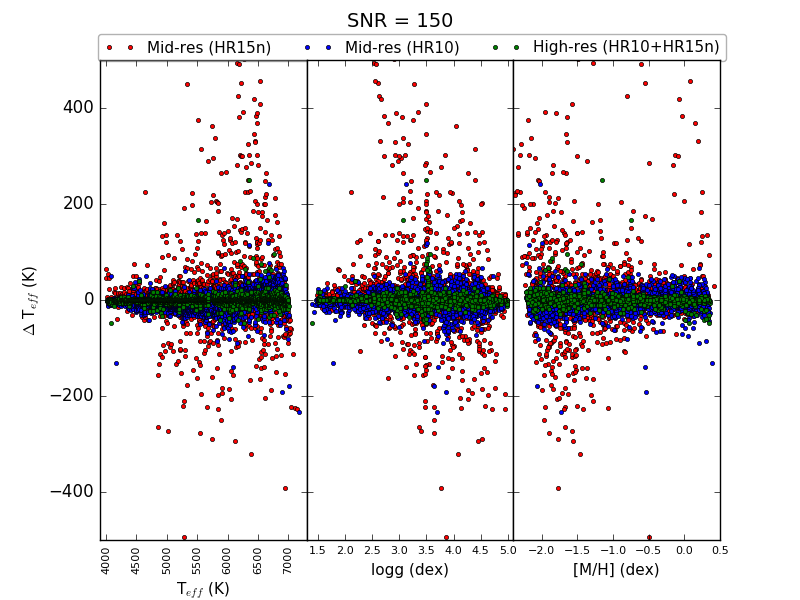} 
 \end{minipage}
 \begin{minipage}{0.33\textwidth}
  \includegraphics[width=6.5cm, height=5.5cm]{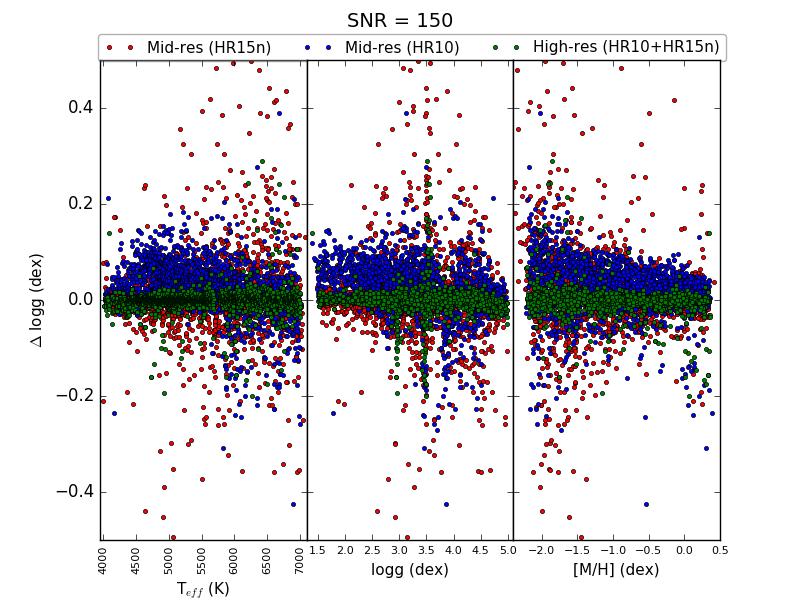} 
 \end{minipage}
 \begin{minipage}{0.33\textwidth}
  \includegraphics[width=6.5cm, height=5.5cm]{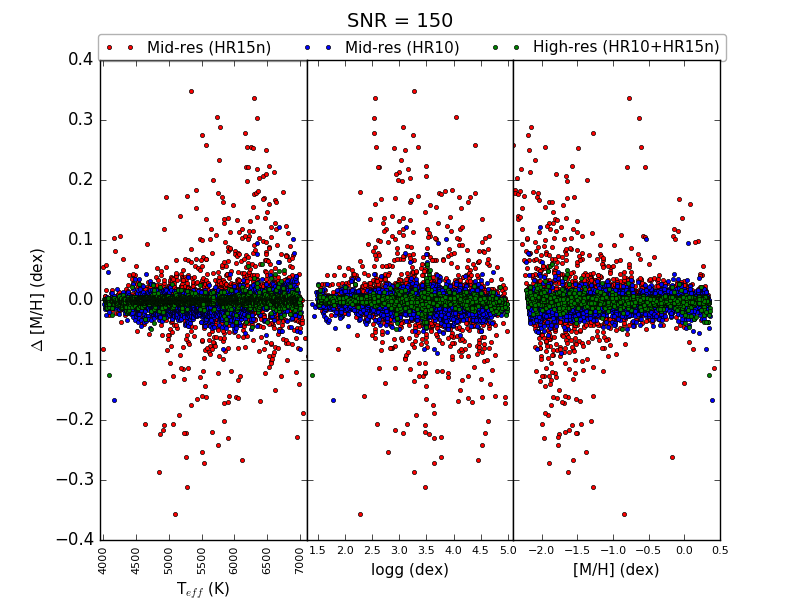} 
 \end{minipage}
 \centering
 \begin{minipage}{0.33\textwidth}
  \includegraphics[width=6.5cm, height=5.5cm]{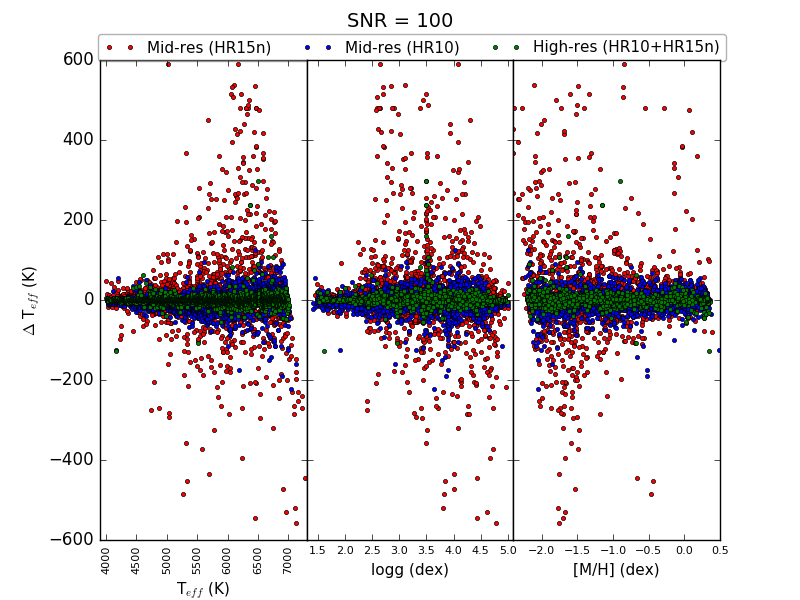} 
 \end{minipage}
 \begin{minipage}{0.33\textwidth}
  \includegraphics[width=6.5cm, height=5.5cm]{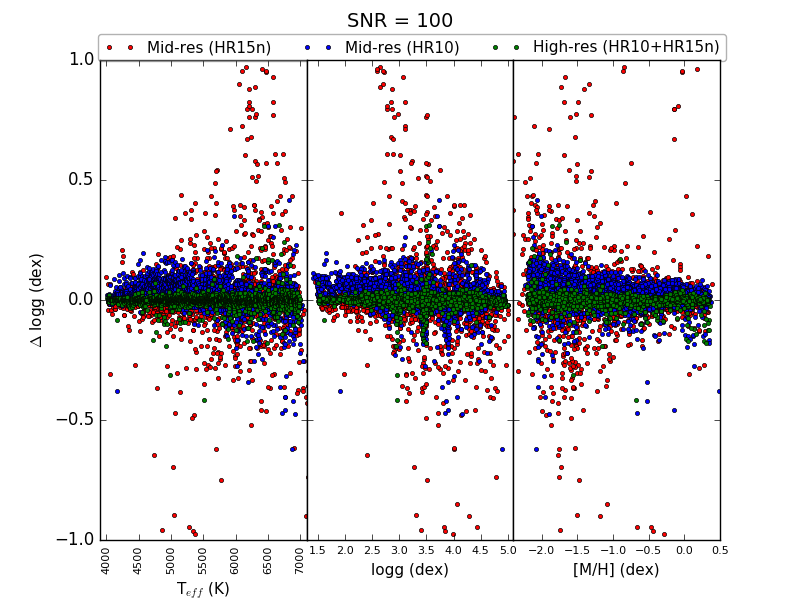} 
 \end{minipage}
 \begin{minipage}{0.33\textwidth}
  \includegraphics[width=6.5cm, height=5.5cm]{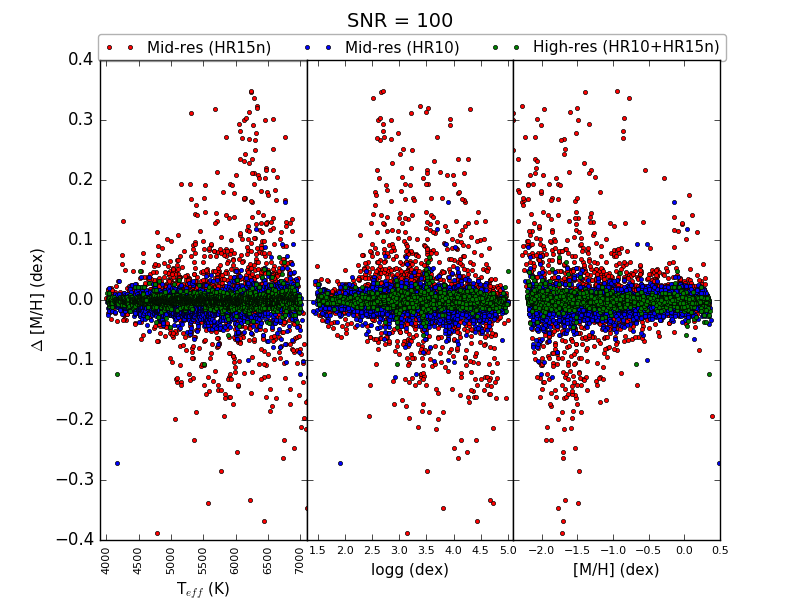} 
 \end{minipage}
 \centering
 \begin{minipage}{0.33\textwidth}
  \includegraphics[width=6.5cm, height=5.5cm]{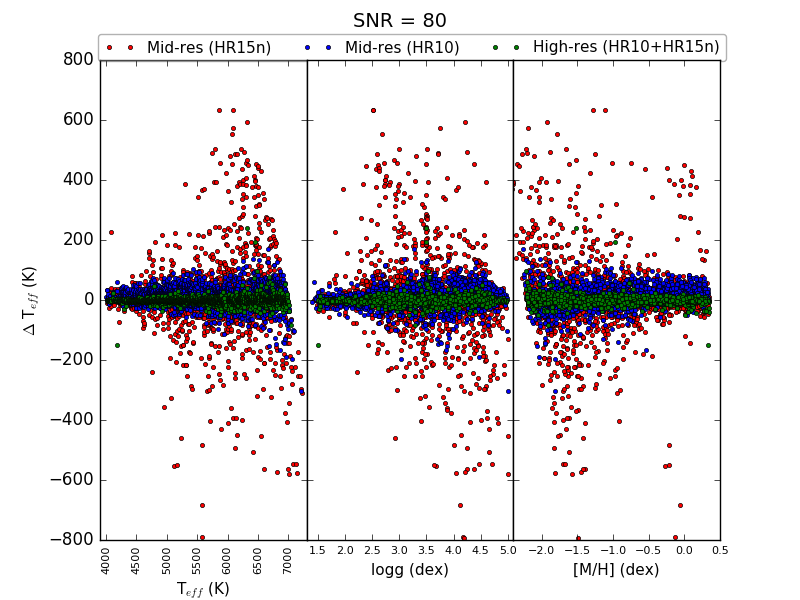} 
 \end{minipage}
 \begin{minipage}{0.33\textwidth}
  \includegraphics[width=6.5cm, height=5.5cm]{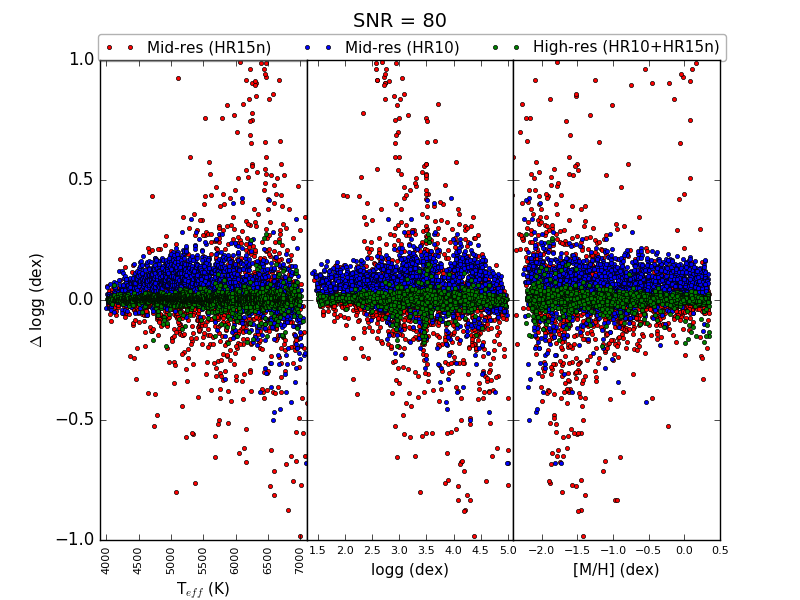} 
 \end{minipage}
 \begin{minipage}{0.33\textwidth}
  \includegraphics[width=6.5cm, height=5.5cm]{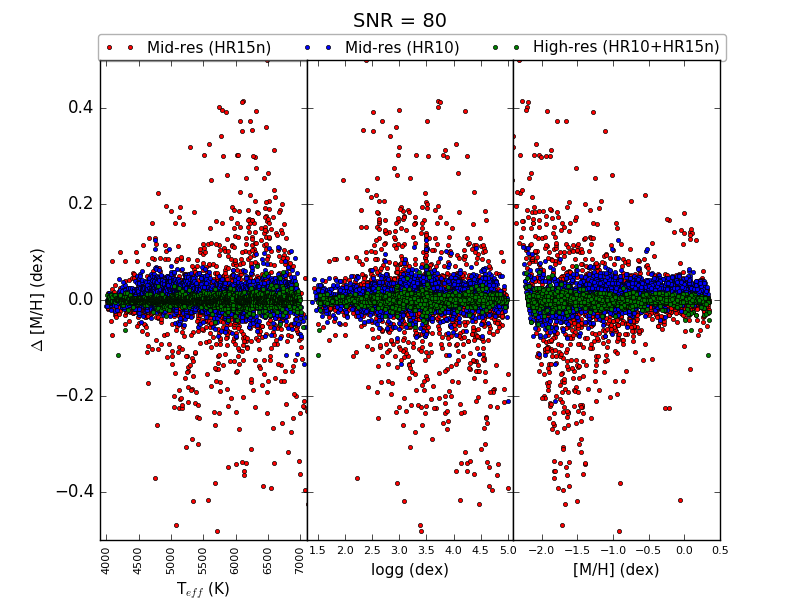} 
 \end{minipage}
 \centering
 \begin{minipage}{0.33\textwidth}
  \includegraphics[width=6.5cm, height=5.5cm]{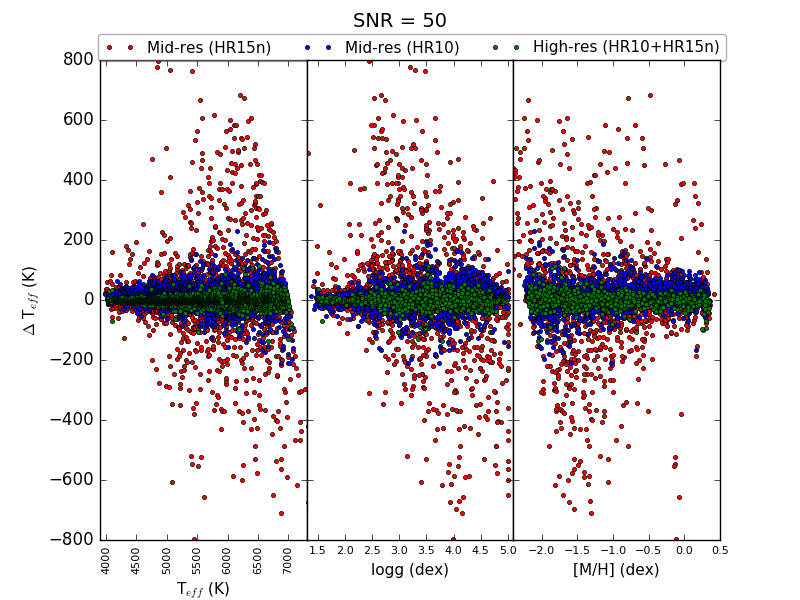} 
 \end{minipage}
 \begin{minipage}{0.33\textwidth}
  \includegraphics[width=6.5cm, height=5.5cm]{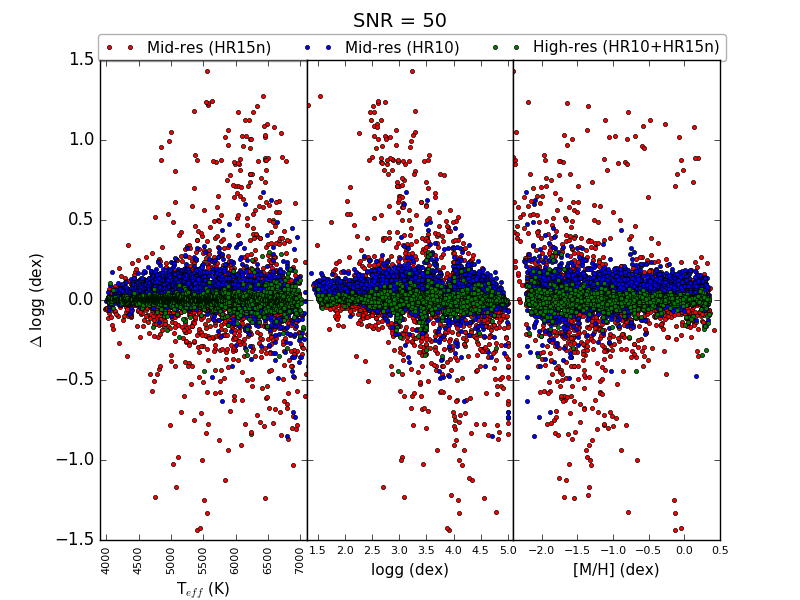} 
 \end{minipage}
 \begin{minipage}{0.33\textwidth}
  \includegraphics[width=6.5cm, height=5.5cm]{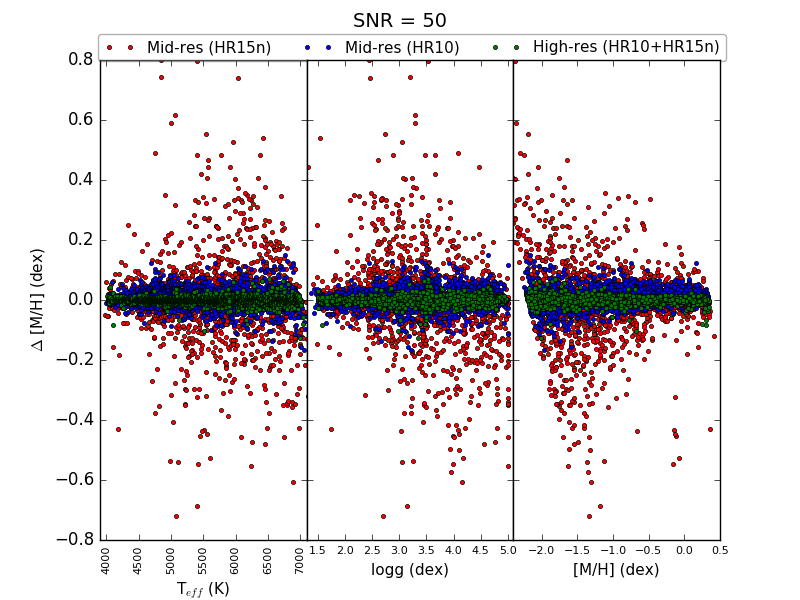} 
 \end{minipage}
\end{figure*}

\clearpage
\newpage

\section{Results for diffent model atmospheres}
\vspace{1cm}
\begin{figure}
  \centering
   \includegraphics[width=1.0\linewidth]{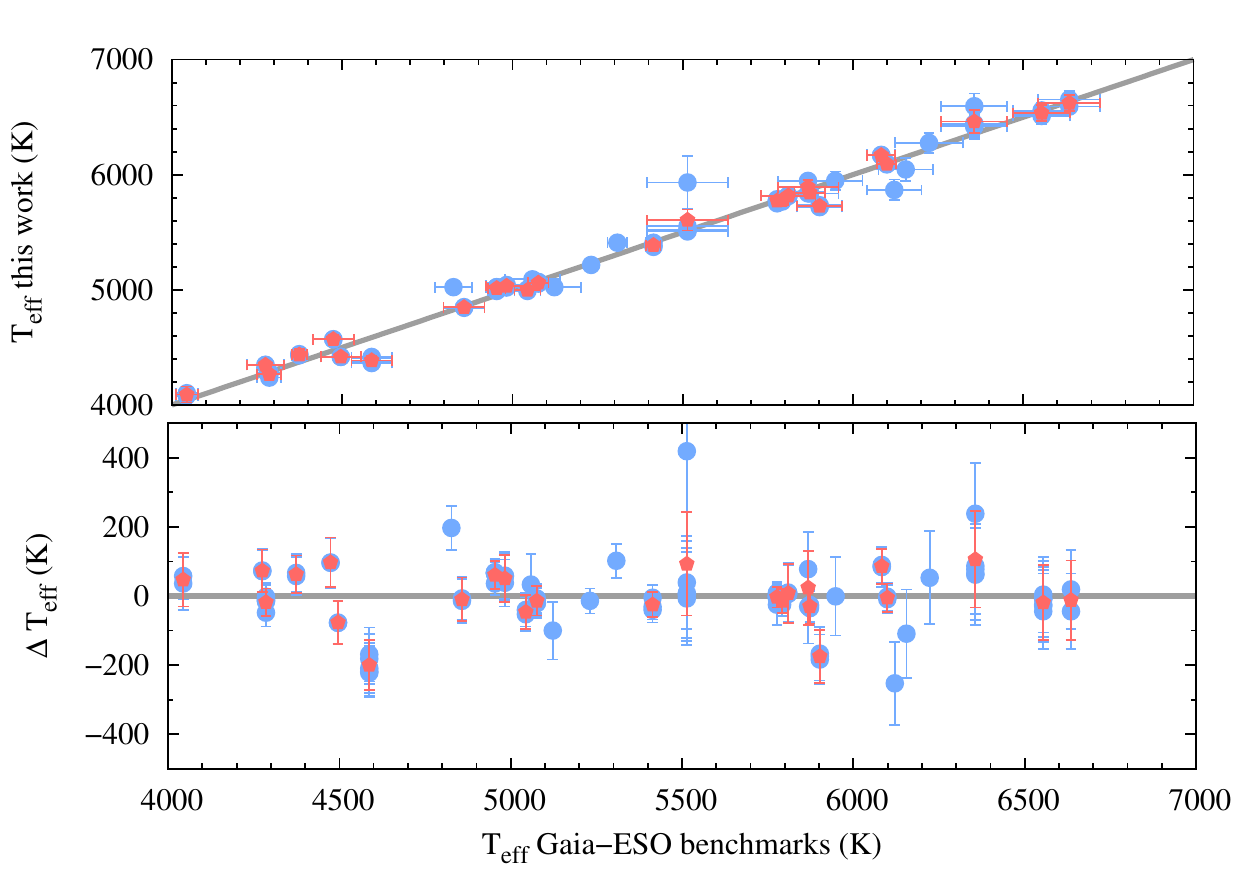} \\
   \includegraphics[width=1.0\linewidth]{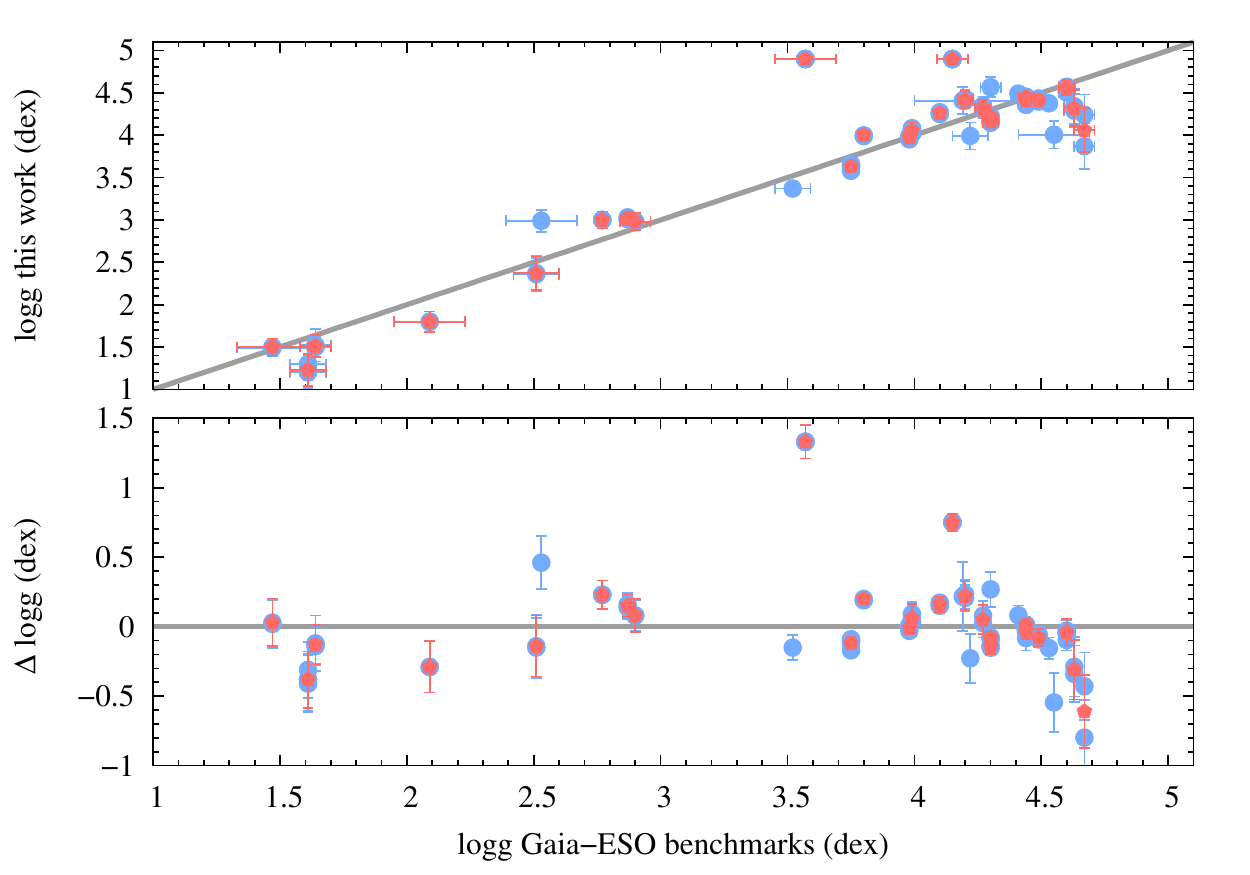} \\
   \includegraphics[width=1.0\linewidth]{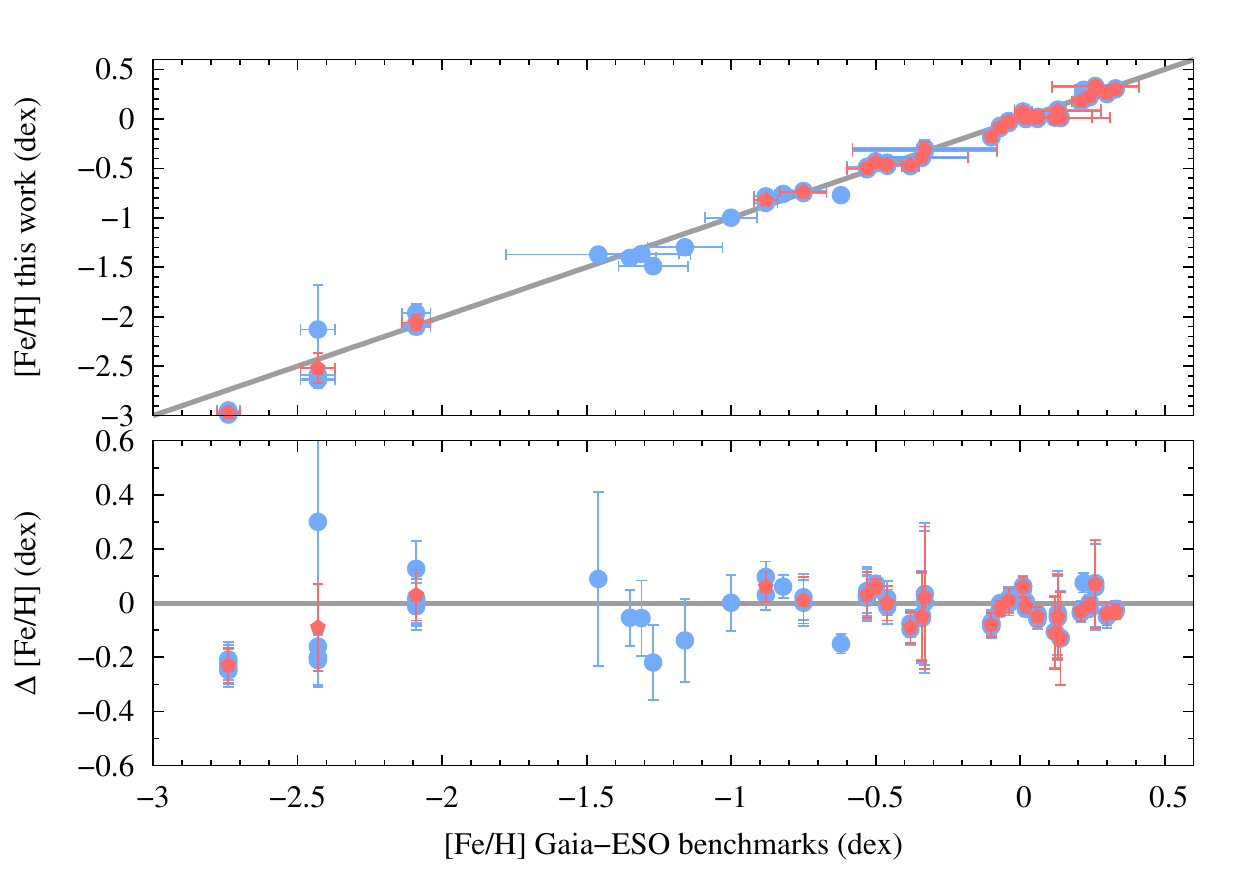}
  \caption{Differences in effective temperature, surface gravity and metallicity between the GES benchmark parameters and this work using the high resolution spectra and ATLAS-APOGEE models 
  (blue squares). For stars with multiple spectra, their mean values are plotted with green circles. }
  \label{high_res_apogee}
  \end{figure}

\begin{figure}
  \centering
   \includegraphics[width=1.05\linewidth]{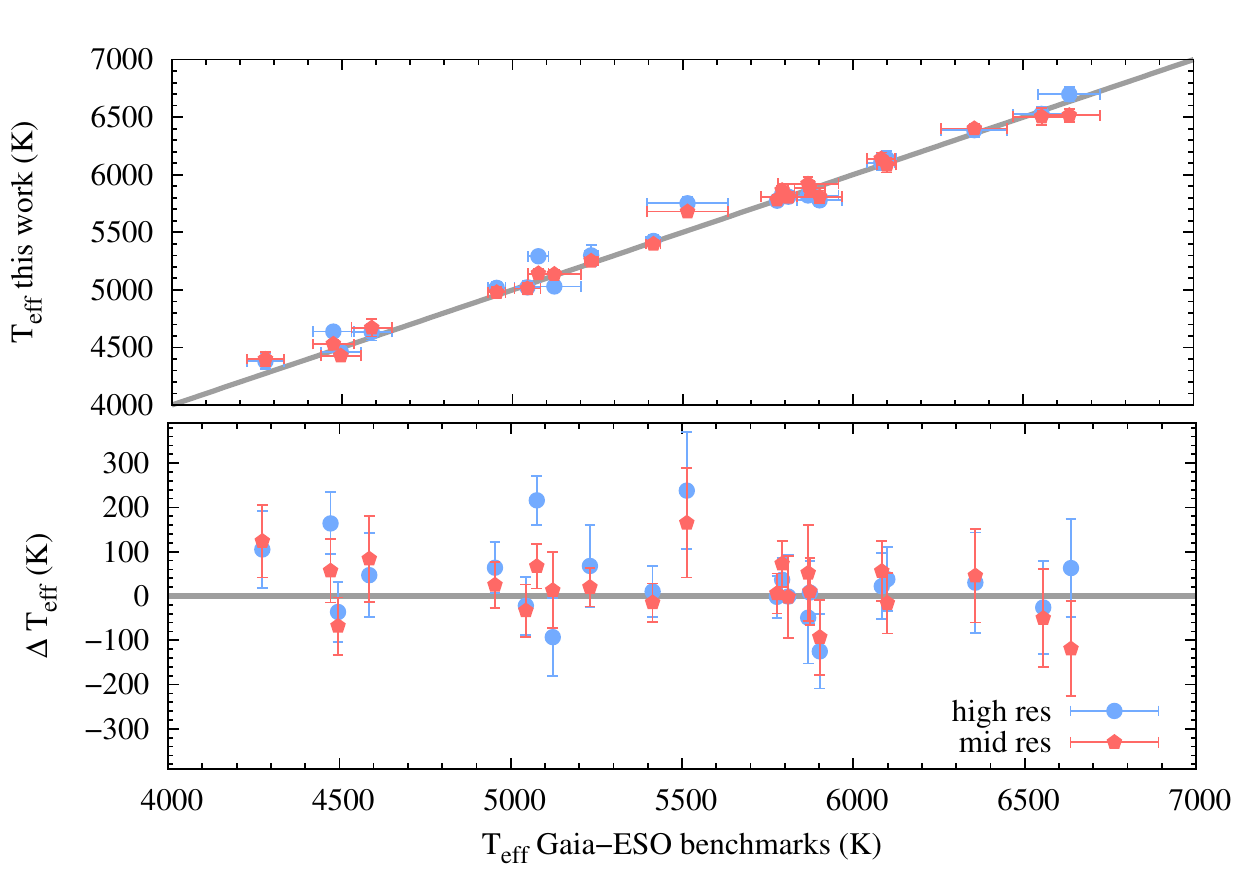} \\
   \includegraphics[width=1.05\linewidth]{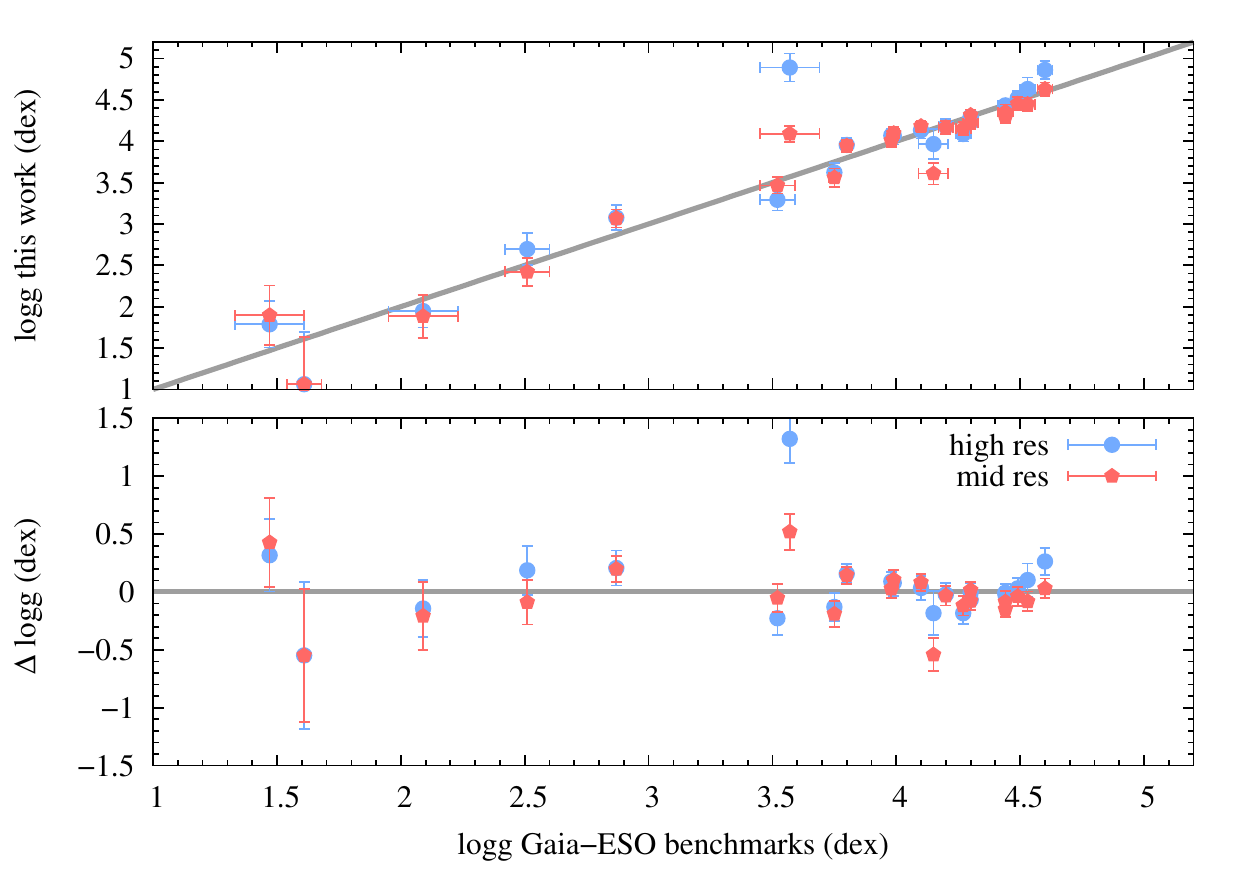} \\
   \includegraphics[width=1.05\linewidth]{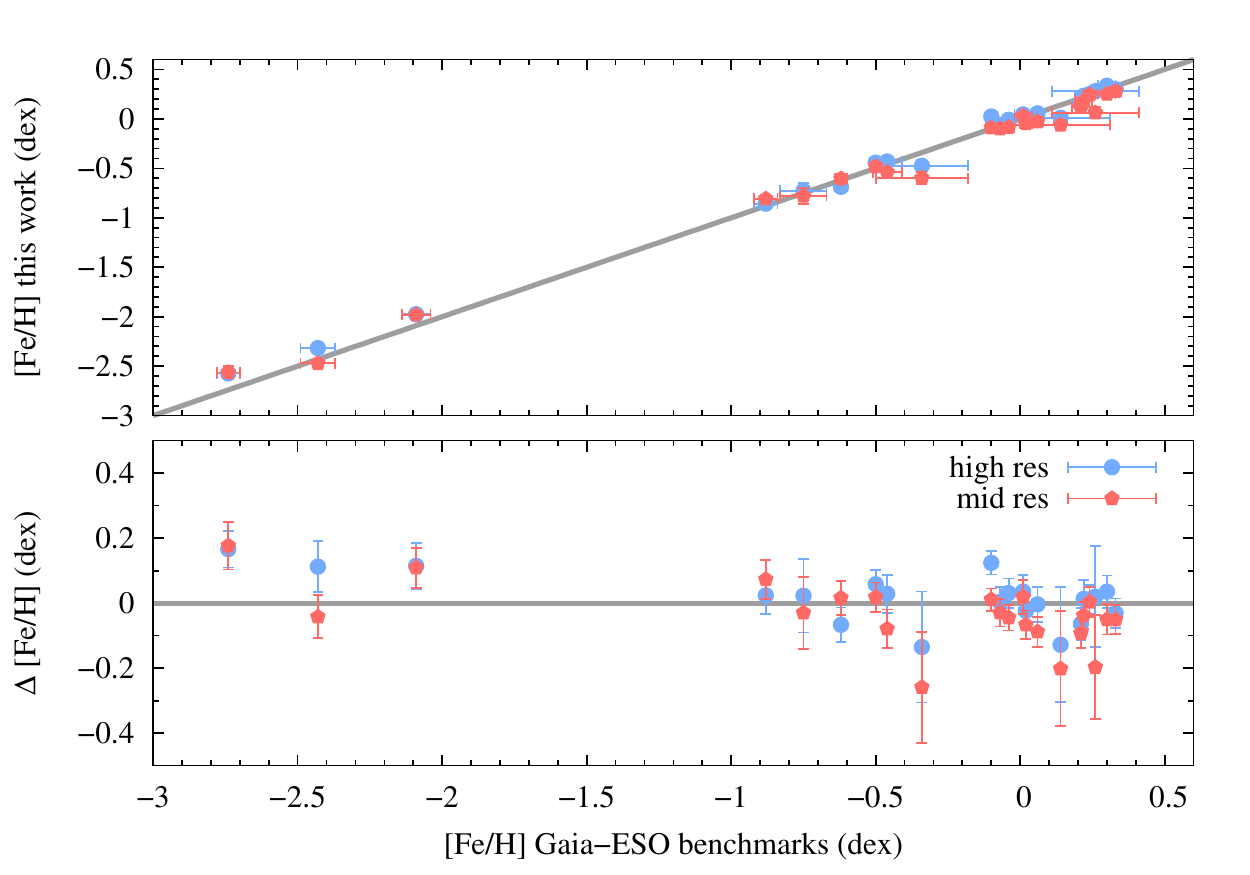} 
  \caption{Differences in effective temperature, surface gravity and metallicity between the Gaia FGK benchmark parameters and this work using the HR10 GIRAFFE spectra for the ATLAS-APOGEE models. For 
  comparison we plot the parameters of high resolution spectra using the same line list.}
  \label{mid_res_hr10_apogee}
  \end{figure}

\begin{figure}
  \centering
   \includegraphics[width=1.05\linewidth]{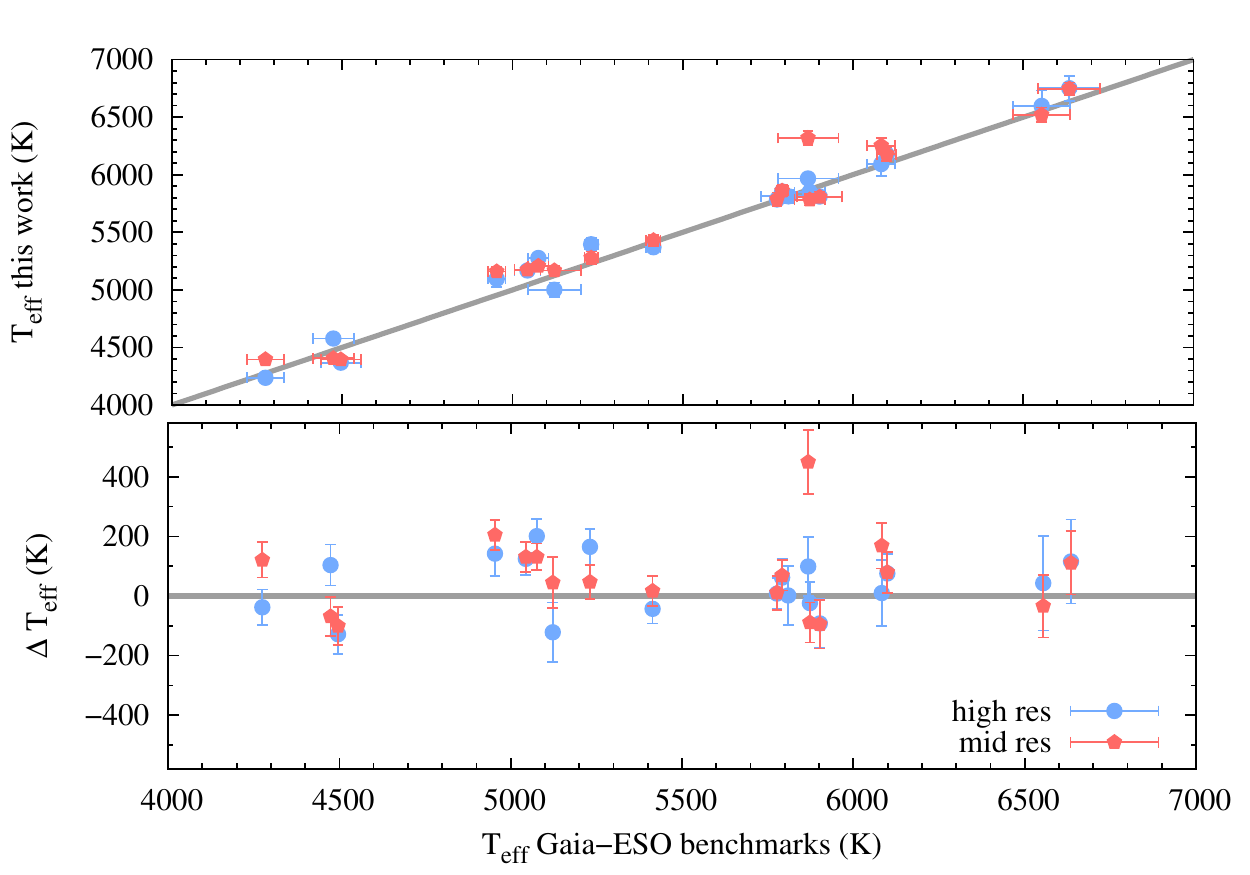} \\
   \includegraphics[width=1.05\linewidth]{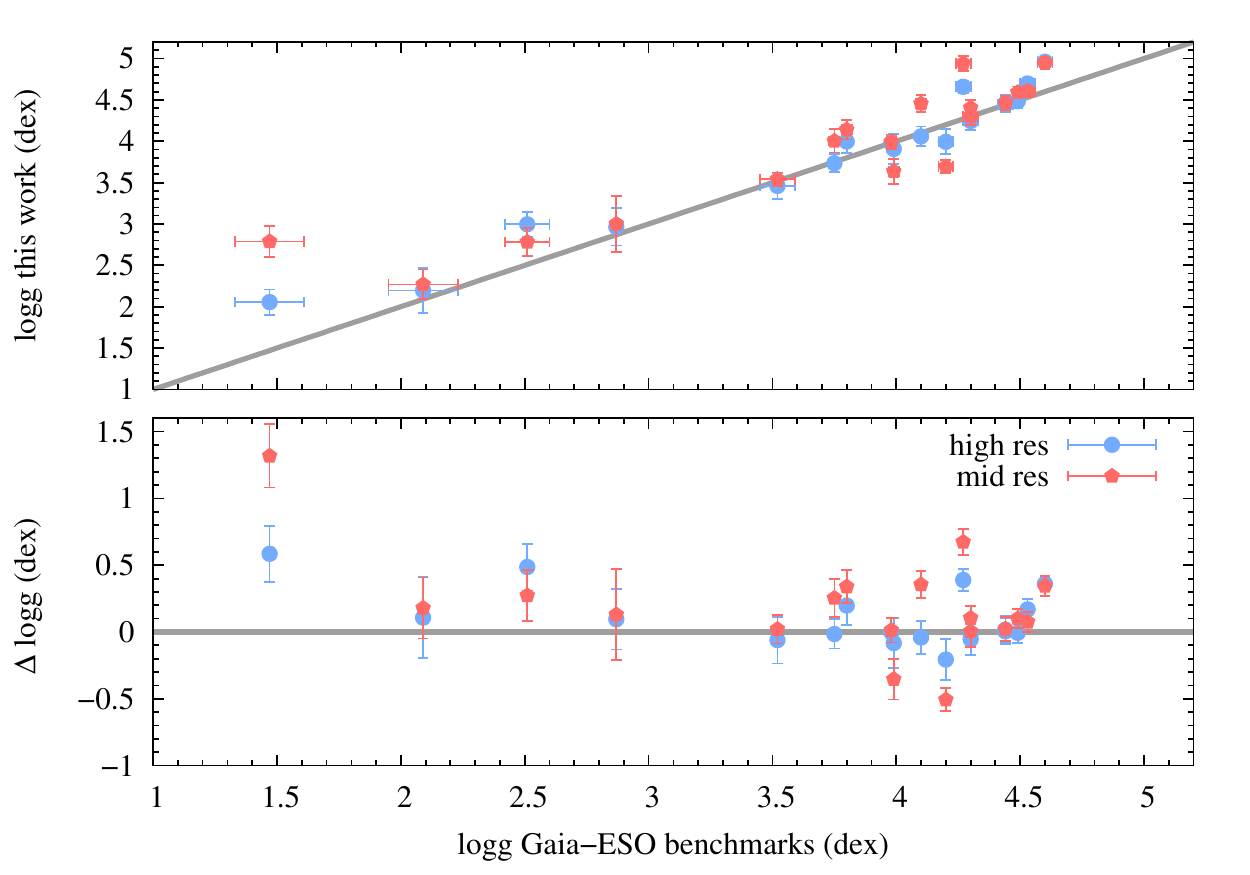} \\
   \includegraphics[width=1.05\linewidth]{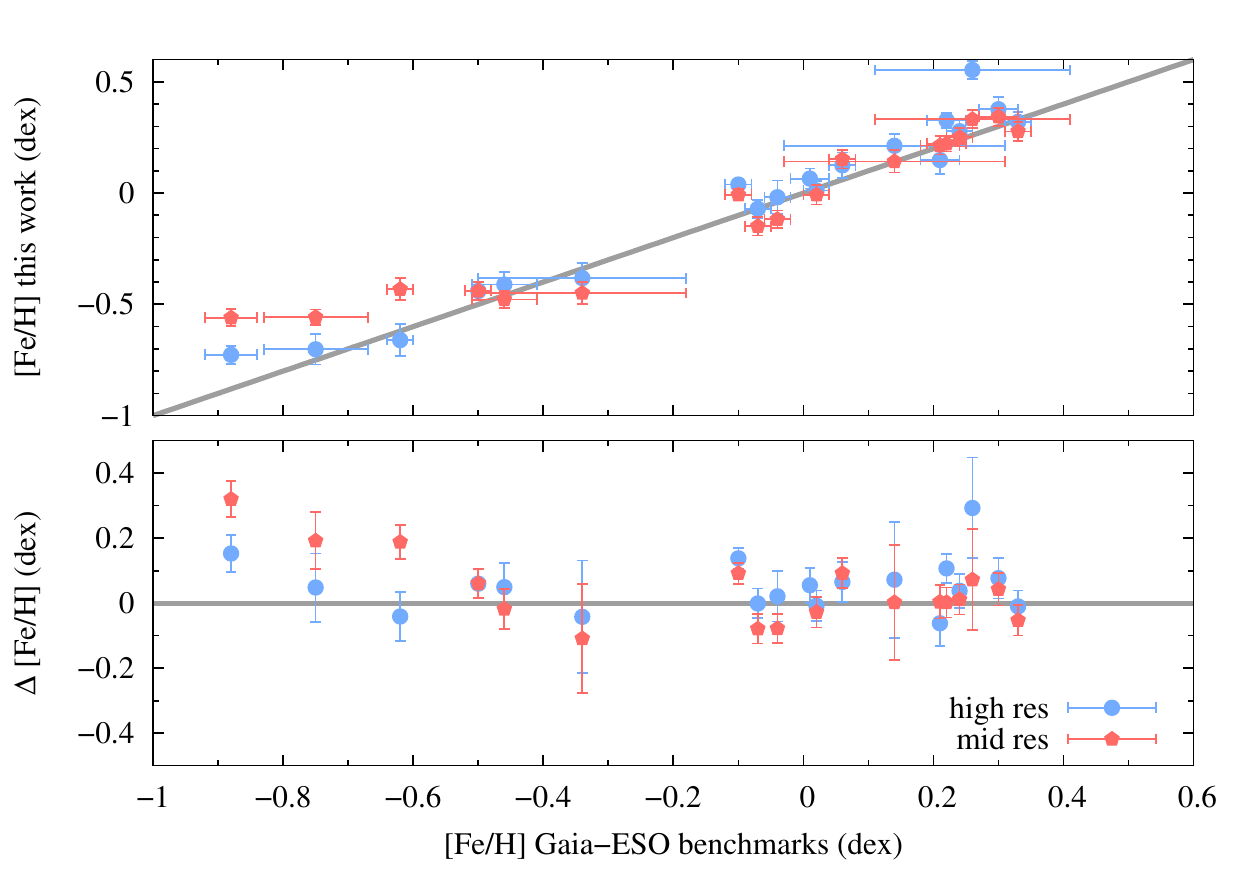}
  \caption{Differences in effective temperature, surface gravity and metallicity between the Gaia FGK benchmark parameters and this work using the HR15n GIRAFFE spectra for the ATLAS-APOGEE models. For 
  comparison we plot the parameters of high resolution spectra using the same line list.}
  \label{mid_res_hr15n_apogee}
  \end{figure}

\begin{figure}
  \centering
   \includegraphics[width=1.05\linewidth]{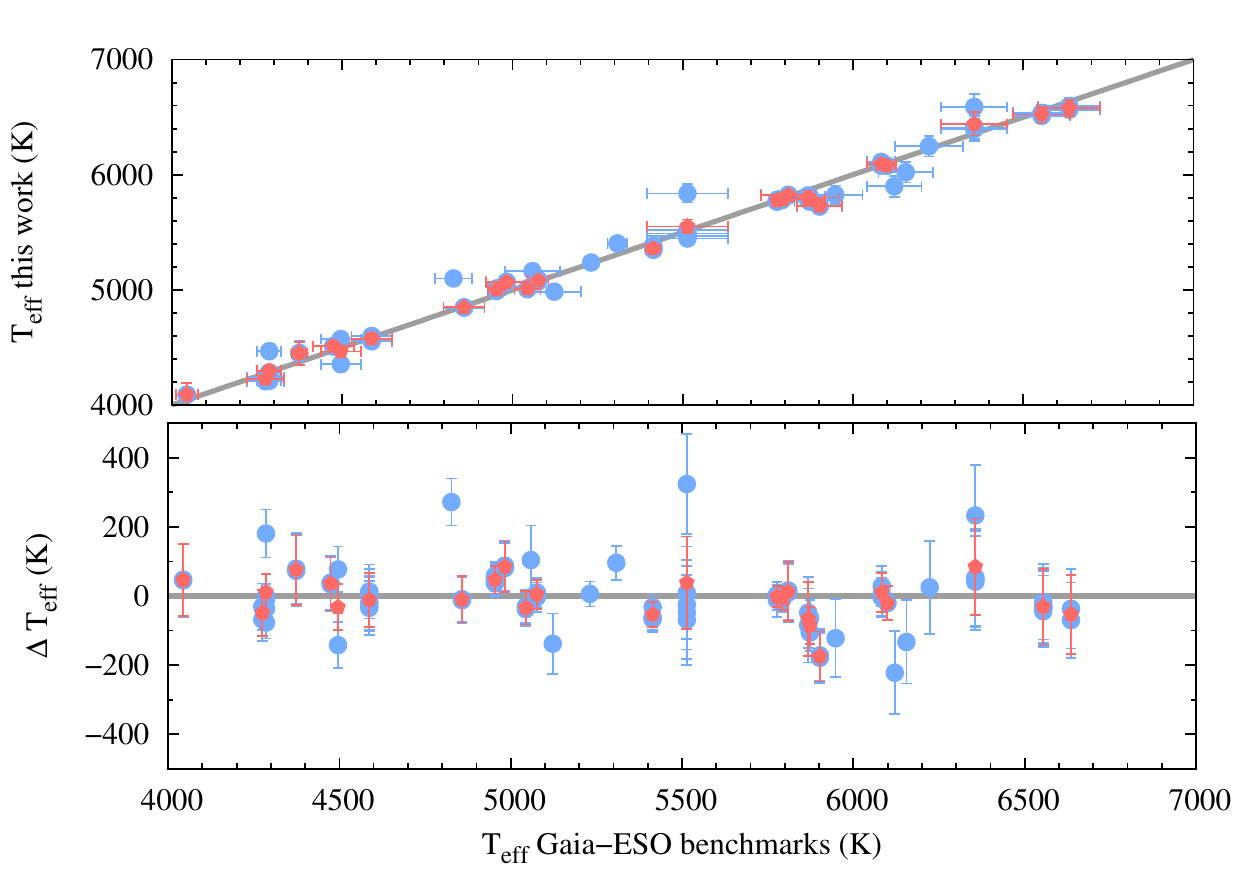} \\
   \includegraphics[width=1.05\linewidth]{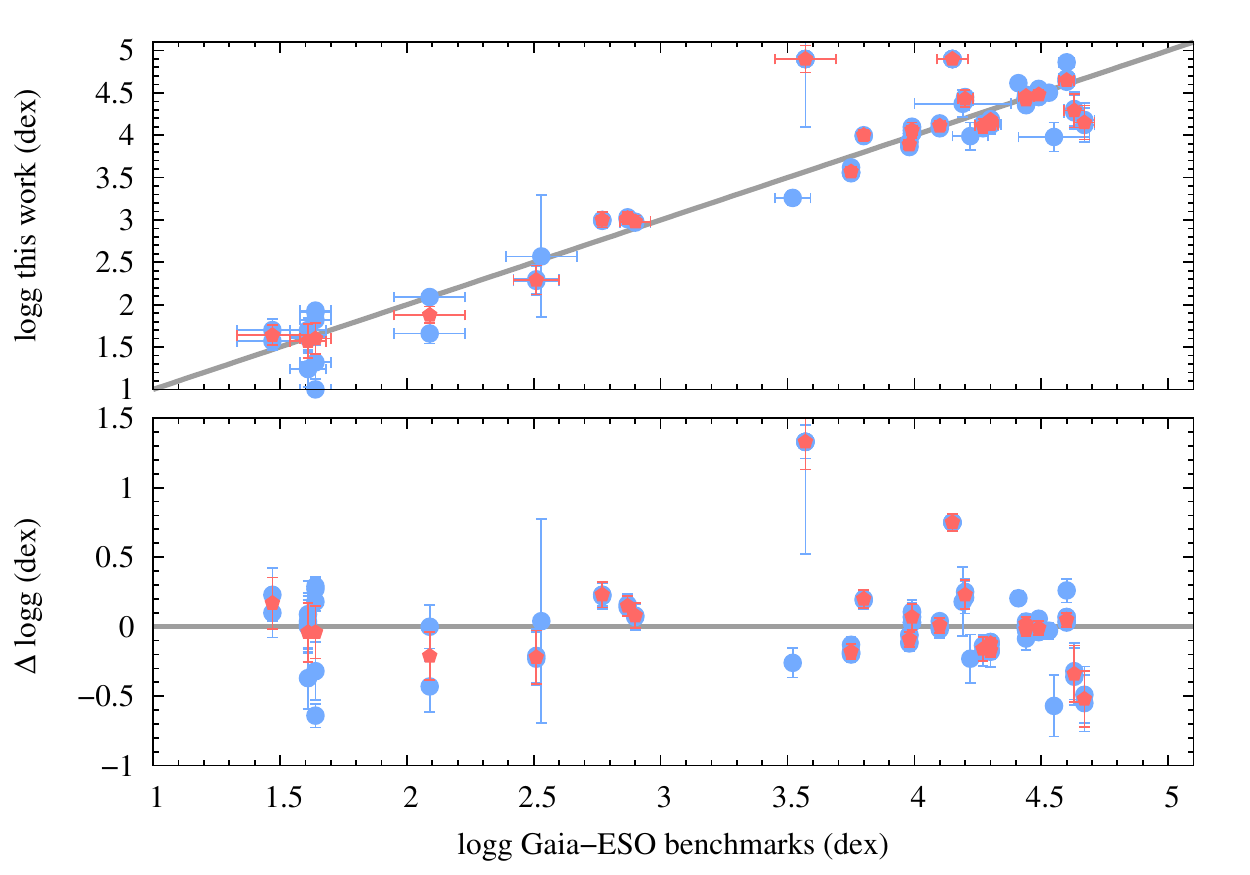} \\
   \includegraphics[width=1.05\linewidth]{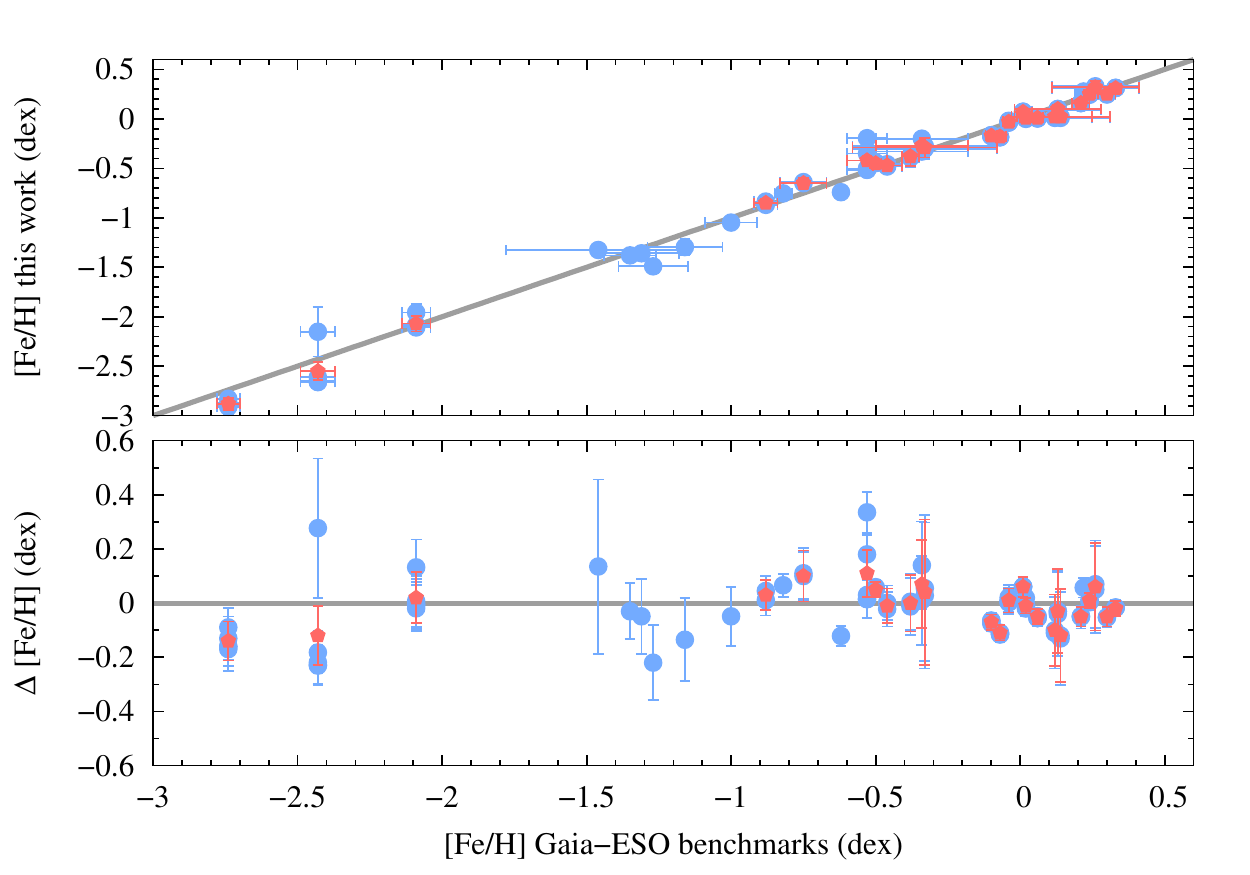}
  \caption{Differences in effective temperature, surface gravity and metallicity between the GES benchmark parameters and this work using the high resolution spectra and MARCS models 
  (blue squares). For stars with multiple spectra, their mean values are plotted with green circles. }
  \label{high_res_marcs}
  \end{figure}
\begin{figure}
  \centering
   \includegraphics[width=1.05\linewidth]{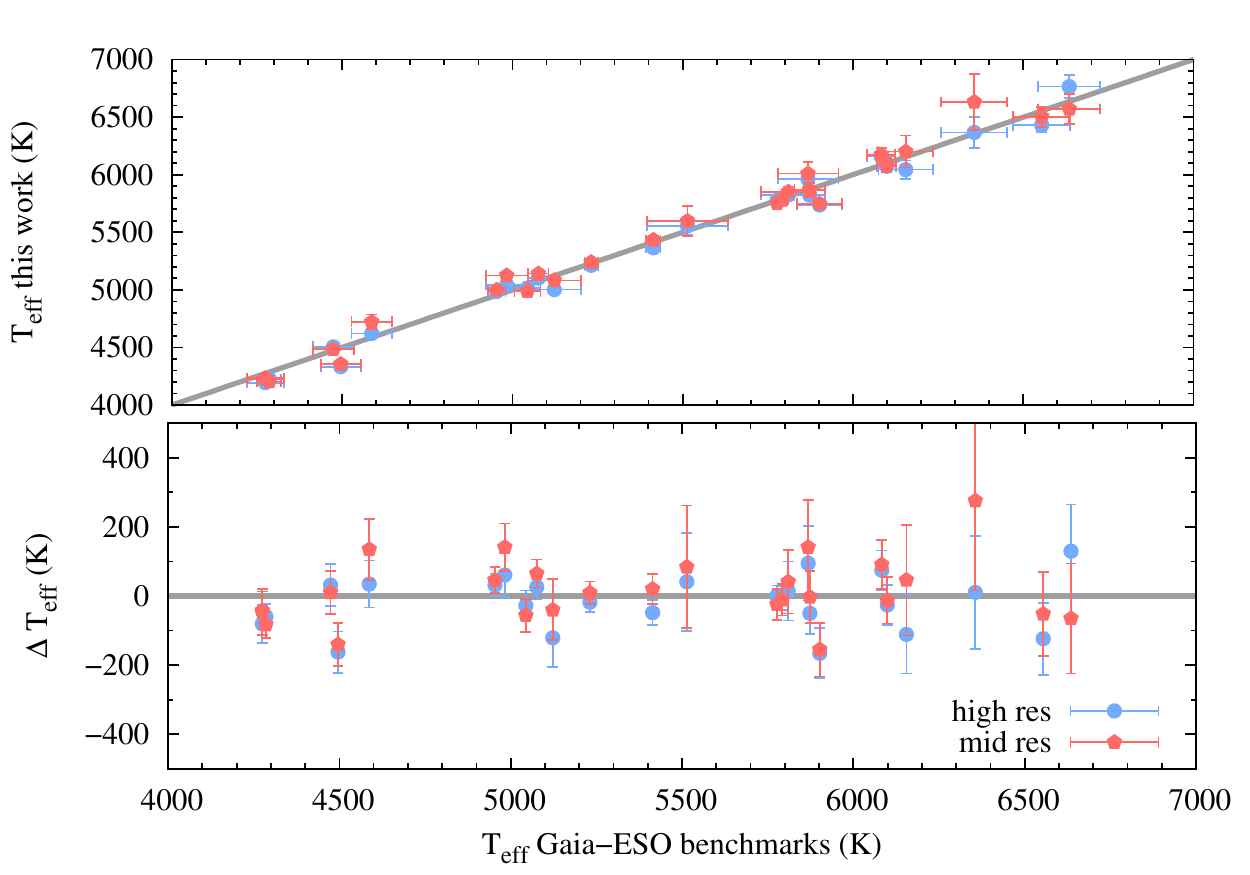} \\
   \includegraphics[width=1.05\linewidth]{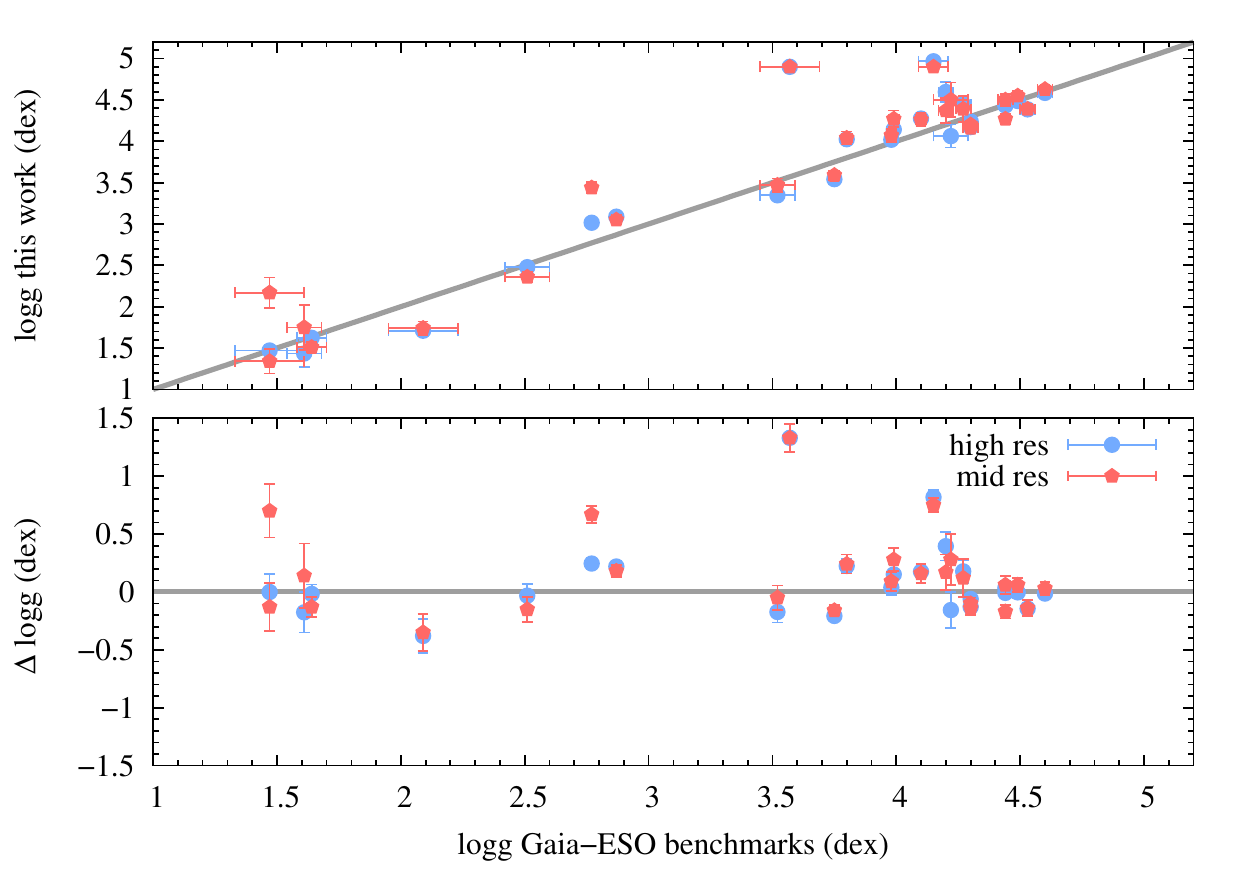} \\
   \includegraphics[width=1.05\linewidth]{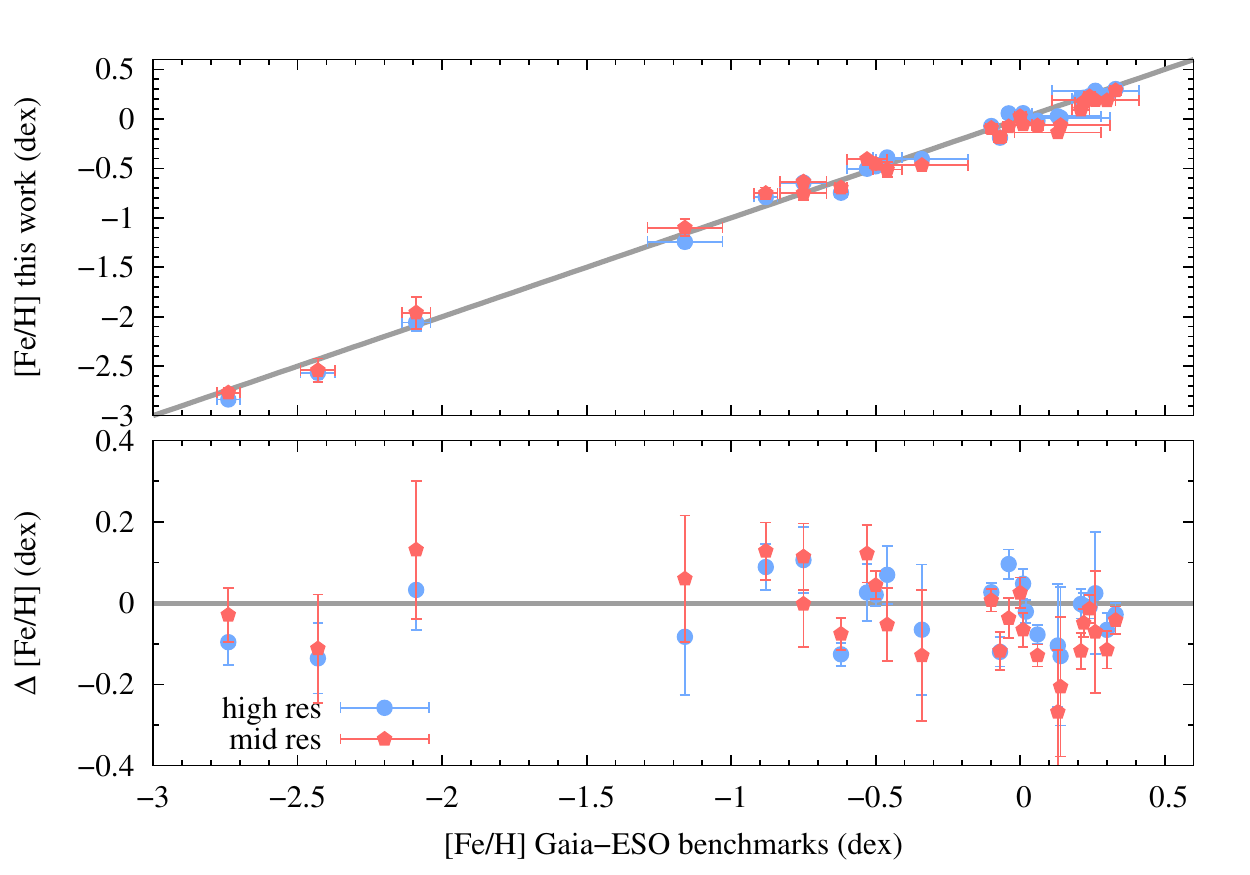}
  \caption{Differences in effective temperature, surface gravity and metallicity between the Gaia FGK benchmark parameters and this work using the HR10 GIRAFFE spectra for the MARCS models. For 
  comparison we plot the parameters of high resolution spectra using the same line list.}
  \label{mid_res_hr10_marcs}
  \end{figure}

\begin{figure}
  \centering
   \includegraphics[width=1.05\linewidth]{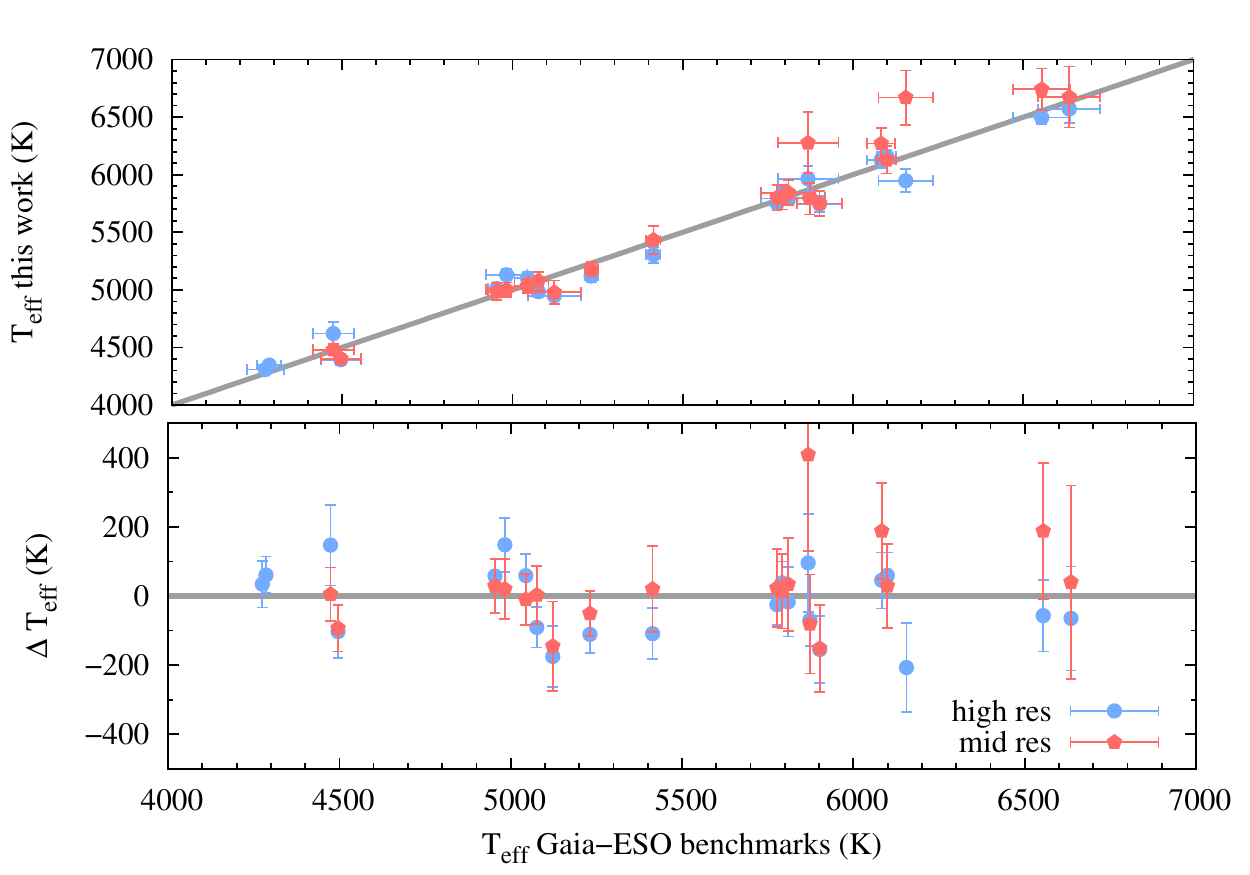} \\
   \includegraphics[width=1.05\linewidth]{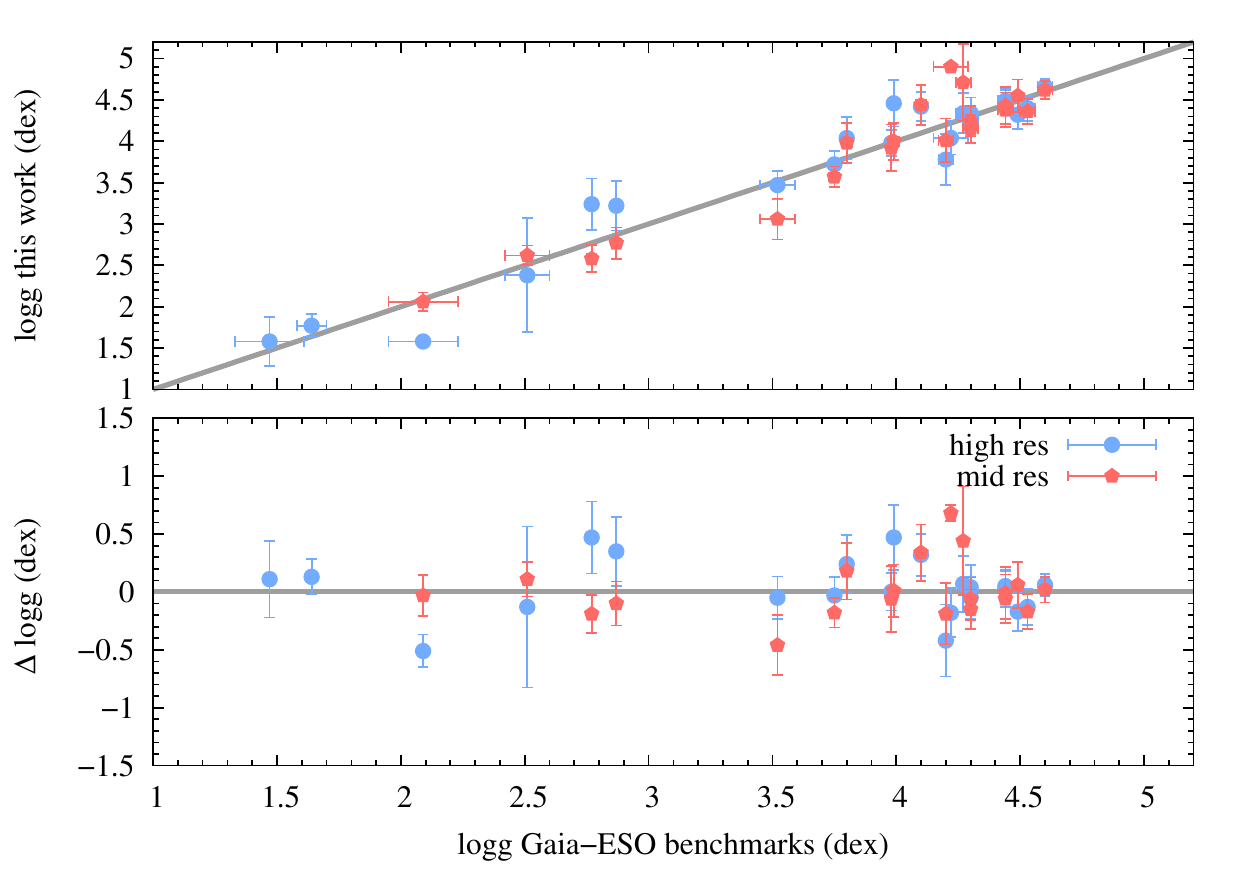} \\
   \includegraphics[width=1.05\linewidth]{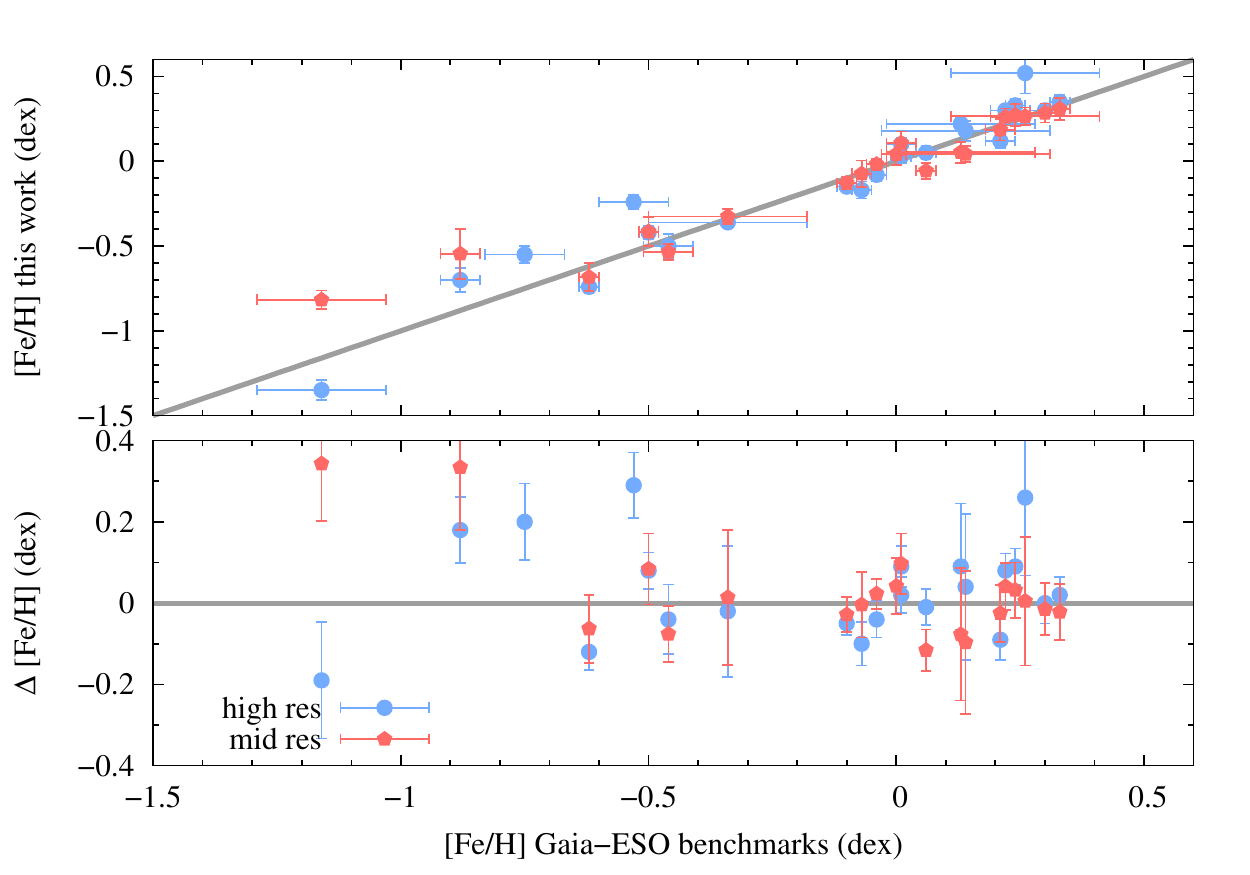}
  \caption{Differences in effective temperature, surface gravity and metallicity between the Gaia FGK benchmark parameters and this work using the HR15n GIRAFFE spectra for the MARCS models For 
  comparison we plot the parameters of high resolution spectra using the same line list.}
  \label{mid_res_hr15n_marcs}
  \end{figure}


\bsp	
\label{lastpage}
\end{document}